\documentclass[]{interact}

\usepackage{epstopdf}
\usepackage{subfig}
\usepackage{graphicx}
\usepackage{adjustbox}

\usepackage[natbibapa,nodoi]{apacite}
\theoremstyle{plain}

\theoremstyle{definition}

\theoremstyle{remark}

\newcommand\Real{\mbox{Re}} 
\newcommand\Da{\mbox{\textit{Da}}}   
\newcommand\Ka{\mbox{\textit{Ka}}}   
\newcommand{\wt}[1]{\widetilde{#1}}
\newcommand\etal{\mbox{\textit{et al.\ }}}

\newcommand\eg{e.g.\ }

\begin{document}

\articletype{ARTICLE TEMPLATE}

\title{Predictive data-driven model based on generative adversarial network for premixed turbulence-combustion regimes}

\author{
\name{T. Grenga\textsuperscript{a}\thanks{CONTACT T. Grenga. Email: t.grenga@itv.rwth-aachen.de} and L. Nista\textsuperscript{a} and C. Schumann\textsuperscript{a} and A.~N. Karimi\textsuperscript{a} and G. Scialabba\textsuperscript{a} and A. Attili\textsuperscript{b} and H. Pitsch\textsuperscript{a}}
\affil{\textsuperscript{a} Institute for Combustion Technology, RWTH Aachen University, Aachen, 52056, Germany; \\ \textsuperscript{b} Institute for Multiscale Thermofluids, School of Engineering, University of Edinburgh, Edinburgh, EH9 3FD, United Kingdom}}

\maketitle

\begin{abstract}
Premixed flames exhibit different asymptotic regimes of interaction between heat release and turbulence depending on their respective length scales. 
At high Karlovitz number, the dilatation caused by heat release does not have any relevant effect on turbulent kinetic energy with respect to non-reacting flow, while at low Karlovitz number, the mean shear is a sink of turbulent kinetic energy, and counter-gradient transport is observed. 
This latter phenomenon is not well captured by closure models commonly used in Large Eddy Simulations that are based on gradient diffusion.
The massive amount of data available from Direct Numerical Simulation (DNS) opens the possibility to develop data-driven models able to represent physical mechanisms and non-linear features present in both these regimes.
In this work, the databases are formed by DNSs of two planar hydrogen/air flames at different Karlovitz numbers corresponding to the two asymptotic regimes.
In this context, the Generative Adversarial Network (GAN) gives the possibility to successfully recognize and reconstruct both gradient and counter-gradient phenomena if trained with databases where both regimes are included.
Two GAN models were first trained each for a specific Karlovitz number and tested using the same dataset in order to verify the capability of the models to learn the features of a single asymptotic regime and assess its accuracy. In both cases, the GAN models were able to reconstruct the Reynolds stress subfilter scales accurately. 
Later, the GAN was trained with a mixture of both datasets to create a model containing physical knowledge of both combustion regimes. This model was able to reconstruct the subfilter scales for both cases capturing the interaction between heat release and turbulence closely to the DNS as shown from the turbulent kinetic budget and barycentric maps.

\end{abstract}


\section{Introduction}
The pathway to carbon-neutral energy systems, as indicated by the European Union~\citeyearpar{EUgreen} and United Nations~\citeyearpar{UNgreen}, necessarily includes the use of hydrogen and other e-fuels, such as ammonia, oxymethylene ethers, and dimethyl ether, as energy carriers to connect locations where renewable energies are available with areas where final consumers are present. 
Thus, combustion will remain one of the main components in energy systems for the coming decades.

Numerical modeling of turbulent combustion through Reynolds Averaged Numerical Simulations (RANS) or Large Eddy Simulations (LES) is essential for computational design of future efficient thermo-chemical energy conversion systems. 
As the most popular closure models for turbulent transport have been developed for non-reacting flows~\citep{zhang1995premixed}, often they are not able to represent the complex interaction between heat release and turbulence.

\subsection{Turbulent Combustion Modeling}\label{class}

The reciprocal actions between heat release and turbulence have been investigated, and different asymptotic regimes in premixed combustion have been determined by the relation between heat release and turbulence length scales.
According to Bilger \etal~\citeyearpar{bilger2004}, heat release affects turbulence when dilatation is larger than the small-scale turbulence-induced strain.
In other terms, the dilatation effects are expected to be important when the length scales of the heat release (a.k.a. the flame) are smaller than the smallest length scales of the turbulence. Similar scaling arguments have been proposed by Veynante \etal~\citep{veynante1997gradient}.
This relation can be expressed as
\begin{equation}
\left(\frac{\epsilon}{\nu}\right)^{1/2} \ll \left(\frac{\rho_u}{\rho_b}-1\right) \frac{S_L}{\delta_F},
\label{scales_rel}
\end{equation}
where $\epsilon$ is the turbulent kinetic energy dissipation rate, $\nu$ is the kinematic viscosity of the fluid, $\rho_u$ and $\rho_b$ are the densities of the unburned reactants and burned products, respectively, $\delta_F$ is the thickness of the laminar premixed flame, and $S_L$ its burning velocity. 
The quantity on the right-hand side of Eq.~\ref{scales_rel} is the order of the dilatation due to heat release, while the one on the left-hand side is the order of the turbulence-induced strain rate.
The Karlovitz number, $Ka$, relates the small scales of turbulence to the flame scales:
\begin{equation}
Ka = \frac{\delta_F}{S_L}\left(\frac{\epsilon}{\nu}\right)^{1/2}.
\label{ka_def}
\end{equation}  
A critical Karlovitz number~\citep{macart17a} can be defined as,
\begin{equation}
Ka_{cr} \equiv \frac{\rho_u}{\rho_b} -1 \gg \frac{\delta_F}{S_L}\left(\frac{\epsilon}{\nu}\right)^{1/2} = Ka,
\label{kacr}
\end{equation}
such that the dilatation effects are expected to be important when the $Ka$ number is less than $Ka_{cr}$.


In order to validate these theoretical scaling arguments as well as to obtain deeper insight into the physical phenomenon, a number of Direct Numerical Simulations (DNS) studies have been conducted for turbulent premixed combustion. 
In this view, several authors \citep{zhang1995premixed, o2017cross} have shown that, at low Karlovitz number, the primary source of turbulent kinetic energy is pressure-dilatation; conversely, at high Karlovitz numbers, the dilatation becomes weaker, and, as in non-reacting flows, the primary source of turbulent kinetic energy becomes the production from large-scale strain~\citep{aspden2011turbulence}. 
Similarly, the spectrum of turbulent kinetic energy has been shown to relate with flame scales rather than turbulence scales at low Karlovitz number~\citep{kolla2014velocity}, while vorticity behaves as in non-reacting flows in turbulent premixed flames at high Karlovitz number~\citep{bobbitt2016vorticity}. 
Analogous trends have also been observed for the scalar variance, that is, the dominance of production from chemistry at low Karlovitz number and the dominance of production from the large-scale scalar gradient at high Karlovitz number~\citep{rogerson2007analysis}.

The analyses of the alignment of the flame normal with the eigenvectors of the strain rate tensor have highlighted further effects of dilatation from heat release on turbulence in premixed flames. 
At low Karlovitz number, the orientation of the flame-normal reverses with respect to eigenvectors of the strain rate tensor and becomes aligned with the eigenvector corresponding to the most extensive eigenvalue of the strain rate tensor~\citep{hamlington2011interactions, steinberg2012statistics}.
On the contrary, at a high Karlovitz number, the flame-normal vector is aligned with the eigenvector of the strain rate tensor corresponding to the most compressive eigenvalue~\citep{hamlington2011interactions, wang2016turbulence}, similar to the alignment of scalar iso-surfaces with eigenvectors of the strain rate in non-reacting flows~\citep{batchelor1952effect}. 

Although the aforementioned DNS studies have provided tremendous insights into some of the phenomenological effects of combustion heat release on turbulence, identifying the asymptotic regimes in which dilatation effects on turbulence are expected to be important, the transfer of that knowledge into models is still missing~\citep{BRAY198587}. 
Indeed, the current capabilities of LES and RANS premixed combustion models do not account for such interactions on both filtered and averaged velocity and scalars fields.

In non-reacting flows, the most common approach for modeling the subfilter scalar flux is the so-called gradient-transport model, in which the subfilter scalar flux is assumed to be aligned with the gradient of the filtered scalar~\citep{moin1991dynamic}.
The subfilter scalar flux is then  given by $- C \overline{\rho} \Delta^2 |\widetilde{S} | \frac{\partial \widetilde{Y}_k }{\partial x_i}$, where the constant $C$ is either specified or determined dynamically by using the information at the smallest resolved scales~\citep{moin1991dynamic, germano1991dynamic, lilly1992proposed}, $\Delta$ is the LES filter size and $\widetilde{S}$ is the magnitude of the Favre-filtered strain rate (note that all Favre-filtered quantities are denoted as $\tilde{\cdot}$ ).
However, in turbulent premixed flames at low $Ka$, counter-gradient-transport, initially predicted theoretically by Libby and Bray~\citeyearpar{libby1981countergradient}, has been observed in a wide variety of both experimental and DNS studies~\citep{lipatnikov2010effects}. 
In the flame normal direction, the flow velocity accelerates due to thermal expansion, so scalars with a positive gradient across the flame have a positive scalar flux and vice versa. 
For an infinitely thin flame (i.e., $Ka \approx 0$)~\citep{libby1981countergradient, bray1981turbulence}, a rigorous model for this counter-gradient-transport is given by
\begin{equation}
\overline{\rho} \widetilde{u_i Y_k} - \overline{\rho} \widetilde{u}_i \widetilde{Y}_k = - \overline{\rho} \frac{ \left(Y_{k,b} -  \widetilde{Y}_k \right)  \left( \widetilde{Y}_k - Y_{k,u} \right)}{Y_{k,b} -Y_{k,u}} \tau S_L n_i
\label{subfilter2}
\end{equation}
where $n_i$ is the flame normal vector oriented from unburned to burned and $Y_{k,b}$ and $Y_{k,u}$ are the scalar values in the burned and unburned gases, respectively. 
To model flames at finite $Ka$, various attempts have been made to linearly combine the gradient- and counter-gradient-transport models with and without an additional coefficient, in both LES and RANS~\citep{veynante1997gradient, fureby2005fractal}.
However, such hybrid models have not been shown to be general because they are essentially an empirical concatenation of models from gradient-transport and counter-gradient-transport limits rather than a physics-based model that truly merges the presumably nonlinear physics of the two extreme regimes~\citep{lipatnikov2010effects}.

A model similar to Eq.~\ref{subfilter2} can be derived for the subfilter stresses in the limit of an infinitely thin flame~\citep{BRAY198587}
\begin{equation}
\overline{\rho} \widetilde{u_i u_j}  - \overline{\rho} \widetilde{u}_i\widetilde{u}_j = \overline{\rho} \widetilde{C} \left( 1 - \widetilde{C} \right) \left( \tau S_L \right)^2 n_i n_j,
\label{13}
\end{equation}
where $\widetilde{C}$ defines a reaction progress variable defined to be zero in the unburned gases and unity in the burned gases. 
However, the model has been shown to be inadequate for statistically non-planar flames and turbulent premixed jet flames~\citep{lipatnikov2010effects, driscoll1988measurement}. 
This model for the subfilter stresses, even when combined with a Smagorinsky model, fails to capture the combined effects of shear and dilatation for finite thickness flames in turbulent premixed flames. As a result of this model failure, a common approach to modeling turbulent premixed flames in LES is to utilize a model for counter-gradient-transport for the subfilter scalar flux with a pure Smagorinsky model for the subfilter stresses~\citep{BRAY198587}, although it leads to a physical inconsistency.

A further fundamental challenge in LES modeling is the presence of backscattering.
This phenomenon may be found also in non-reacting flow, but it becomes statistically relevant for reacting flows.
Indeed, heat release occurring at the scales of the flame, which are unresolved in LES, affects the large-scale, resolved turbulence~\citep{o2017cross}.
Such processes transfer energy from small to large scales, in a backward energy cascade, which is different from the view of the classical turbulence energy cascade~\citep{richardson2007weather} transferring energy from large to small scales.
In reacting flows, when dilatation effects on turbulence are important, the backward cascade is statistically relevant~\citep{o2017cross, OBrien2014} and requires specialized models.
This challenge is far more difficult for backward cascade problems since the physics affecting the resolved scales are completely unresolved. 
There exist empirical-statistical models~\citep{leith1990stochastic, mason1992stochastic, schumann1995stochastic}, in which heat release effects are correlated with resolved quantities, although this is devoid of any real physics. 
Data-driven approaches and Machine Learning (ML) algorithms present the capabilities to overcome this challenge because the features of the physical phenomena to be modeled have been previously \emph{learned} by the artificial neural network (ANN). 
The Generative Adversarial Network (GAN) algorithm used in this study, will not create a model to solve the turbulent and combustion closure problems for directly estimating the subfilter quantities. 
Conversely, it will reconstruct the unresolved data on the basis of resolved filtered (or rather LES) fields and the DNS datasets used for the training, as described in the following sections.
Subsequently, the unclosed subfilter terms will be evaluated from their definitions.

\subsection{Machine Learning for Turbulence and Combustion Closure Modeling}

Several researchers have exploited the use of ANN in combustion modeling, initially restricted to chemical kinetics and later extended to turbulent combustion closure. 
Flemming \etal~\citeyearpar{flemming2005using} and later Ihme \etal \citeyearpar{ihme2009optimal} used ANN for the representation of flamelet tables. 
The considered - rather simple - ML framework was not suitable to represent the strongly non-linear character of high-temperature reacting flows with sufficient accuracy~\citep{ihme2008generation}, but was shown to improve the high-frequency part of the spectrum of the chemical source term because of the smooth representation compared to linear interpolation in a conventional flamelet table~\citep{IHME20091545}. 
As a result, it led to more accurate predictions of direct combustion noise compared with conventional flamelet table approaches.
In order to improve and optimize the choice of meta-parameters of the ANN, Ihme \etal \citeyearpar{Ihme_NC} also devised a surrogate-based optimization method demonstrating that the optimal structure of the ANN strongly depends on represented quantity. 

For applications in turbulent combustion, Convolutional Neural Networks (CNNs) are interesting because they introduce the notion of parameter sharing: instead of having to learn the relationships between input and output everywhere separately, CNNs learn spatial features, which is very useful for representing turbulence ~\citep{lapeyre2019training}.

CNNs were successfully employed to model the flame surface density~\citep{lapeyre2019training} using as inputs the filtered progress variable values. Moreover, CNNs were used to solve the deconvolution problem~\citep{nikolaou2018modelling, nikolaou2019progress}, e.g. they were trained to estimate the unfiltered progress variable field from the knowledge of the filtered field and model the scalar variance. A CNN, trained with a finely-resolved LES database, was also used to model both the source and scalar flux terms of the filtered scalar transport equation ~\citep{seltz2019direct}.

Reconstructing the fully-resolved flow or subfilter quantities from large-scale or coarse-grained data has significant applications in various domains, \eg particle image velocimetry measurements~\citep{cao2000piv}, or  LES for weather predictions~\citep{rotunno2009large}, where deep learning networks can play a significant role in the representation of complex non-linear relations.
The reconstruction of subfilter information with deep learning networks is a promising approach to link the large-scale results obtained from experiments or filtered equations to the actual flow fields.

In recent work, a GAN architecture has been used to reconstruct fully-resolved 2-D and 3-D velocity fields from filtered data, aiming to close the filtered Navier-Stokes equations.
Deng \etal~\citeyearpar{deng2019super} applied a similar network structure, the super-resolution GAN (SRGAN)~\citep{SRGAN} and enhanced SRGAN (ESRGAN)~\citep{ESRGAN} to reconstruct the wake flow around a single-cylinder in 2-D. To generate high-resolution (HR) data for training, PIV measurements were performed, while low-resolution (LR) data were provided by means of bicubic downsampling. 
 
For LES simulations, the super-resolution approach could be used to reconstruct unresolved high-frequency data, and thus provide a means to close the filtered Navier-Stokes equations. Indeed, closure modeling is recognized as a key application of ML in the fluid dynamics community~\citep{Brenner2019}.

Fukami \etal~\citeyearpar{fukami2019super} were perhaps the first to apply a deep-learning super-resolution approach to 2-D decaying isotropic turbulence. Their CNN architectures outperformed bicubic interpolation in perceptive quality, as this classical interpolation algorithm yielded overly smooth fields, especially at higher upsampling factors. 

Liu \etal~\citeyearpar{Liu2020} compared two deep-learning frameworks with bicubic interpolation in 2-D in the context of forced isotropic turbulence and wall-bound turbulent channel flows taken from the Johns-Hopkins turbulence database (JHTDB)~\citep{JHTDB_1, JHTDB_2}. 
With respect to the energy spectrum, both networks were able to improve upon the bicubic interpolation, but the one that next to the spatial also considered temporal data was able to reconstruct accurately to a higher wavenumber. 

Pant \etal~\citeyearpar{pant2020deep} trained a deep CNN for the task of super-resolution of forced isotropic turbulence in 2-D, arguing that it is computationally a less burdensome network than SRGAN or ESRGAN. 
The trained model was able to improve the peak signal to noise ratio and structural similarity index measure by a small amount, while there was a greater improvement to be seen at larger filter sizes. Conversely, turbulent kinetic energy, turbulent velocity distribution, and the probability density function (PDF) of the vorticity were better predicted at smaller filter sizes.

Kong \etal~\citeyearpar{Kong2020} applied two super-resolution models to a 2-D temperature field of supersonic combustion. High-resolution data were obtained from a 3-D RANS simulation. They employed a standard SRCNN and a multiple path super-resolution CNN (MPSRC). 

Subramaniam \etal~\citeyearpar{subramaniam2020turbulence} argued that super-resolution solutions may not abide the physical laws and hence proposed physically founded loss functions. In particular, they investigated a 3-D CNN and an SRGAN-based GAN solution with a physics loss that is based on the residuals of the continuity and pressure Poisson equations. The models were trained on DNS data of forced homogeneous isotropic turbulence. 
The GAN outperformed the CNN as the physics loss converged to a lower value for the GAN. 
The ML solutions' energy spectra were found to be consistent with the DNS beyond the cut-off wavenumber but diverged to lower values than the DNS and eventually overpredicted the energy density with increasing wavenumber. 

Starting from a similar observation, Kim \etal~\citeyearpar{Kim2021} proposed unsupervised learning with a 2-D cycle-consistent Wasserstein GAN (cycleGAN) with gradient penalty, that is trained without matched DNS labels. HR data from DNSs of the JHTDB were used as reference, and LR data were obtained by applying a top-hat filter. Forced HIT and turbulent channel flow were considered. In comparison with other models, only the GANs were able to add small scales, high wavenumber features at higher upscaling factors, as was demonstrated by the velocity and vorticity fields, the PDF of vorticity, and energy spectra. 

Bode \etal~\citeyearpar{bode2021using} applied a physics-inspired GAN network to 3-D HIT. 
Their network is based on a non-upsampling ESGRAN \citep{ESRGAN} but extended to 3-D fields and with a loss function including gradients and residual of the continuity equation. DNS data were used as HR labels, whereas LR data was provided by means of Gaussian filtering with kernel size $64$. 
It was found that when trained on lower $Re_{\lambda}$ than the model was tested on, the network added insufficient features. 
However, when trained on a higher $Re_{\lambda}$ than evaluated on, the network added the desired quantity of small-scale features. Indeed, the model was able to precisely reconstruct the energy spectra up to very high wavenumbers and the residual of the continuity equation was only of order $10^{-8}$. When applied as an a-posteriori model in an LES simulation using the same filter size, there was a very good temporal agreement of turbulent kinetic energy and dissipation with the DNS solution. Even though the model was trained on HIT, when applied to reacting jet flow, it was able to predict fuel mass fraction successfully.



In summary, using deep learning to super resolve turbulent flow fields is a promising approach, consistently outperforming classical interpolation significantly. A sufficiently deep model must be used to learn turbulent features, though, typically comprised of residual blocks and multiple paths in combination with non-linear activation functions. 
The reason why the applications to turbulent combustion are sparse is likely due to the increased complexity associated with reacting flows.
When a GAN architecture is compared to traditional CNNs, the GAN performs typically better, thus it was selected for the present study.  
Furthermore, little attention has been devoted to the universality of the models proposed. 
However, to be applicable as an LES closure model, the model needs to perform well in a variety of physical regimes and thus requires generalization capabilities, which will be explored in this work for different Karlovitz number regimes.

\section{DNS Datasets}
\label{data}

Two spatially-developing turbulent premixed planar jet flames at $Re_{0} = 5000$ with different Karlovitz numbers~\citep{macart17a} were considered in this work.
They are composed of a central jet with bulk velocity $U_0$ and width $H_0$, which is separated by thin walls from coflow jets of bulk velocity $U_c$ and width $H_c$. 
The values are reported in Table~\ref{tbl.sim} along with flames and simulations parameters.
The inlet flow in the central jet was previously computed in a DNS of a fully-developed turbulent channel flow. 
For these cases, fully-developed laminar velocity profiles were specified at the inlet for the primary coflow jets.
A region of constant low velocity isolates the coflow jets from the domain boundaries.

\begin{table}
  \begin{center}
    \def~{\hphantom{0}}
  \begin{tabular}{l c c}
    \hline
    Case & $K1$ & $K2$ \\
    \hline
    $H_{0}$ (mm) & 4.32 & 1.08  \\
    $U_{0}$ (m/s) & 23.36 & 93.44 \\
    $H_{c}$ (mm) & 6.18 & 1.54  \\
    $U_{c}$ (m/s) & 6.02 & 24.11 \\
    $\Real_0$ & 5,000 & 5,000 \\
    $\Da_0$ & 0.99 & 0.06 \\
    $\Da_{\wt{C}=0.5}$ & 0.60 & 0.05 \\
    $\Ka_0$ & 10.9 & 43.5 \\
    $\Ka_{\wt{C}=0.5}$ & 2.6 & 32.0 \\
    $u'/s_L$ & 1.25 & 7.00 \\
    Domain ($x,y,z$) & $12H_0\times24H_0\times3H_0$ & $24H_0\times16H_0\times3H_0$ \\
    Grid size & $768\times586\times256$ & $1536\times576\times256$ \\
    \hline
  \end{tabular}
  \caption{Simulation parameters for the low- ($K1$) and high-Karlovitz number ($K2$) datasets. Karlovitz number is reported at the centerline and at $\widetilde{C}=0.5$ for axial location $x/H_0=3$.}
  \label{tbl.sim}
  \end{center}
\end{table}

The central jet consists of a gaseous mixture of hydrogen and oxygen at stoichiometric equivalence ratio, diluted 80.9 \% by mass with nitrogen, at $T_0 = 300$ K and $p_0 = 1$ atm. 
Equilibrium products of combustion of the same mixture issue from the coflow jets at $T_c = 2047.5$ K and $p_c = 1$ atm. 
A nine-species hydrogen chemical kinetic model \citep{davis2005optimized} was used.
For this mixture, the laminar flame thickness is $\delta_F=0.435$ mm and the laminar flame speed is $s_L=1.195$ m/s, from which the critical Karlovitz number is estimated to be $Ka_{cr}=6.7$.

\begin{figure}[tbh]
\begin{center}
\includegraphics[width =0.46\textwidth]{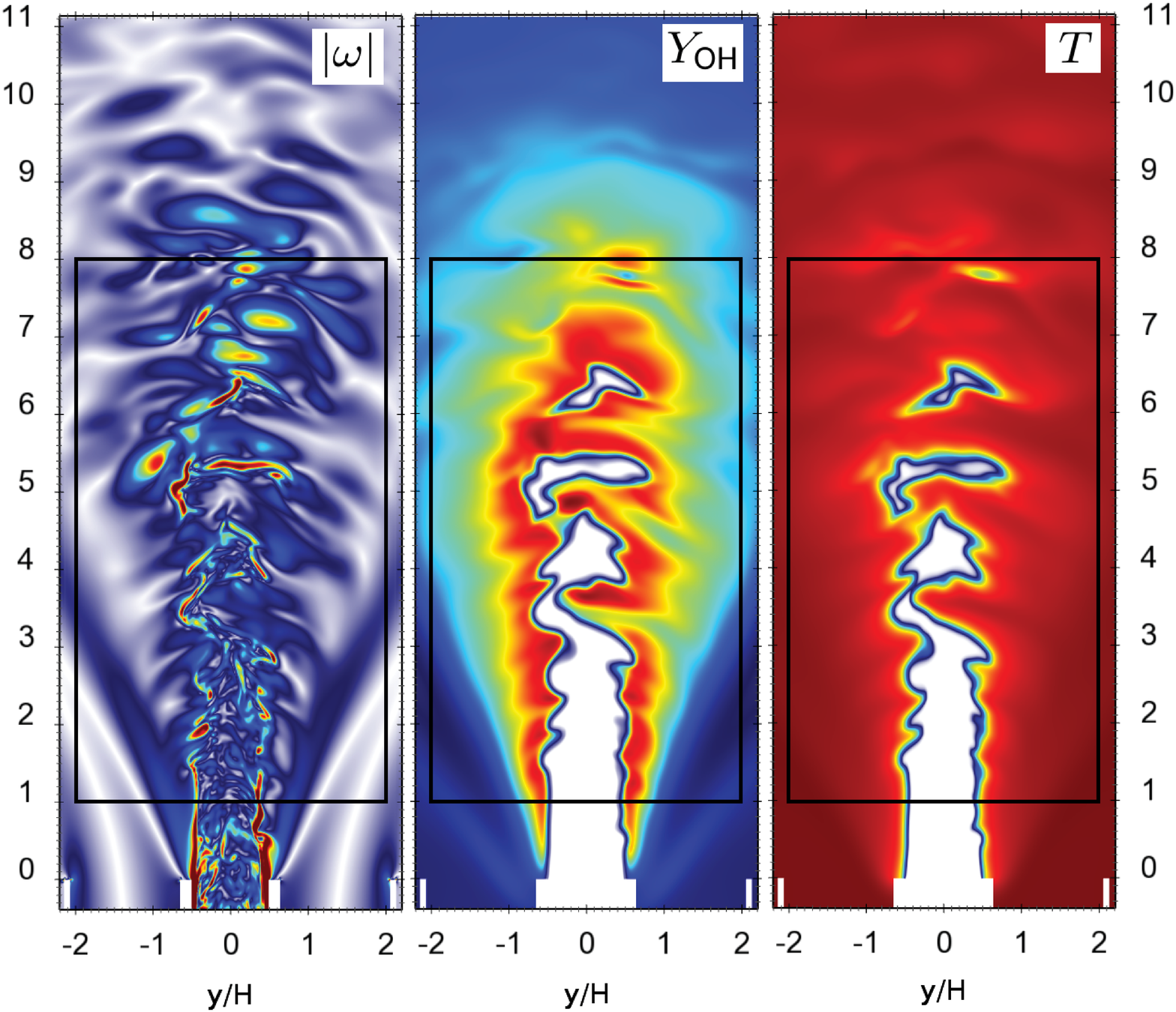}
\includegraphics[width =0.53\textwidth]{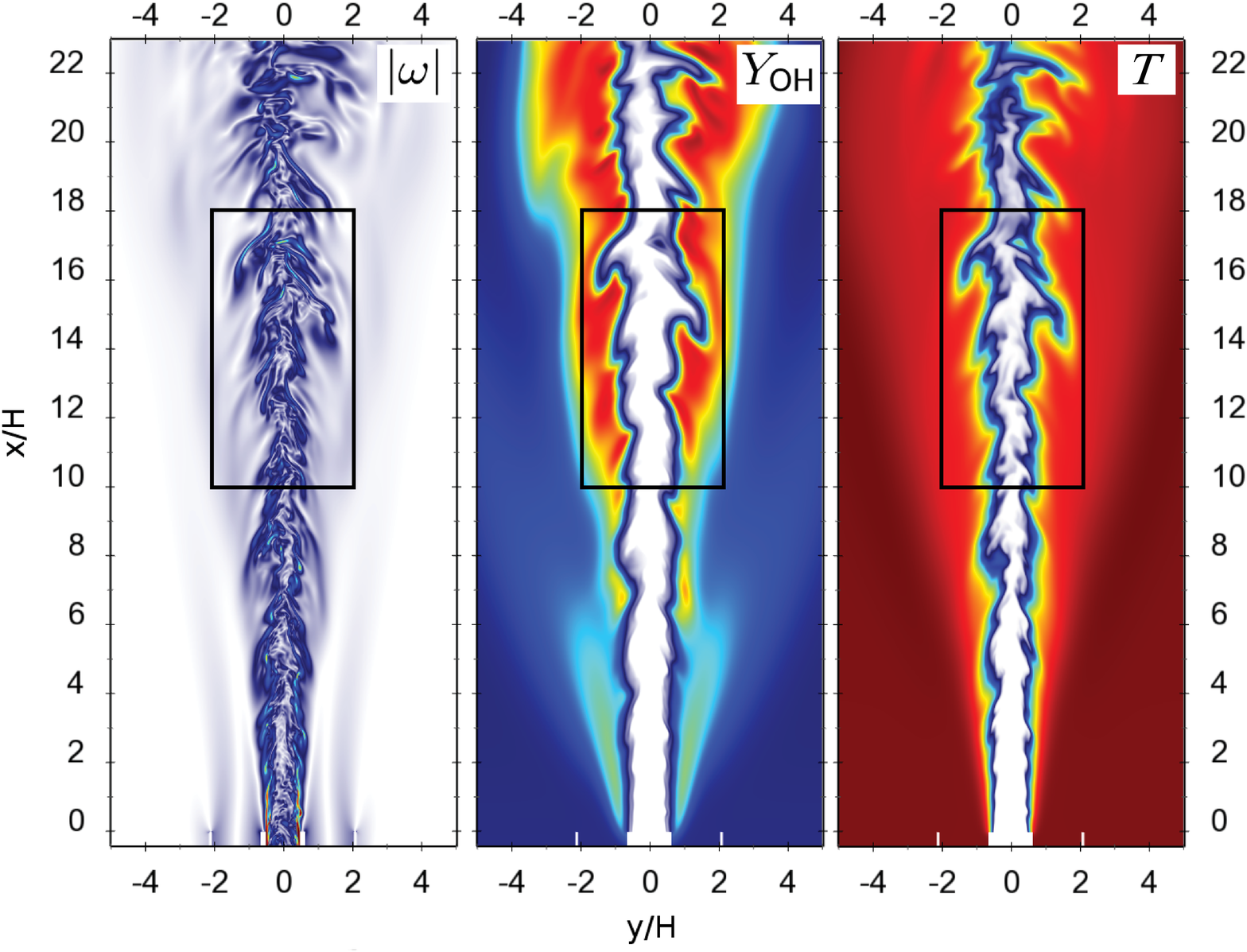}
\caption{Instantaneous images of vorticity, OH mass fraction, and temperature across a slice in the x-y plane for the cases $K1$ (left) and $K2$ (right). The black boxes indicate the sub-domains analyzed~\citep{grenga_dmdbook}.}
\label{field1}
\end{center}
\end{figure}

The two datasets used within this work feature one with a Karlovitz number below ($K1$) and one above ($K2$) the critical value. The bulk Reynolds number is kept constant, also the ratios $U_0/U_c$ and $H_0/H_c$ were kept fixed, while the Karlovitz number was varied by modifying the turbulence strain rate $U_0/H_0$. 
Further characterization of the turbulence statistics of both configurations may be found in~\citep{macart17a, macart17b, grenga2018dynamic, grenga_dmdbook}.
Figure \ref{field1} shows the instantaneous vorticity, OH mass fraction, and temperature fields for the x-y plane at the center of the domain, the rectangles represent the subdomains considered in the present work.

To generate the DNS database, the Navier-Stokes equations were solved applying the low-Mach number numerical formulation using a semi-implicit iterative algorithm by Desjardins \etal~\citeyearpar{Desjardins2008} implemented in the code NGA. The species equations were solved with a monolithic scheme using an approximately factorized exact Jacobian \citep{MacArt2016}. 

The $K1$ domain has dimensions $12H_0\times24H_0\times3H_0$ in the streamwise ($x$), cross-stream ($y$), and spanwise ($z$) directions, respectively. The computational grid has $768\times586\times256$ points.  The domain for $K2$ has dimensions $24H_0\times16H_0\times3H_0$, and a finer grid with $1536\times576\times256$ points. 
The boundary conditions are, in both cases, inflow on the $-x$ face, outflow on $+x$ face, free slip on $\pm y$ faces, and periodic in the $z$-direction. 

In the present work are used data from subdomains containing the core portion of the flames without the regions close to the nozzle and the burned gas on the side, where the heat release is limited.
These fully contain the interactions between heat release and turbulence, so that the size of the dataset is limited yet packed with meaningful quantities.
The subdomain considered for $K1$ has dimensions $7H_0\times4H_0\times H_0$, or rather $390\times316\times86$ grid points, in the $x$-, $y$-, and $z$-directions, respectively. 
The subdomain contains about $10.6\times10^6$ grid points, so each snapshot contains about $3.3 \times 10^7$ values, as only three variables (the three velocity components $u$, $v$, $w$) were considered. 
For the training of the GAN, $401$ different snapshots with a time spacing of $4 \mu$s have been used. 
Subsequently, about 12.7 billion data values or rather more than 300 GB were used to train, test, and verify the GAN with the $K1$ dataset. 

In $K2$, the subdomain considered has dimensions $8H_0\times4H_0\times H_0$, or rather $454\times310\times86$ grid points, containing about $12.1\times10^6$ grid points. The amount of data for each snapshot is roughly $3.6 \times 10^7$, so 14.6 billion values were used by the GAN for the complete analysis of the $K2$ dataset. The snapshots have been taken with an interval of $\Delta t = 3.25 \mu$s.
The total time window considered is on the order of two integral time scales, while the sampling frequency is an order of magnitude lower than the Kolmogorov timescale.

\section{Methodology and Neural Network Architecture}

\begin{figure}[t]
    \centering
    \includegraphics[width=\textwidth]{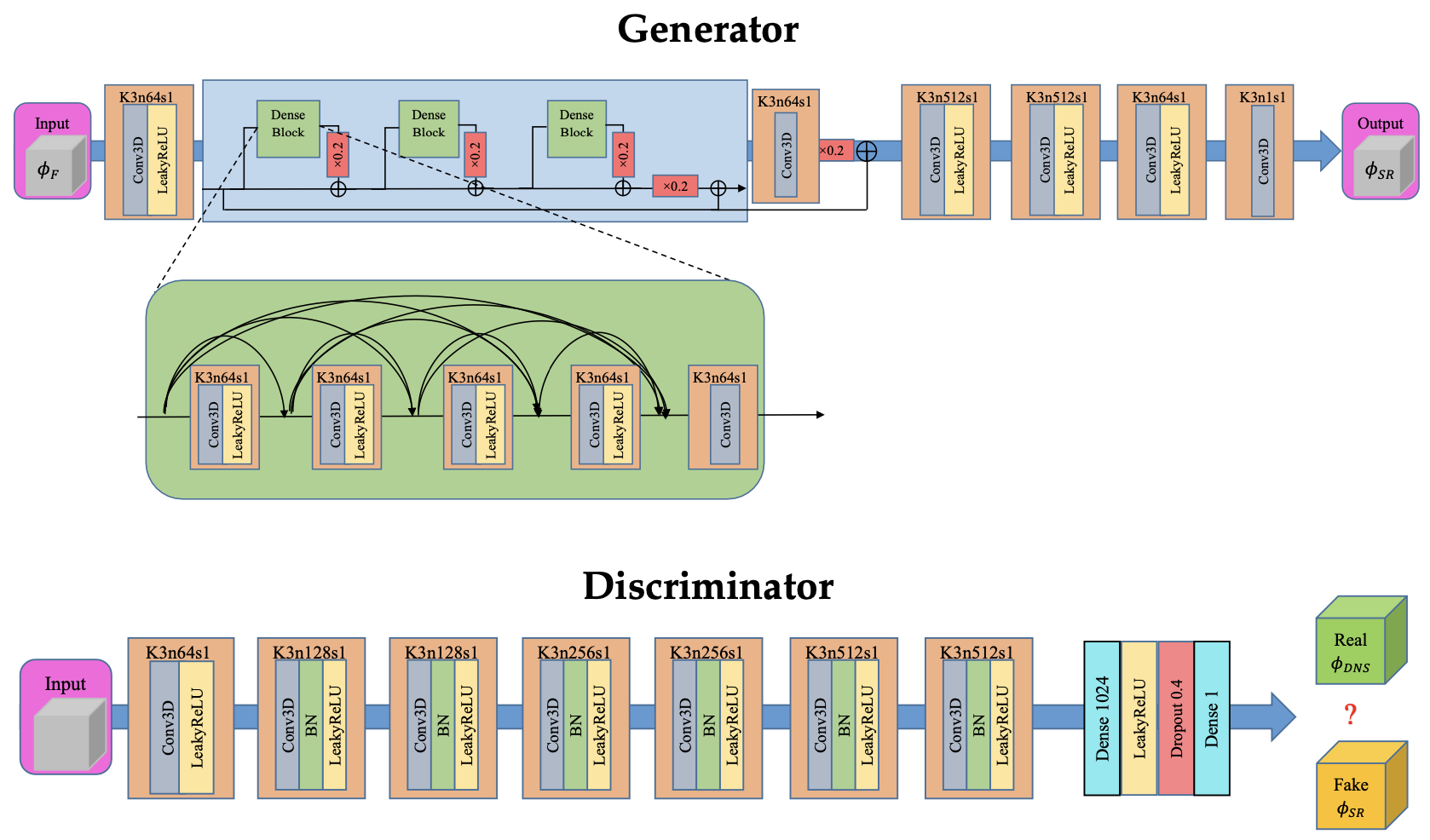}
    \caption{The structure of the generator (above) and the discriminator (below) of the GAN architecture.}
    \label{fig:TSRGAN_architecture}
\end{figure}

The proposed networks used in this work are based on the architecture developed by Bode \etal~\citeyearpar{bode2021using}, schematically represented in Fig. \ref{fig:TSRGAN_architecture}, which was originally inspired by the ESRGAN architecture~\citep{ESRGAN} and adapted for small-scale turbulence reconstruction. 

In general, GAN architectures consist of two competing networks: a generator and a discriminator. In this work, the generator is partially adapted from the original SRResNet network~\citep{he2016deep} where it makes use of three-dimensional convolutional layers with leaky rectified linear units (LReLu)~\citep{Maas2013, geron2019hands} as activation functions. The residual-in-residual dense blocks (RRDB)~\citep{ESRGAN} contain fundamental architecture components, such as skip-connections and dense blocks, enabling the generation of super-resolved data through a very deep network capable of learning and modeling all relevant complex transformations. The main difference with the original ESRGAN structure is the lack of the upsampling layers, as the network is developed to add small-scale turbulent features without performing upsampling operations~\citep{bode2021using}. Hence, the input and output of the generator hold the same dimensions, but the energy distribution is enriched towards high-wavenumber frequencies. The total number of trainable parameters of the generator is around 19 million.

The discriminator is a deep deconvolutional architecture of fully connected layers with binary classification output, giving the probability for a fake generation or a ground truth prediction. It differs from the original ESRGAN discriminator by the introduction of a dropout layer needed to prevent overfitting~\citep{geron2019hands}. The total number of trainable parameters of the discriminator is around 15 million.

\subsection{Training Strategy and Loss Function Definition}

Super-resolution reconstruction was applied to the datasets described in Sec~\ref{data}.
In order to obtain LES-like data, both datasets were filtered. A box filter of width $\Delta=16\,dx$ was considered for the low $K1$ datasets ensuring an averaged resolved energy of $79.63\%$ with respect to the DNS field, a typical value for a well-resolved LES analysis \citep{pope2000turbulent}.
Analogously, a filter width of $\Delta=10\,dx$ was considered for the $K2$ dataset reducing the resolved energy by approximately 20$\%$ (averaged resolved energy of $\approx$ 81.68$\%$ with respect to the DNS field).

Because of GPU memory limitations, the networks were trained using sub-boxes of a size that depends on the filter width in a way that the 3D filter kernel fits inside the sub-box 8 times. 
In order to avoid performance drops during the initial loading of the training dataset, a staged training approach was introduced. 
Each stage consisted of boxes from 8 different snapshots, which were randomly selected out of all available snapshots, shuffling them before the usage in the GAN. 
The snapshot used for testing was previously removed from the selection. 
For the $K1$ dataset, a total of 2880 boxes of size $32 \times 32 \times 32$ per stage were considered.
A total of 10120 boxes of size $20 \times 20 \times 20$ per stage were extracted for the $K2$ dataset.
Each box included three physical fields, namely the three velocity components ($u$, $v$, $w$). 
Following the usual approach established in the literature~\citep{geron2019hands} to improve the network’s performance, each of the variables in the input is normalized with its global maximum and minimum. 
For one stage - or chunk of the data - the model was trained for 10 epochs before the next stage was loaded in. 
To improve the training stability, the generator was trained alone for a certain number of stages (called pre-training), then the GAN, as a combination of the generator and discriminator, was trained for the same number of stages. 
An initial learning rate for the pre-training of $10^{-4}$ and the use of the ADAM optimizer~\citep{geron2019hands} were selected based on previous investigations~\citep{nista2021turbulent}. 
The same initial learning rate was used for the discriminator during the GAN training, while the initial learning rate for the generator in the GAN training section was decreased by one order of magnitude relative to the initial learning rate of the discriminator. To aid convergence to a local minimum, the learning rate was halved every 5 epochs of each stage. 

Given the large datasets employed and the deep convolutional frameworks (entirely based on TensorFlow v2~\citep{abadi2016tensorflow}), the training strategy was parallelized on multiple GPUs to train the models faster. An efficient data-parallel approach based on the Horovod library ~\citep{sergeev2018horovod} was employed. To circumvent memory limitations associated with batch size, the network was replicated across several workers (GPUs), splitting the training stages among these units and updating the gradients synchronously at the end of each batch. The investigations were performed on the RWTH Aachen University cluster (CLAIX18$-$GPU) using two nodes, where each host has two NVIDIA Tesla V100 16GB GPUs. This allowed an overall speed-up around a factor of four, relative to the training time on a single GPU.

The original perceptual loss presented on the ESRGAN implementation \citep{ESRGAN} was replaced by a combination of three loss functions: the adversarial loss $\mathrm{L_{RADG}}$~\citep{jolicoeur2018relativistic}, the pixel loss $L_{\mathrm{pixel}}$, and the gradient loss $\mathrm{L_{gradient}}$. The generator loss is then defined as:
\begin{equation}
\mathcal{L}_{\mathrm{gen}} =  \beta_1 \, L_{\mathrm{pixel}} + \beta_2 \, \mathrm{L_{gradient}} + \beta_4 \, \mathrm{L_{RADG}}
\label{eqn:loss_gan}
\end{equation}
where $\beta_i$ are the respective weighting coefficients. In this work, the coefficients $\beta_1=5.0$, $\beta_1=0.1$, and $\beta_1=10^{-5}$ were used. During the pre-training, i.e. when the generator was trained in supervised mode, only the pixel loss was employed. To train the discriminator, the logistic loss based on predicted labels of ground truth and generated field was considered. 

First, the neural network was trained on one of the simulation datasets, e.g. either at low $\mathrm{Ka}$ or at high $\mathrm{Ka}$ and applied to the same flame condition. That was essential to understand if the model has learned the general physical behavior. 
Furthermore, the architecture was trained with a mixture of the datasets of the two flames and subsequently applied to both flames to investigate if it is advantageous for the model to learn with more diverse data and if it can capture the peculiarities distinctive to the two regimes.

\section{Results}

\begin{figure}[t]
  \centering
  \subfloat[Filtered DNS]{\adjincludegraphics[width=0.28\linewidth,trim={{0.291\width} {.333\width} {.295\width} {.437\width}}, clip=true]{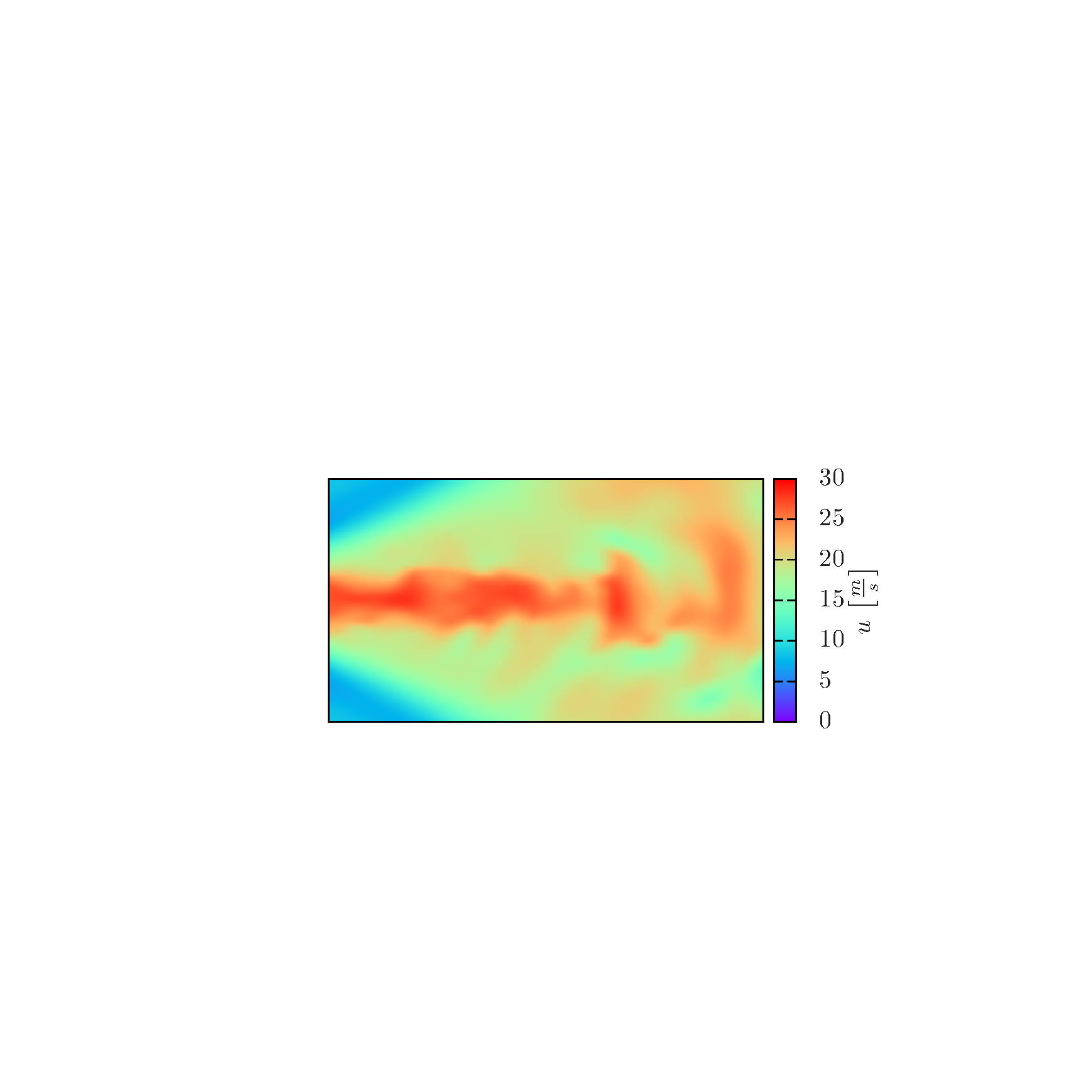}}
  \subfloat[GAN]{\adjincludegraphics[width=0.28\linewidth,trim={{.291\width} {.333\width} {.295\width} {.437\width}}, clip=true]{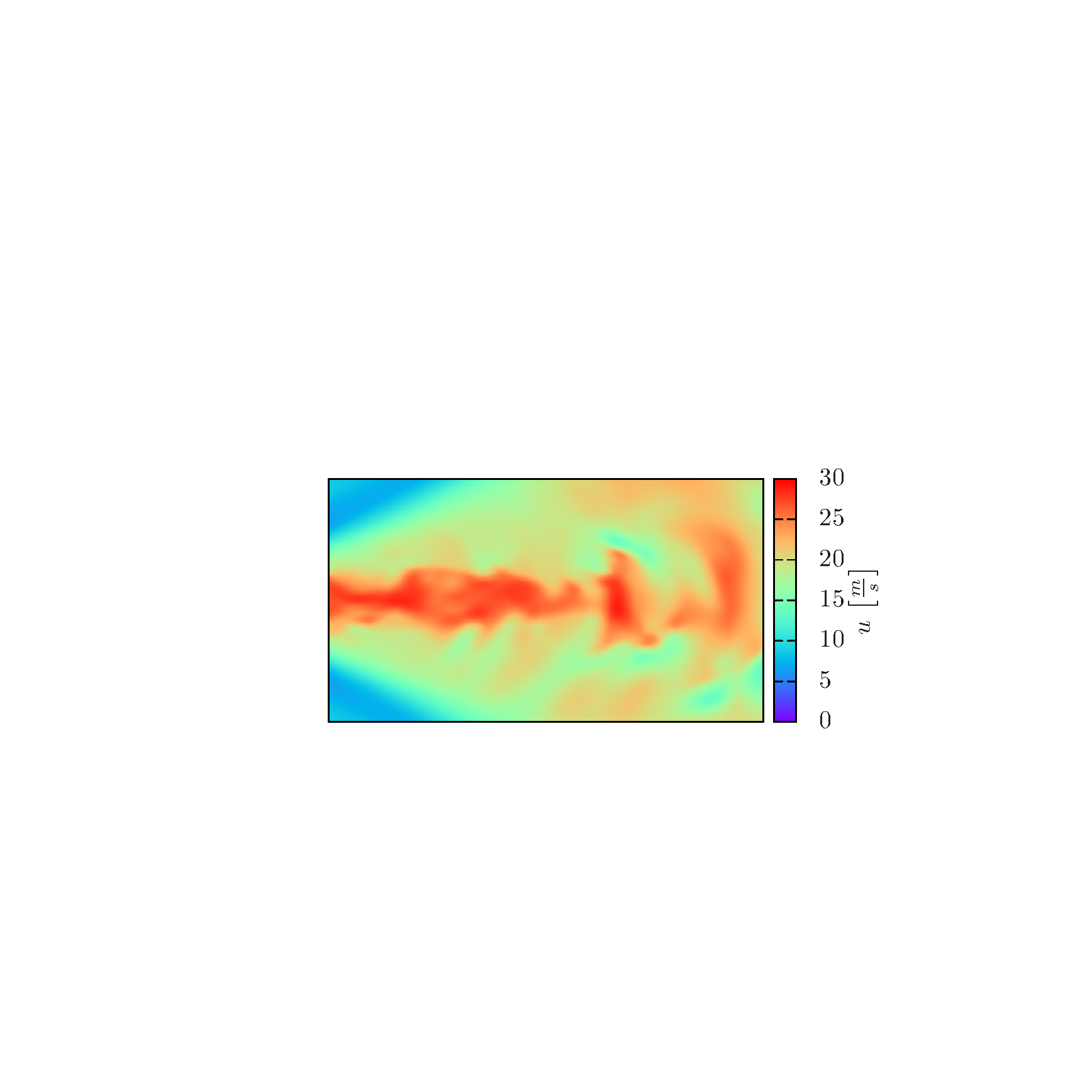}}
  \subfloat[DNS]{\adjincludegraphics[width=0.348\linewidth,trim={{.291\width} {.333\width} {.194\width} {.43\width}}, clip=true]{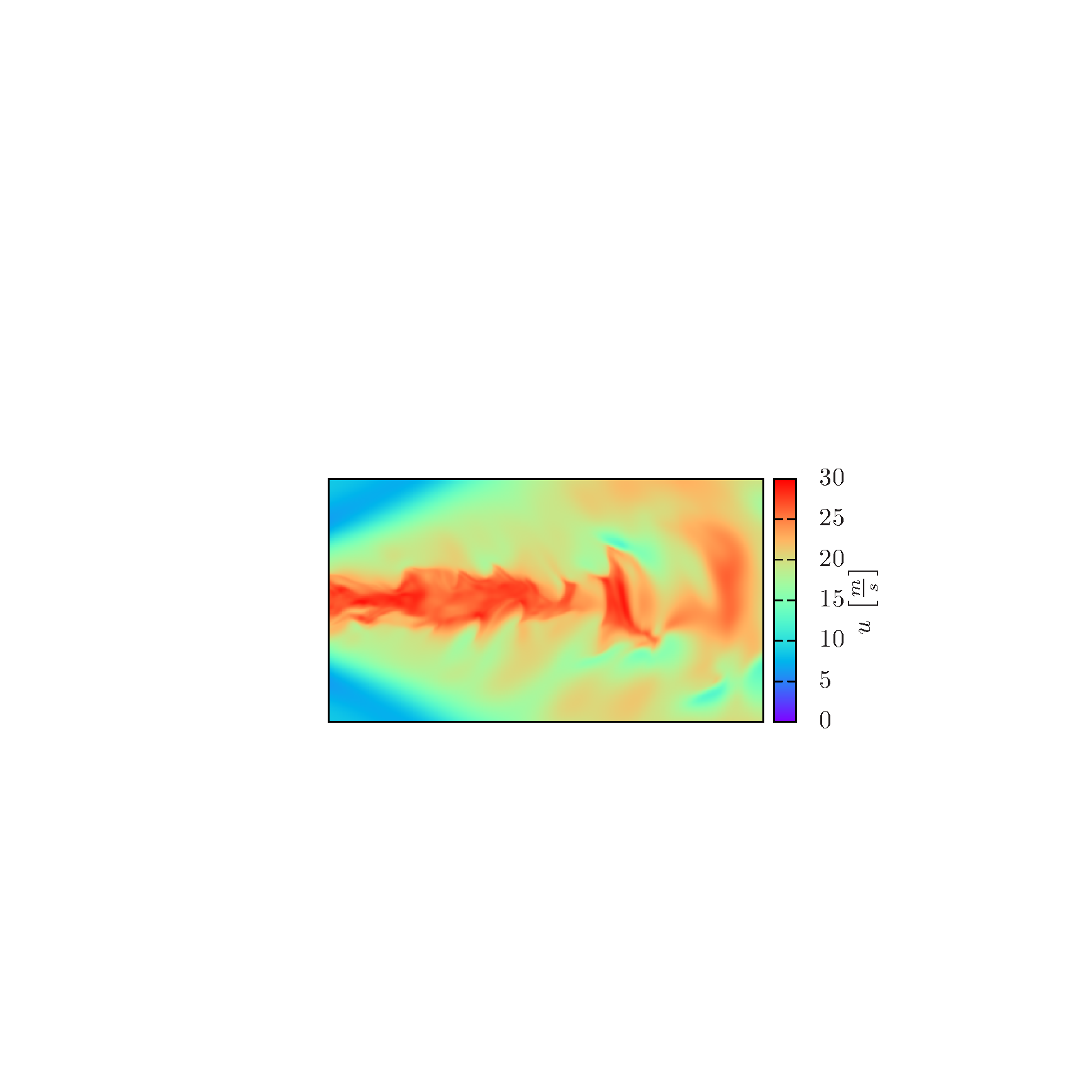}}
  \caption[$u$ contour, $K1$]{Contour plot of the velocity component $u$ for the filtered DNS field, super-resolved field and the DNS for the $K1$ case.}
  \label{fig:ContourUK1}
\end{figure}

\begin{figure}[t]
  \centering
  \subfloat[Filtered DNS]{\adjincludegraphics[width=0.28\linewidth,trim={{0.294\width} {.333\width} {.295\width} {.437\width}}, clip=true]{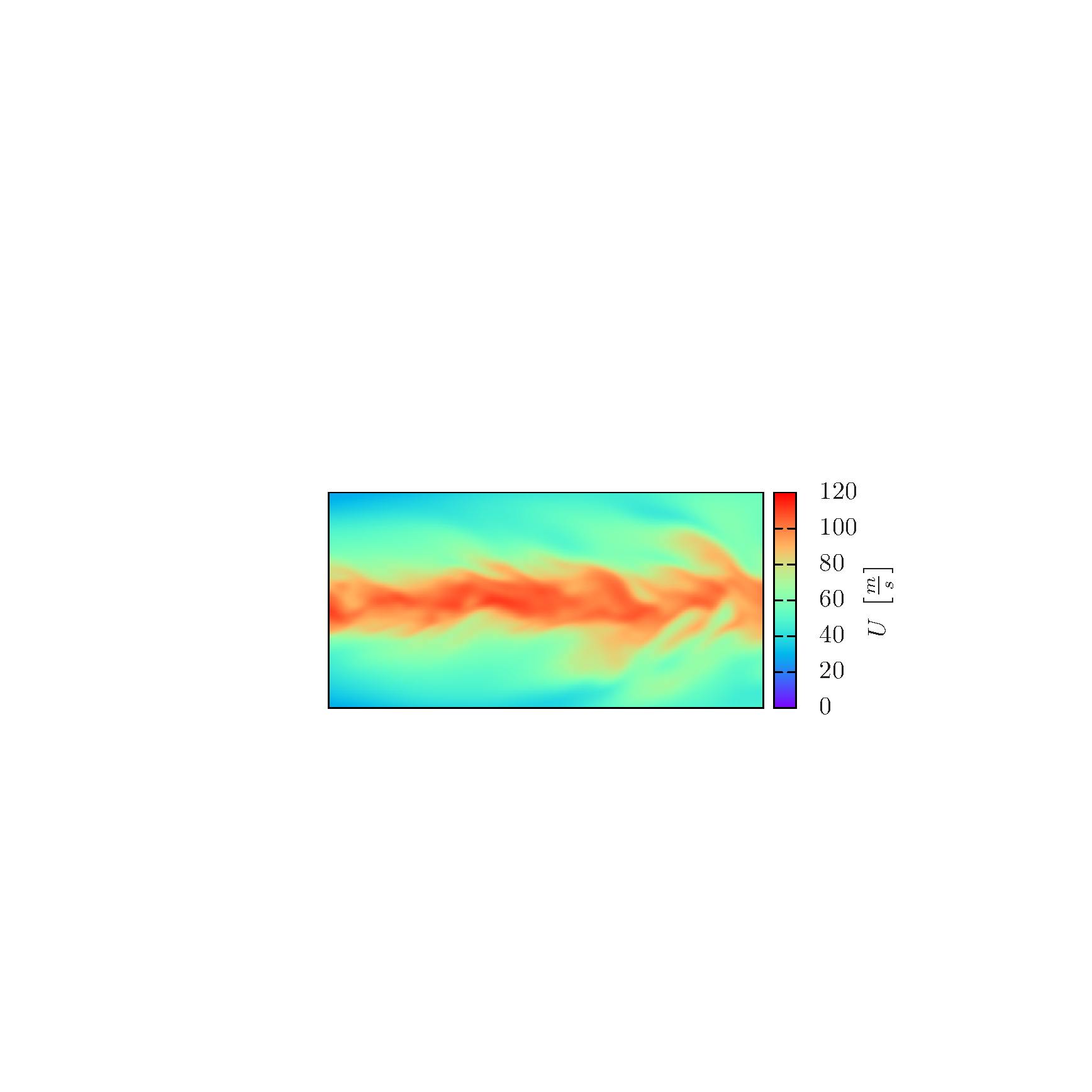}}\hspace{1mm}
  \subfloat[GAN]{\adjincludegraphics[width=0.28\linewidth,trim={{.291\width} {.333\width} {.295\width} {.437\width}}, clip=true]{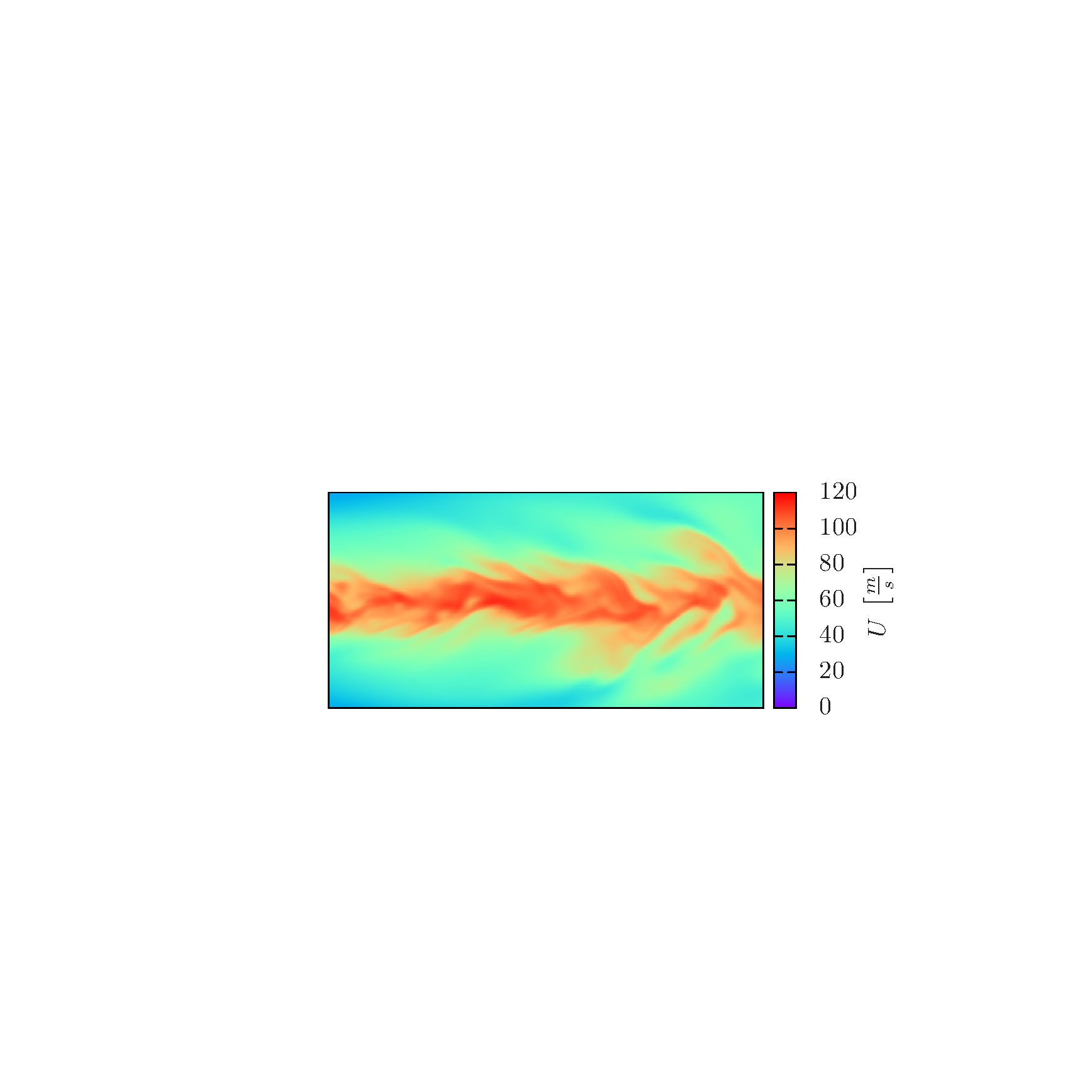}}\hspace{1mm}
  \subfloat[DNS]{\adjincludegraphics[width=0.358\linewidth,trim={{.291\width} {.333\width} {.18\width} {.43\width}}, clip=true]{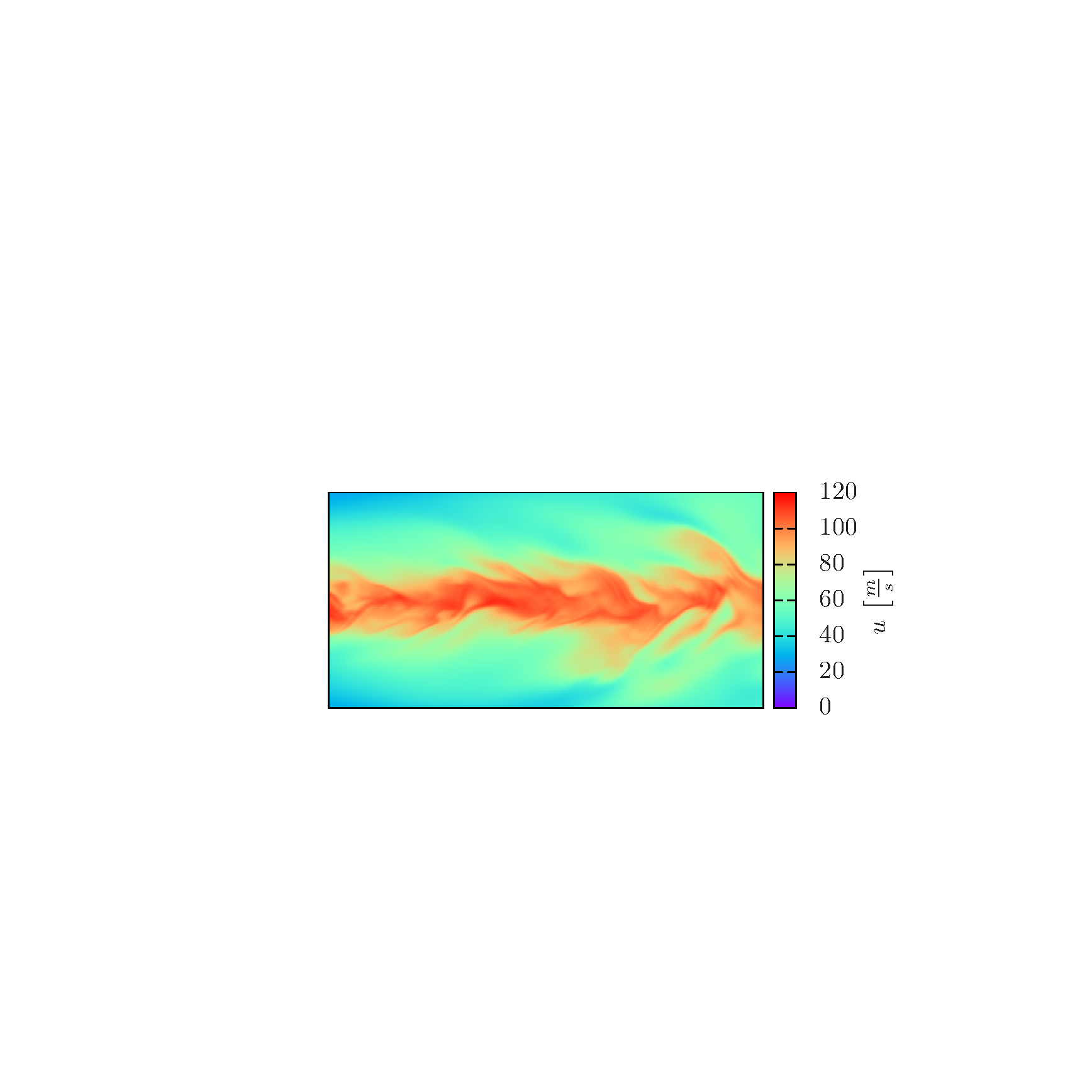}}\hfill
  \caption[$u$ contour, $K2$]{Contour plot of the streamwise velocity component for the filtered DNS field, super-resolved field and the DNS for the $K2$ case.}
  \label{fig:ContourUK2}
\end{figure}

Figure \ref{fig:ContourUK1} depicts contour plots of the streamwise velocity component in the $xy-$plane at the central spanwise position for the $K1$ dataset filtered with a box filter, the reconstructed dataset using the network trained on the $K1$ dataset, and the DNS (ground truth). The plot shows the $xy-$plane at the center of the spanwise direction. Comparing visually the filtered and the super-resolved fields, the network adds some features and increases the magnitude of the velocity component resulting in an image that is perceived as noticeably sharper. 
However, when the field reconstructed by the network is juxtaposed with the DNS field, significant differences become apparent. 
The inaccuracies are less marked for the coflow, while they are more notable for the jet region, particularly at the interface between the jet and the coflow.
The model is realizing features smaller than the filter size. Medium-scale oscillations absent in the filtered data are reconstructed, although with some inaccuracies, while larger-scale oscillations are captured well, particularly toward the outflow boundary of the domain. 
However, in several instances, the model is not quite able to capture the influence of medium scales correctly and, therefore, there must be a difference in the filtered field of the DNS and the GAN. 
Overall, the model is clearly adding information about subfilter scales in the form of kinetic energy, but it is not able to reconstruct a field that resembles the DNS perfectly or could be misconstrued as a field generated with a direct numerical simulation.

Figure \ref{fig:ContourUK2} depicts the contour plots of the streamwise velocity component for the $K2$ dataset for the $xy-$plane at the central spanwise position. The filtered field is significantly more blurred than the DNS field. Comparing the field produced by the GAN trained on $K2$ data to the other fields, the network's prediction looks analogous to a fully resolved field. Quite clearly, the magnitude of velocity is increased or decreased resulting in a field that looks visibly sharper, and some features are enhanced. 
In detail, it adds some subfilter structures such that the GAN field still looks like a filtered field with a much smaller filter size. 
The network seems to under-resolve thin or fine features especially, which are bulkier and less detailed than in the DNS.
In the first $2/3$ of the domain, there is some overshoot of the model relative to the DNS. In the last $1/3$ though, the model exhibits a tad of undershooting. This is indicative of the model not learning the specifics of the velocity over the domain, which is intuitive as the network was trained with subdomains, or boxes, much smaller than the domain shown. Therefore, it likely learned an average of the over and undershoot of what is shown in the plot so that the error is minimized. 
Overall, the prediction of the model does not seem too displeasing as the prediction is favorable compared to the $K1$ prediction. Large and medium-scale fluctuations are mostly captured. 
It is clear that the model does add significant kinetic energy that emulates closely the turbulent kinetic energy-resolved by the DNS. The field obtained by applying the network to the filtered $K2$ dataset can be described as DNS-like, and there exist some differences between the ground truth and the prediction.

\begin{figure}[t]
  \centering
  \subfloat[$\tau^r_{11}$, Static Smagorinsky]{\adjincludegraphics[width=0.45\linewidth, clip=true]{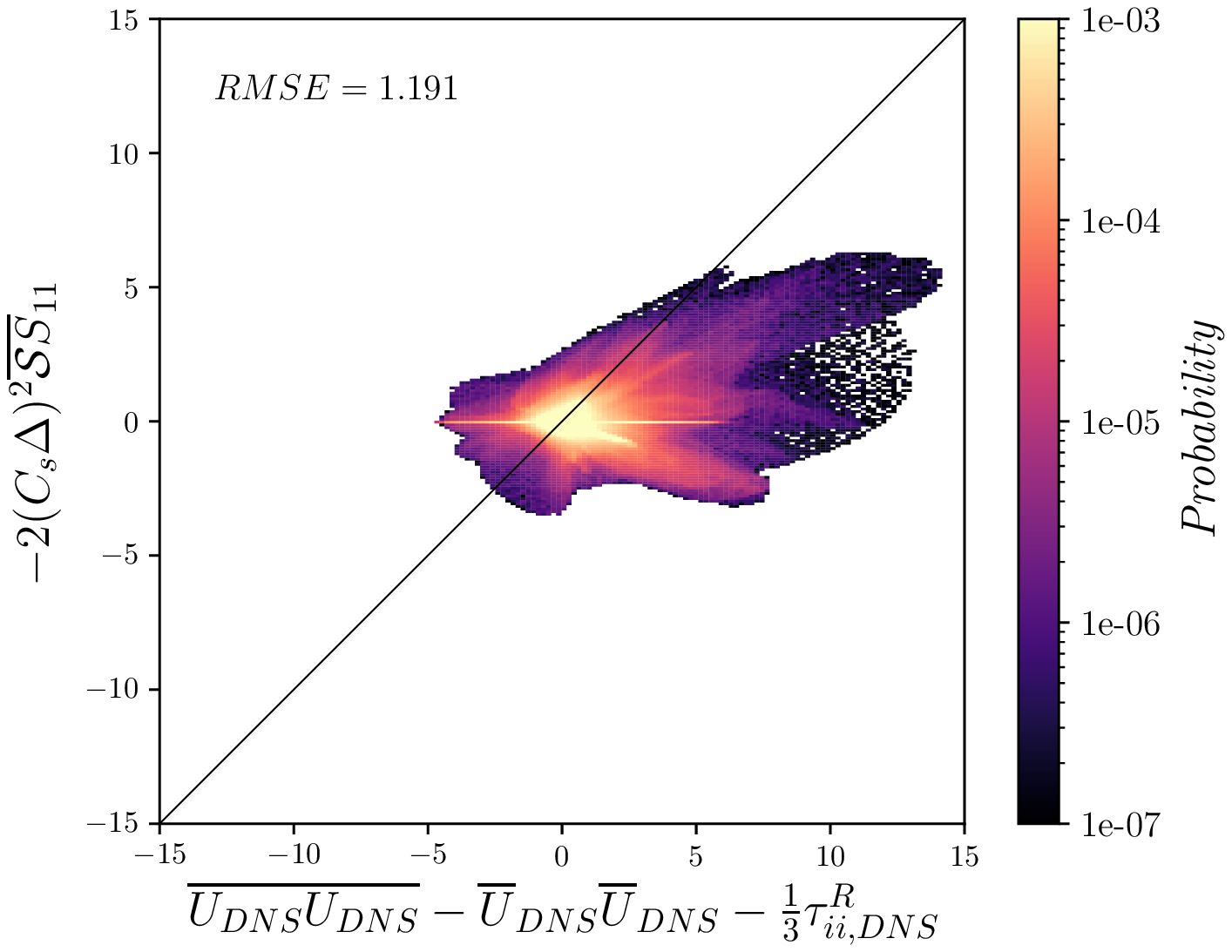}}\hspace{1mm}
  \subfloat[$\tau^r_{12}$, Static Smagorinsky]{\adjincludegraphics[width=0.45\linewidth, clip=true]{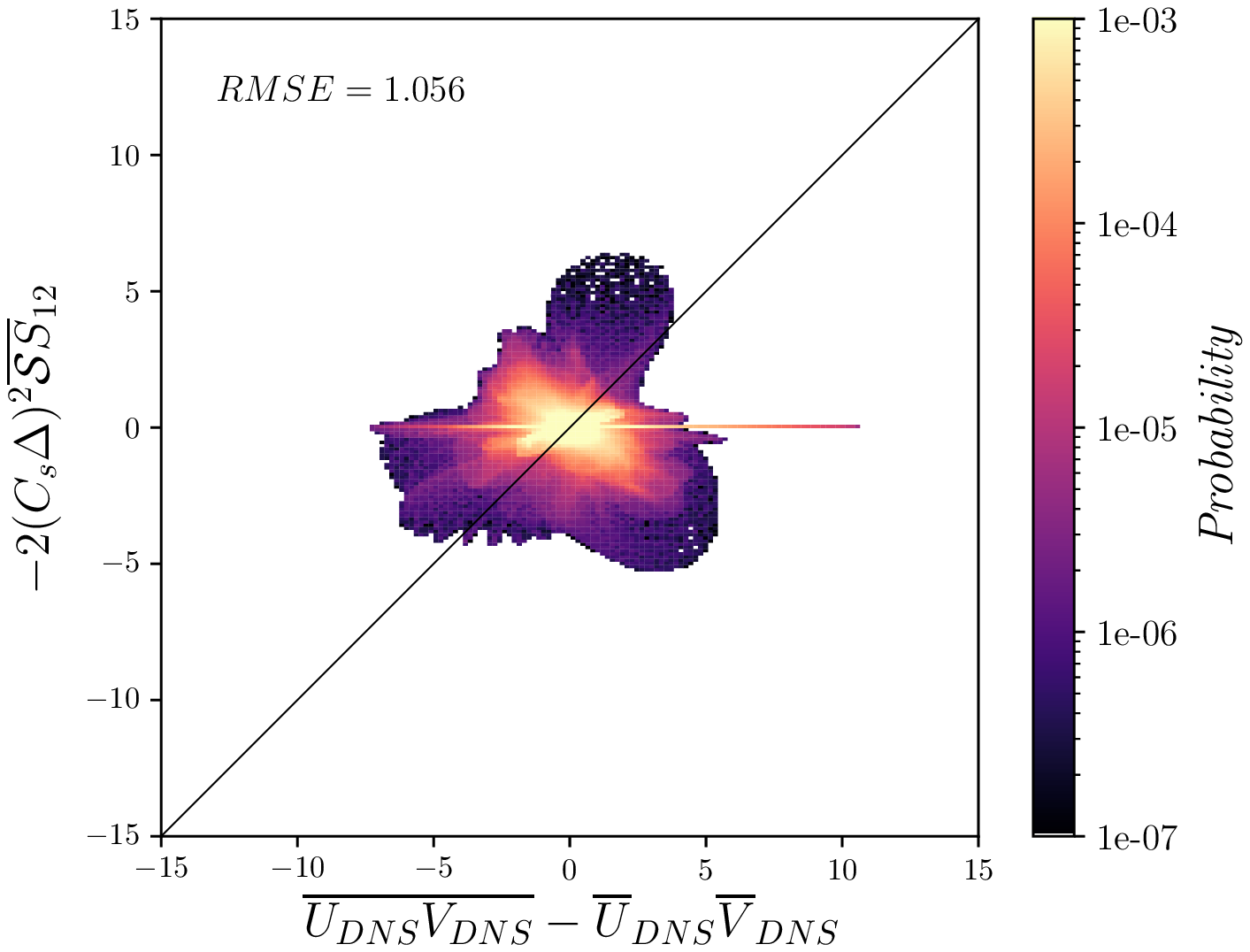}}\hfill
  \subfloat[$\tau^r_{11}$, GAN]{\adjincludegraphics[width=0.45\linewidth, clip=true]{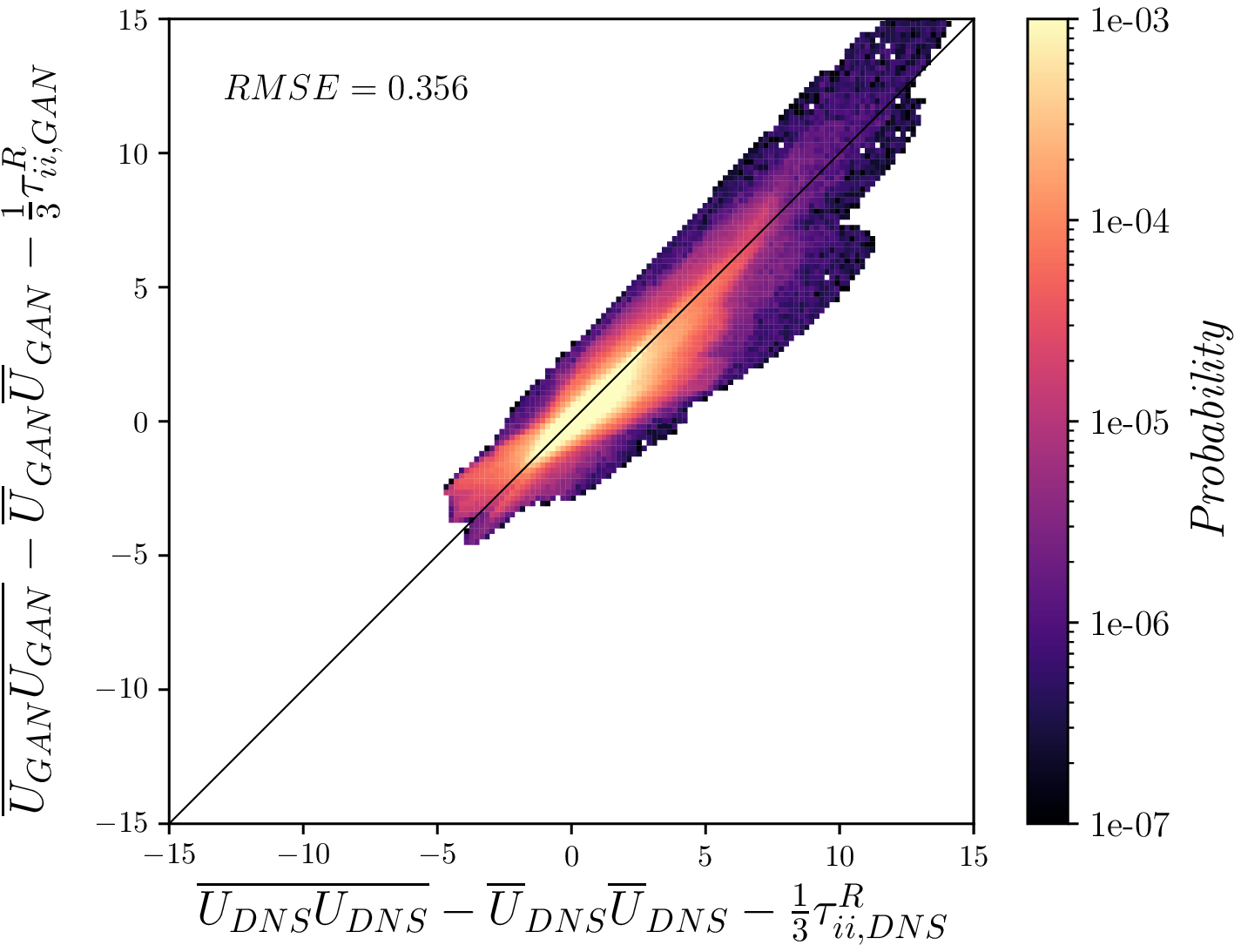}}\hspace{1mm}
  \subfloat[$\tau^r_{12}$, GAN]{\adjincludegraphics[width=0.45\linewidth, clip=true]{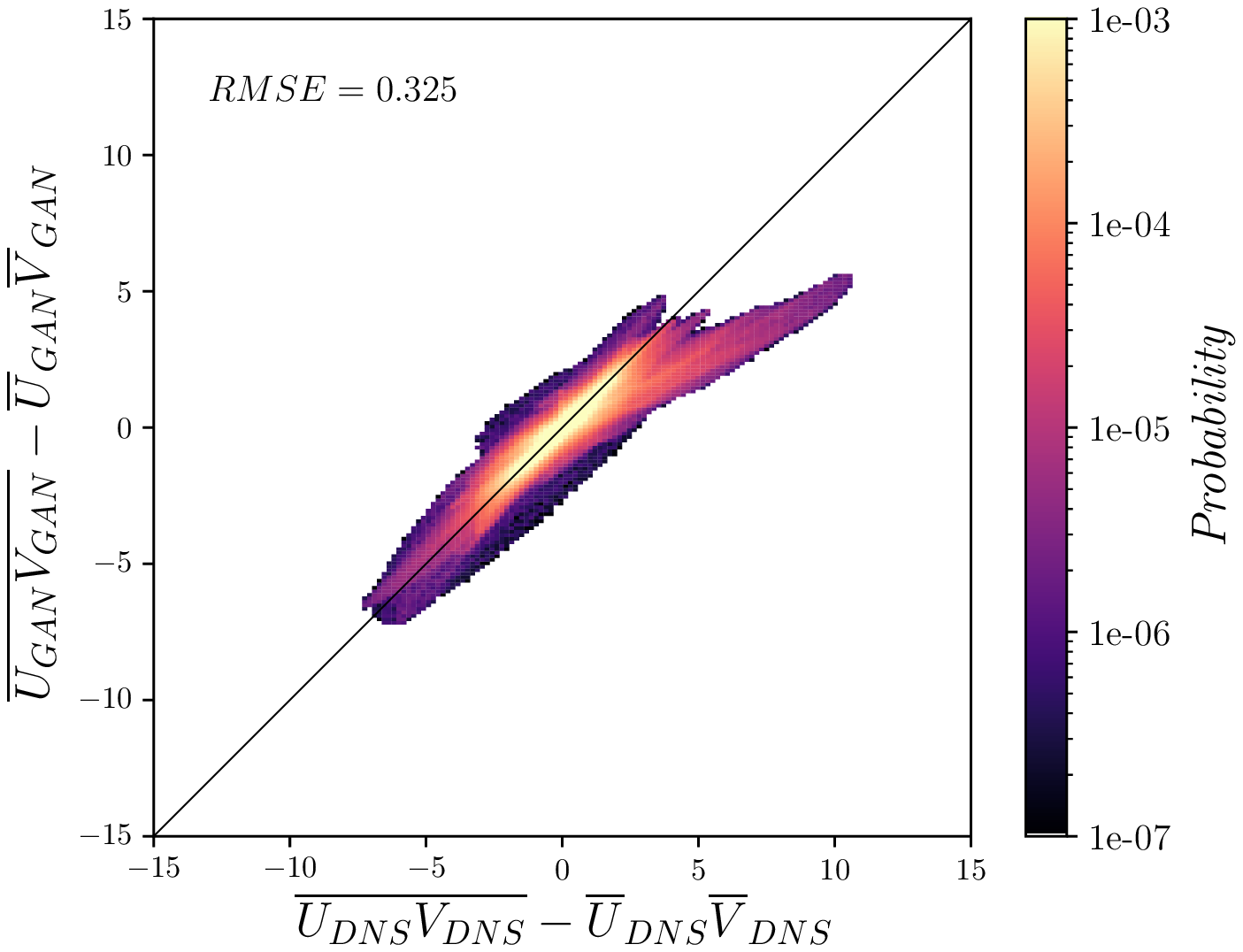}}\hfill
  \caption[jPDF subfilter-scales stress $K1$]{Joint PDF plots for the subfilter-scale stress tensor components $\tau^r_{11}$ and $\tau^r_{12}$ for the $K1$ case. On the horizontal axis, the subfilter-scale stresses are evaluated from the DNS data, while on the vertical axis the same quantities are evaluated with the static Smagorinsky model (top panels) and GAN data (bottom panels). RMSE indicates the Root Mean Squared Error.}
  \label{fig:jpdfLowKa}
\end{figure}

In order to judge the capability to close the equation of momentum in LES, the subfilter-scale stress tensor is, perhaps, the more important quantity to look at. 
In the panels of figure~\ref{fig:jpdfLowKa}, joint PDFs of the first diagonal and the first off-diagonal components of this tensor computed from velocity components of the DNS dataset, the GAN dataset, and the static Smagorinsky model are compared with.
The static Smagorinsky model predictions are worst than those of the network for the $K1$ dataset, as is shown in figure \ref{fig:jpdfLowKa}. 
The alignment of the prediction of the GAN with the diagonal is better and, the error of the subfilter-scale stress is significantly lower compared with the static Smagorinsky model. 
The cross-correlation with respect to the DNS exceeds $90\%$ for all components:
\begin{equation}
CC_{GAN,K1} = \begin{pmatrix}
0.941 & 0.938 & 0.923\\
0.938 & 0.940 & 0.926\\
0.923 & 0.926 & 0.934
\end{pmatrix}.
\end{equation}

\begin{figure}[t]
  \centering
  \subfloat[$\tau^r_{11}$, Static Smagorinsky]{\adjincludegraphics[width=0.45\linewidth, clip=true]{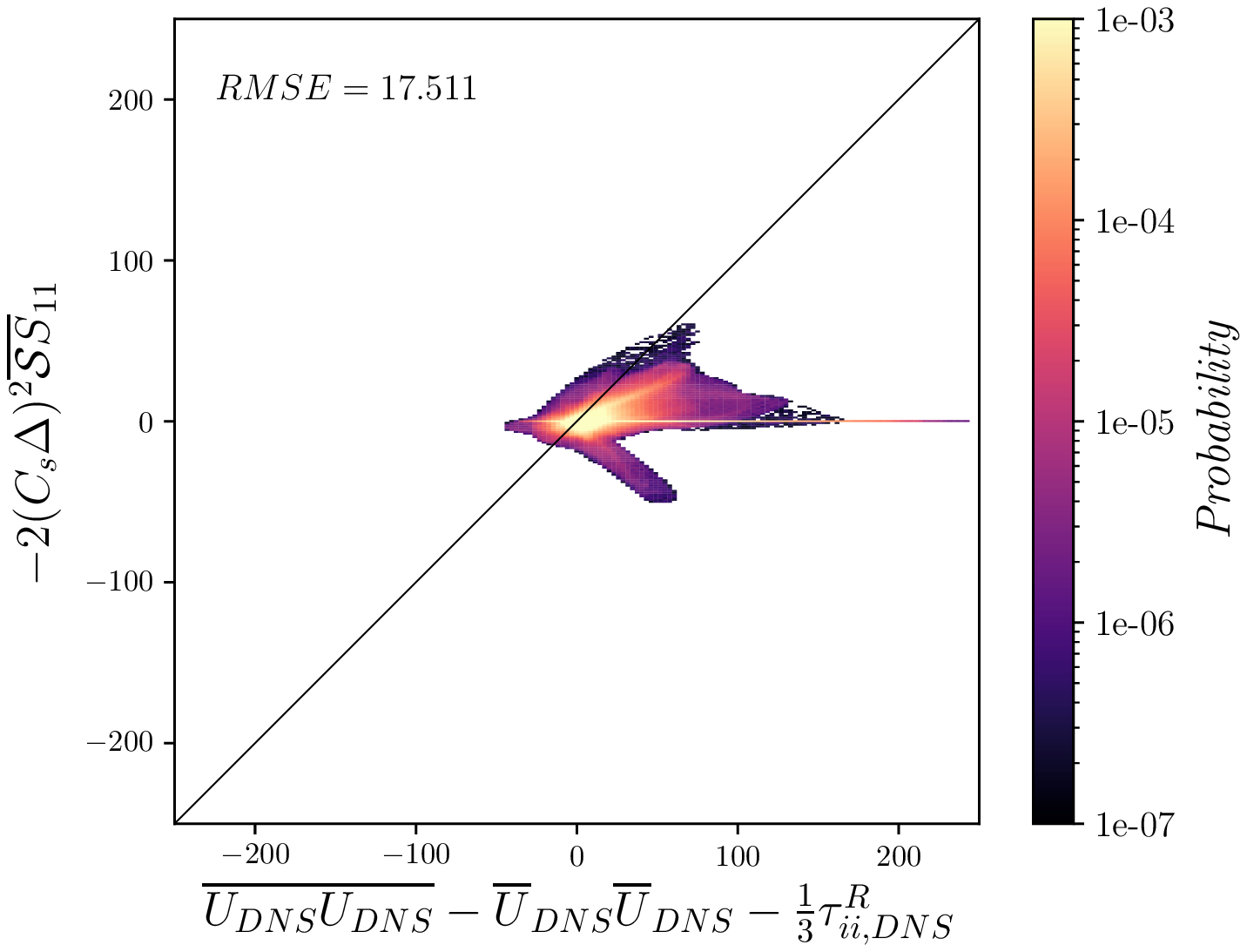}}\hspace{1mm}
  \subfloat[$\tau^r_{12}$, Static Smagorinsky]{\adjincludegraphics[width=0.45\linewidth, clip=true]{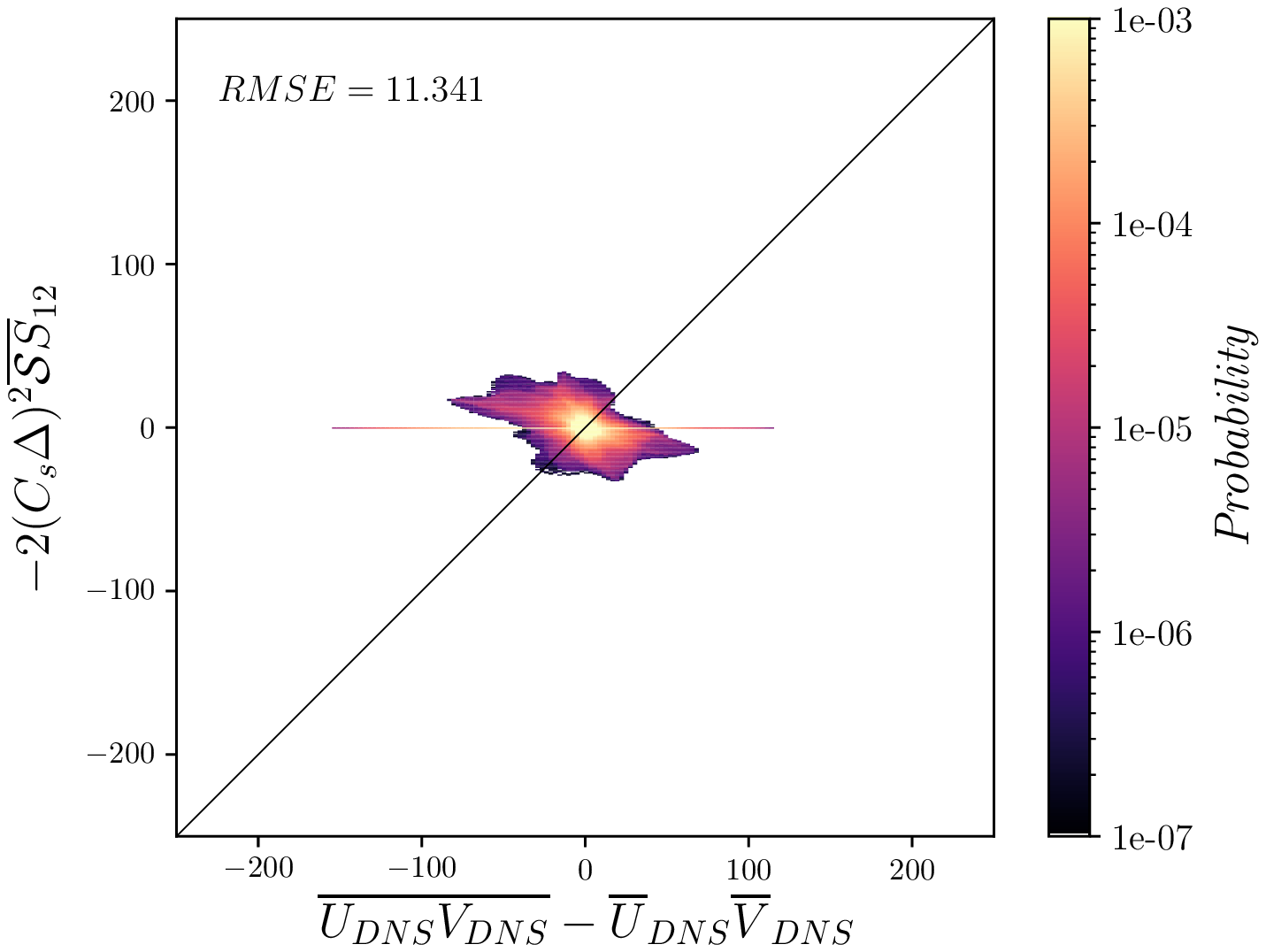}}\hfill
  \subfloat[$\tau^r_{11}$,  GAN]{\adjincludegraphics[width=0.45\linewidth, clip=true]{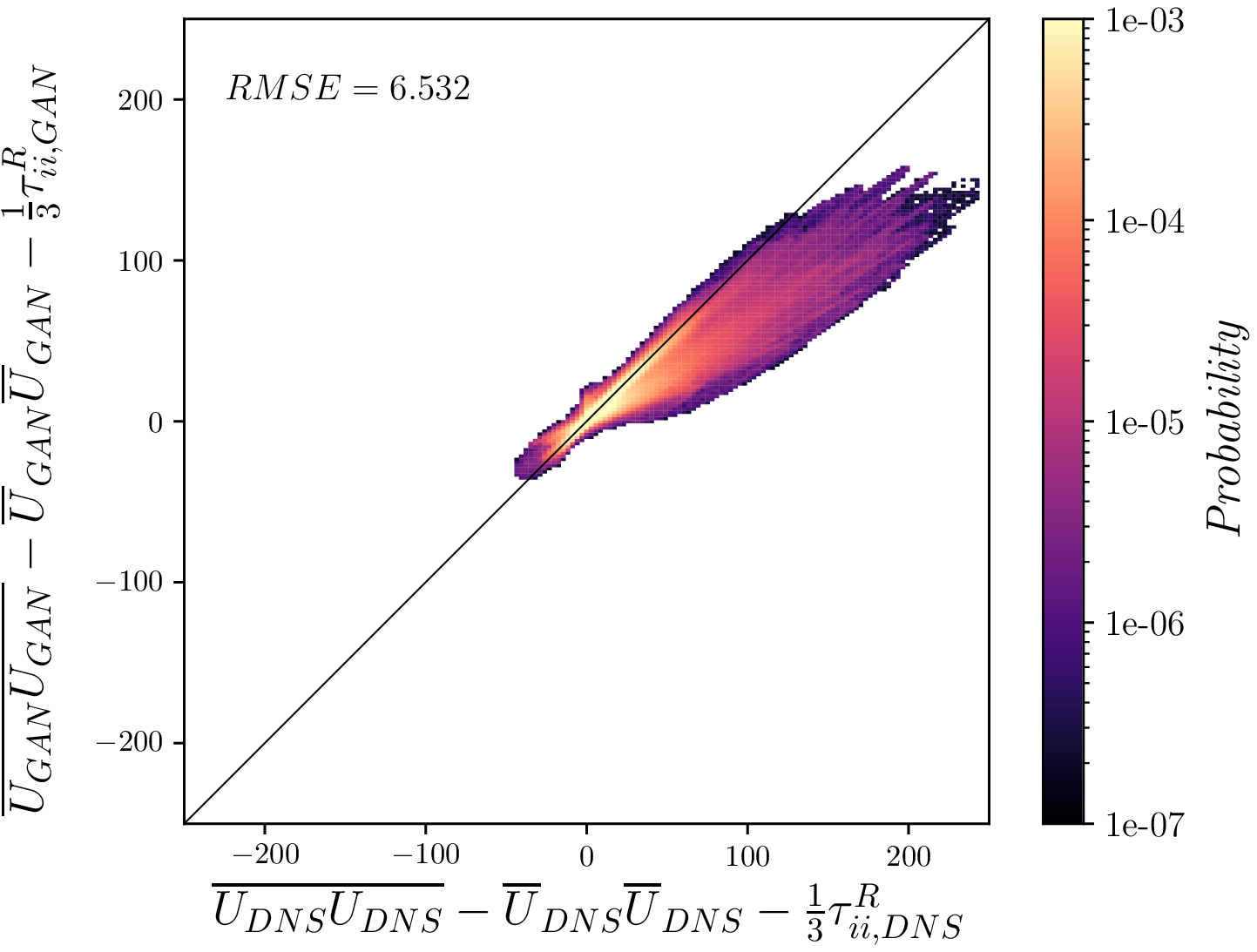}}\hspace{1mm}
  \subfloat[$\tau^r_{12}$, GAN]{\adjincludegraphics[width=0.45\linewidth, clip=true]{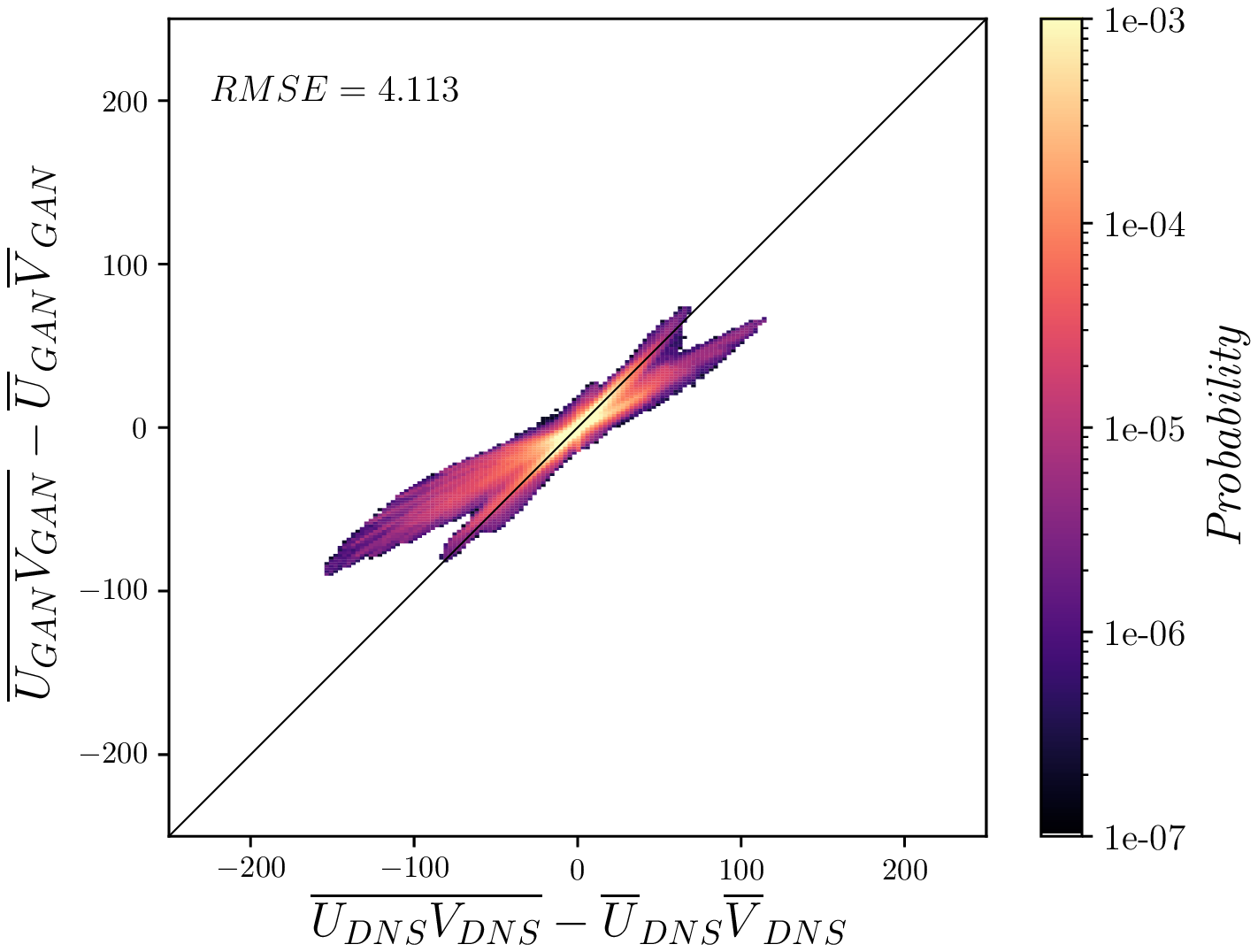}}\hfill
  \caption[jPDF subfilter-scales stress $K2$]{Joint PDF plots for the subfilter-scale stress tensor components $\tau^r_{11}$ and $\tau^r_{12}$ for the $K2$ case. On the horizontal axis, the subfilter-scale stresses are evaluated from the DNS data, while on the vertical axis the same quantities are evaluated with the static Smagorinsky model (top panels) and GAN data (bottom panels). RMSE indicates the Root Mean Squared Error.}
  \label{fig:jpdfHighKa}
\end{figure}

Figure \ref{fig:jpdfHighKa} depicts the joint PDFs of the subfilter stress tensor components for the case $K2$. 
The static Smagorinsky model does not align well with the subfilter scale stress evaluated for the DNS. 
In fact, the mean cross-correlation is barely above $5\%$. Nonetheless, the distribution around the diagonal is not severely bulky, which translates into a reasonably low mean error. 
Yet, the Smagorinsky model mostly underpredicts and fails to predict the values of high stress. 
The stress tensor computed based on the GAN prediction strongly outperforms the static Smagorinsky model. Not only does the alignment of the jPDF with the diagonal improve strongly, but the values further distanced from the center are predicted with much greater accuracy. This behavior is quantified by the cross-correlation matrix:

\begin{equation}
  CC_{GAN, K2} = \begin{pmatrix}
  0.951 & 0.937 & 0.941\\
  0.937 & 0.944 & 0.935\\
  0.941 & 0.935 & 0.946
  \end{pmatrix},
\end{equation}

\noindent which indicates a mean cross-correlation of $\left< CC \right>_{GAN, K2}=94.1\%$, and values of the mean error which are substantially lower relative to the static Smagorinsky model. 

Overall, a considerable improvement over the static Smagorinsky model is achieved. These results are consistent with the analysis of case $K1$, suggesting that this might be a universal behavior of the network. It has been noted that the GAN learns also how to modify the filtered field as part of the training process. Thus, the GAN appears to be able to learn the effects of subfilter-scales to large scale, or rather to potentially model the backward energy cascade.

\begin{figure}[t]
  \centering
  \subfloat[GAN vs DNS]{\adjincludegraphics[width=0.485\linewidth, trim={{0.15\width} {.1\width} {0\width} {.2\width}}, clip=true]{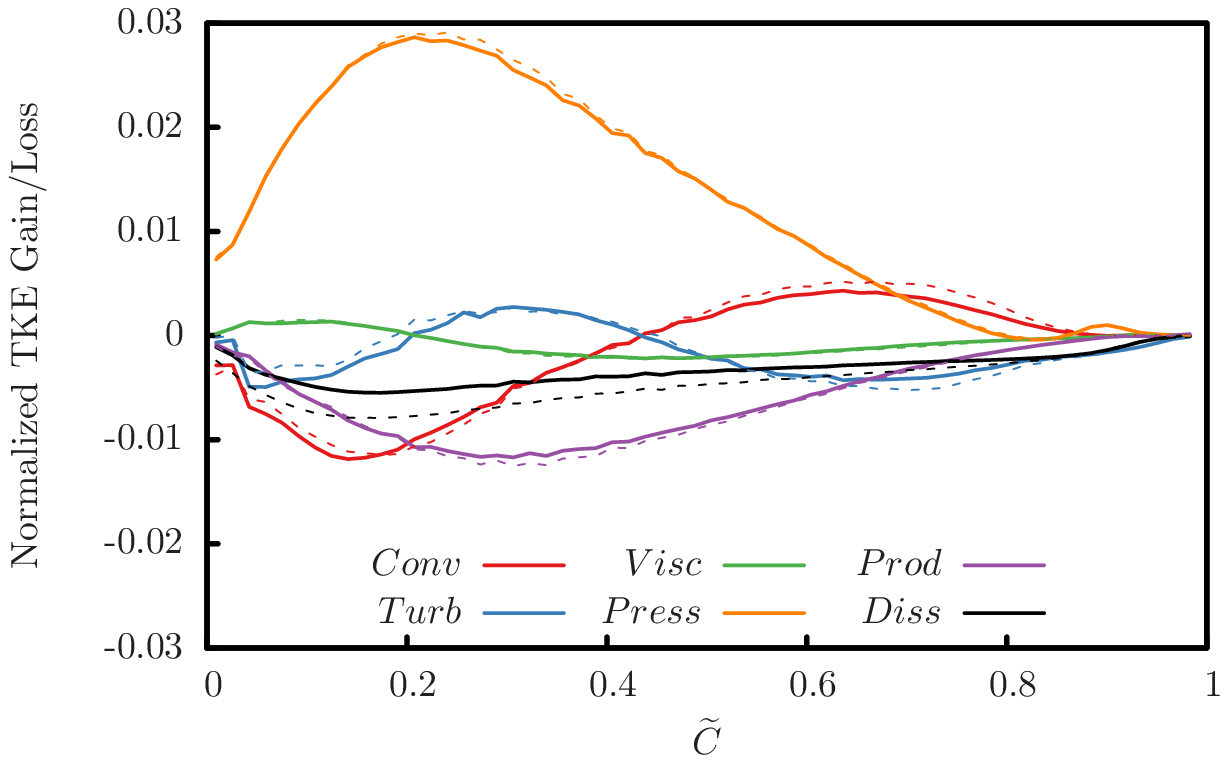}}
  \quad
  \subfloat[Filtered vs DNS Data]{\adjincludegraphics[width=0.485\linewidth, trim={{0.15\width} {.1\width} {0\width} {.2\width}}, clip=true]{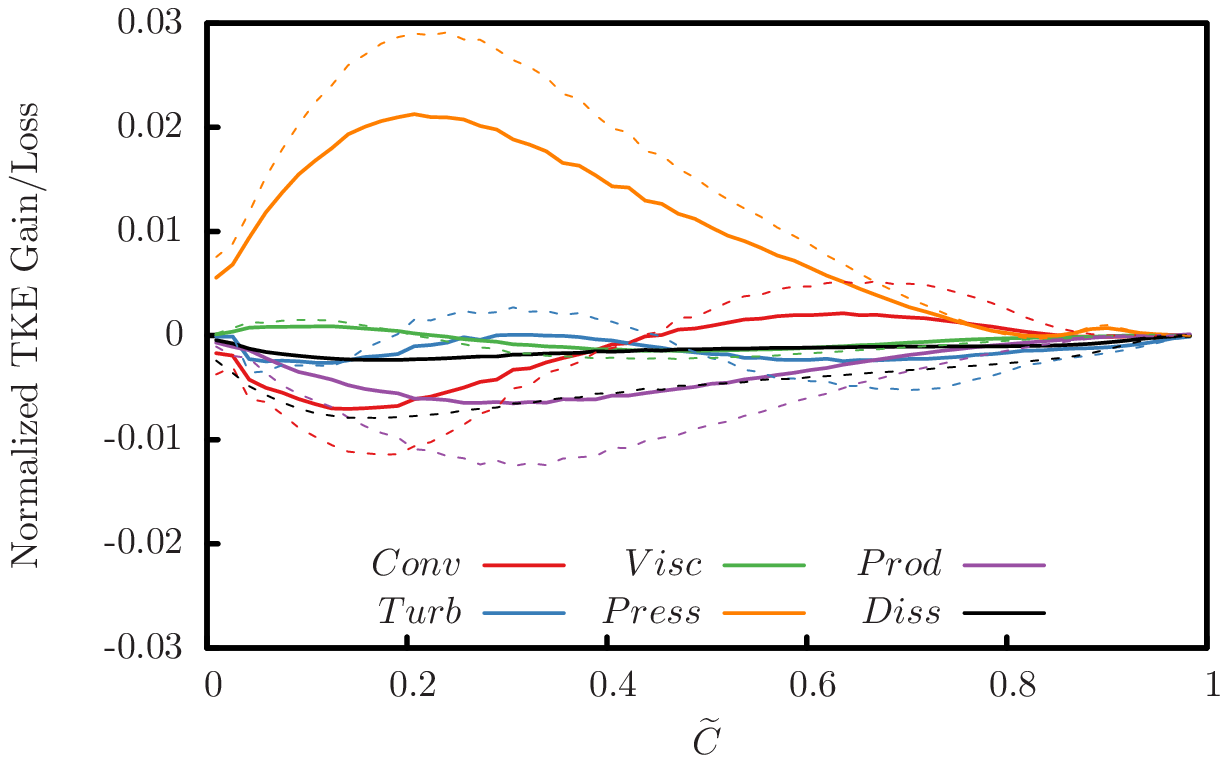}}
  \caption[Turbulent kinetic energy budgets, $K1$]{Normalized budgets of the turbulent kinetic energy for the model trained on $K1$ and applied to $K1$ at $x/H_0=3$. The left plot depicts the network's prediction relative to the DNS, the right plot is the reference computed from the filtered data. The DNS data is plotted with dashed lines and the GAN/Filtered-DNS data are plotted with solid lines. $C$ is the mean convective transport, $T$ the turbulent transport, $V$ the viscous transport, $P$ the velocity-pressure gradient correlation, $\mathcal{P}$ the production by the mean shear and $\widetilde{\varepsilon}$ the viscous dissipation.}
  \label{fig:TKE_Eq_LowLow}
\end{figure}

As previously discussed, the main source of turbulent kinetic energy in flames below the critical Karlovitz number is not mean shear production, but pressure dilatation. In fact, mean shear "production" drains from the turbulent kinetic energy in this regime~\citep{macart17a}. 
Figure \ref{fig:TKE_Eq_LowLow} shows the turbulent kinetic energy budgets normalized by the centerline density, centerline velocity, and local jet half-width at an upstream position of $x/H_0=3$ over the Favre-averaged progress variable. 
As evident from the figure, the agreement of the network's prediction with the DNS is extremely good. 
At low $Ka$, the velocity-pressure gradient correlation is the main source of turbulent kinetic energy, while the mean shear is a sink of energy across the entire flame brush. The lines of the DNS and GAN are virtually indistinguishable for these terms, conversely the filtered data represent them only qualitatively. 
Overall, the prediction of the dissipation with the GAN deviates from the ground truth to the largest degree in comparison with other components, as the network has the most issues predicting the smallest turbulent length scales at which dissipation takes place. Although the dissipation exhibits the largest gap in the network's prediction relative to the DNS, there is nonetheless a marked difference to the filtered field. This suggests that small scales are at least partially learned by the network as well. It should be noted though that this analysis only utilized the velocity statistics from the network, as density, pressure, and progress variable were unavailable as an outcome of the training and therefore taken as the DNS values. Nonetheless, the network seems to learn correctly how to fill the gap of the magnitude of the budgets between filtered and DNS data. It can hence be concluded that the network is able to correctly learn the behavior associated with the production of turbulent kinetic energy below the critical Karlovitz number. Furthermore, the model is also able to predict the magnitude of the budgets with nearly perfect accuracy.

\begin{figure}[t]
  \centering
  \subfloat[GAN vs DNS]{\adjincludegraphics[width=0.485\linewidth, trim={{0.15\width} {.1\width} {0\width} {.2\width}}, clip=true]{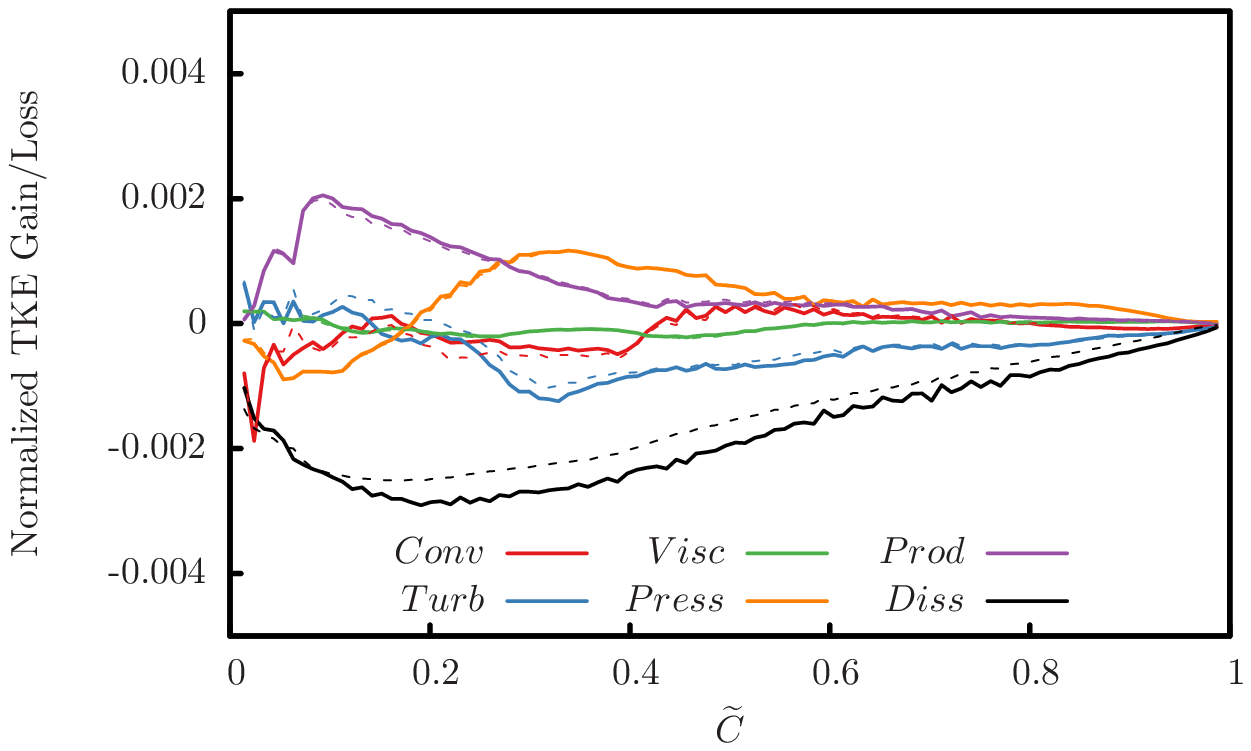}}
  \quad
  \subfloat[Filtered Data vs DNS]{\adjincludegraphics[width=0.485\linewidth, trim={{0.15\width} {.1\width} {0\width} {.2\width}} , clip=true]{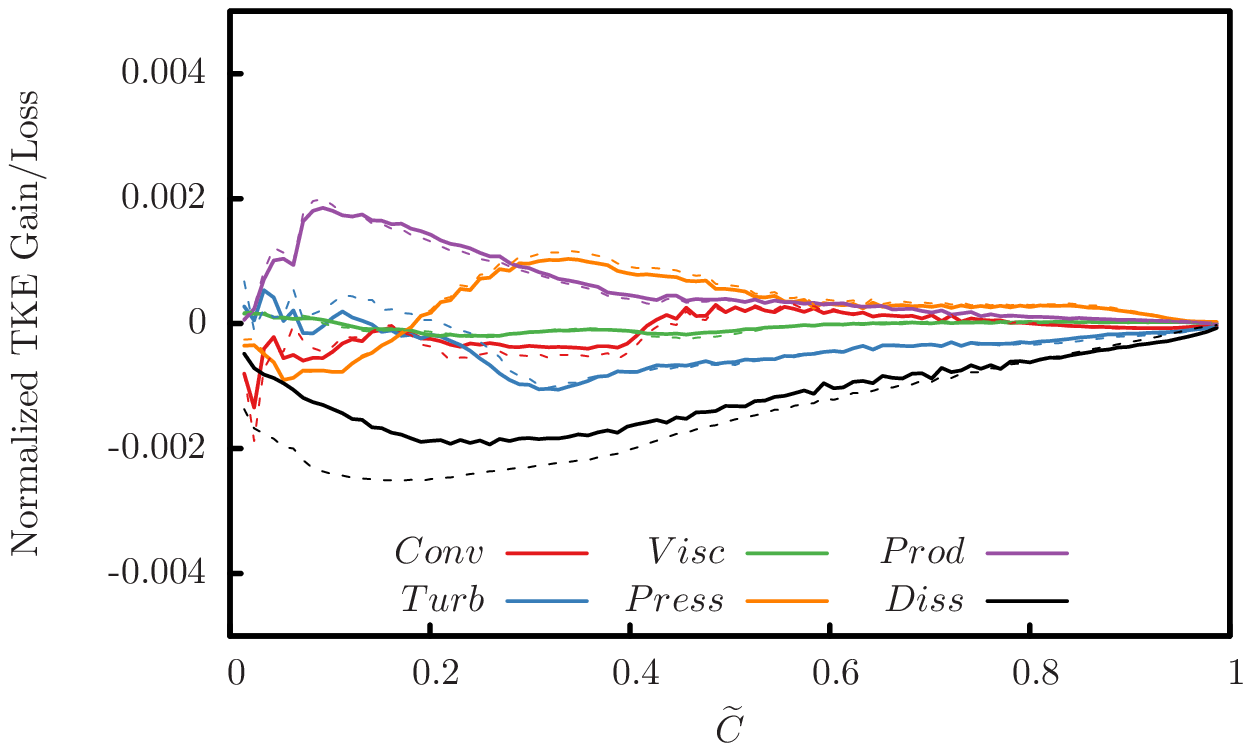}}
  \caption[Turbulent kinetic energy budgets, $K2$]{Budgets of the turbulent kinetic energy for the model trained on $K2$ and applied to $K2$ at $x/H_0=11$. The left plot depicts the network's prediction relative to the DNS, the right plot is the reference computed from the filtered data. The DNS data are plotted with dashed lines and the GAN/Filtered-DNS data are plotted with solid lines. $C$ is the mean convective transport, $T$ the turbulent transport, $V$ the viscous transport, $P$ the velocity-pressure gradient correlation, $\mathcal{P}$ the production by the mean shear and $\widetilde{\varepsilon}$ the viscous dissipation.}
  \label{fig:TKE_Eq_HighHigh}
\end{figure}

Above the critical $Ka$, the energy budgets behave similarly to non-reacting turbulence~\citep{macart17a, grenga_dmdbook}.
Therefore, the mean shear production should be positive and balanced by dissipation, while the velocity-pressure gradient should not play a significant role. 
Figure~\ref{fig:TKE_Eq_HighHigh} depicts this behavior for a plane at $x/H_0 = 11$, where the flame is fully developed (Fig. \ref{field1}). 
They are also present in the filtered field, as shown in the right panel of figure~\ref{fig:TKE_Eq_HighHigh}, but with some expected differences with respect to the DNS.
The GAN model predicts with high accuracy the magnitude of all terms.   
The magnitude of dissipation is overpredicted by the network and the strongest deviation from the DNS statistics is observed for this term. This is similar to what was observed for the $K1$ case, although it is more evident above the $Ka_{cr}$ as the term has a larger impact on the energy budget. The deviation is consistent with the network's limitation to accurately predict small-scale fluctuations. 
The velocity-pressure correlation is negative at low values of the progress variable but becomes positive in regimes where combustion takes place. Overall, it is less significant than the mean shear, though it cannot be neglected. As with the $K1$ case, the turbulent transport is larger in magnitude than the viscous transport, but these terms are relatively insignificant when it comes to gain or lose of turbulent kinetic energy. Apart from the dissipation, the network not only captures the general trend of the budgets over the progress variable, but also captures the magnitude with great accuracy. Compared to the filtered data, which is already close to the DNS data, the most noticeable difference is the dissipation which is underpredicted in the filtered data and overpredicted by the network. Overall, the network certainly improves the agreement with the DNS.

\subsection{Mixed training}

A data-driven model for LES closure has necessarily to be general. Therefore, applying the network to the same dataset it was trained on does not adequately represent the model’s performance as an LES model. To judge whether the GAN may be applied universally, and to understand whether it is learning statistics specific to the different physical regimes, it was trained on random data from both datasets.

For this purpose, the stages used during training were created with shuffled data from either the $K1$ and $K2$ datasets. Four randomly selected snapshots from each of the datasets comprise one stage. In one batch, the size of the input to the network must be consistent. Therefore, the size of the boxes of the $K2$ dataset was increased to match the box size of the $K1$ dataset at $32 \times 32 \times 32$. This means that the ratio of the box to filter size was increased and the boxes used for the $K2$ dataset were not the same used previously. 
Due to the larger domain size of the $K2$ case, there are also approximately $15\%$ more boxes of this dataset in one stage, and therefore there are potentially more grid points for the network to learn the regime above the critical $Ka$. There are two ways in which the dataset can be normalized. As the velocity magnitude is nearly $4\times$ higher for the $K2$ case, the mixed datasets may be normalized consistently with the $K2$ dataset. However, this was found to adversely affect the prediction of the $K1$ data. Likely, this is a consequence of suboptimal usage of the range of the normalization as only a subset is used. At the same time, because the loss function is computed on the normalized datasets, \textit{absolute} errors of the $K1$ prediction are not punished as severely as for the $K2$ case, where the same \textit{relative} error results in increased loss values. Therefore, the datasets were individually normalized to a range between $0$ and $1$. To account for the additional data, the network was trained for a total of $8$ stages, which is double compared to the independent trainings.

The $K2$ case barely differs from the individual training except for the small scales, and the reconstruction quality relative to the $K1$ case is superior. Yet, as with the individual training, there remains a gap between the DNS data and the field produced by the network.
Similarly, the $K1$ case does not suffer considerably from the mixed training but the deviation from the DNS data is, anyway, more evident. Nonetheless, this finding is in line with the consistent training and application on the $K1$ dataset. 
This is bolstered by the mean root-mean-square-error of the velocity components,  which increases by a minor $310$ basis point compared to the GAN model trained with the same dataset. The findings for the $K2$ case are consistent. Compared to the GAN model for $K2$ trained only with $K2$, there is a $10.2\%$ increase in the root-mean-square-error. This effect cannot be attributed to the amount of data trained for, as it did not differ between the applications. In fact, the number of boxes of the $K2$ case and thus grid points available to the network is slightly larger. However, the ratio of grid points of the cases is still close to unity and thus should not have a profound impact. One difference that could be responsible for the relatively inferior prediction of the $K2$ snapshot is the low batch size due to the large boxes employed in the training. Conversely, one could argue that larger scales are contained in the boxes which should aid the network in reconstruction.

\begin{figure}[t]
  \centering
  \subfloat[$K1,\: \tau_{11}^r$]{\adjincludegraphics[width=0.45\linewidth, clip=true]{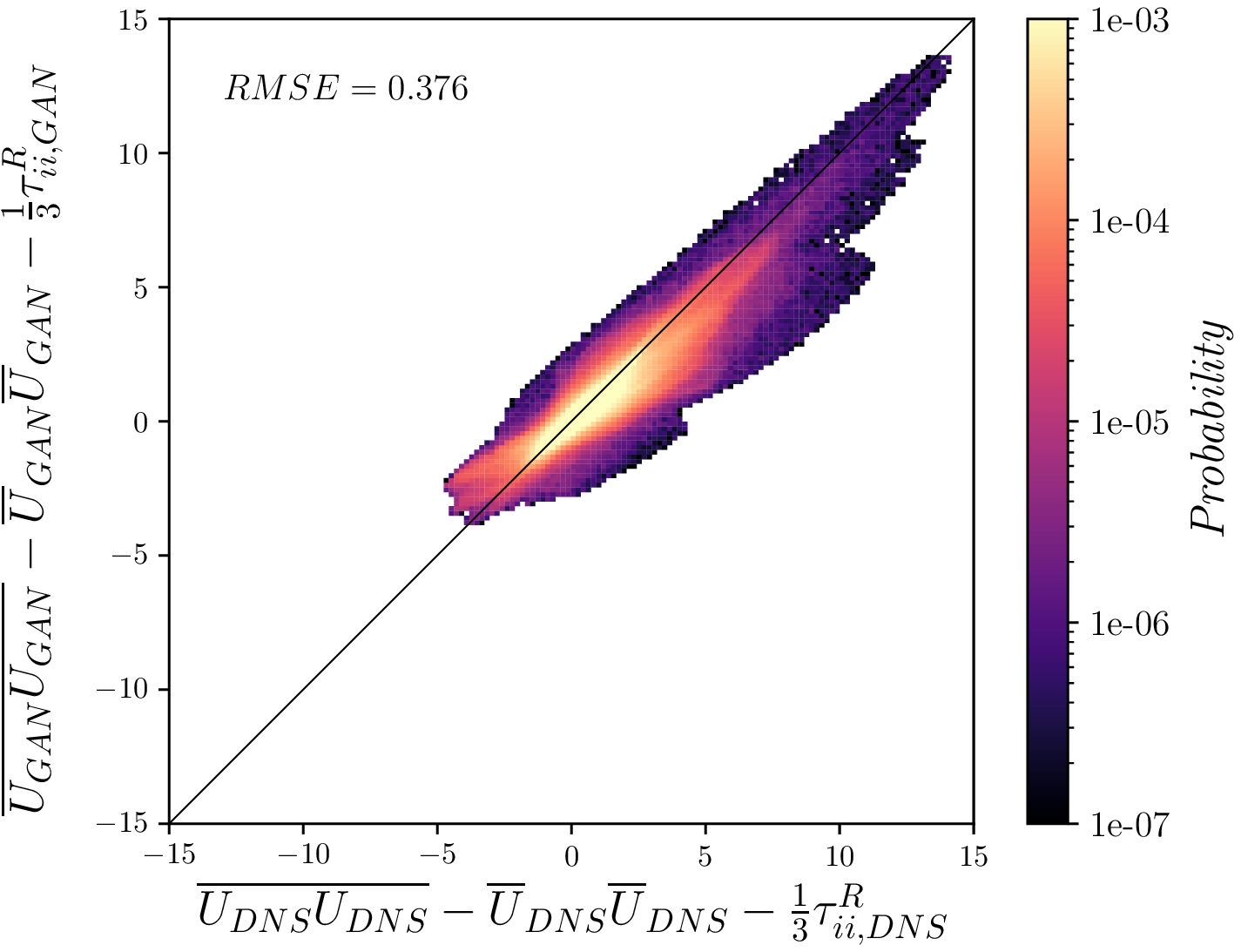}}\hspace{1mm}
  \subfloat[$K1,\: \tau_{12}^r$]{\adjincludegraphics[width=0.45\linewidth, clip=true]{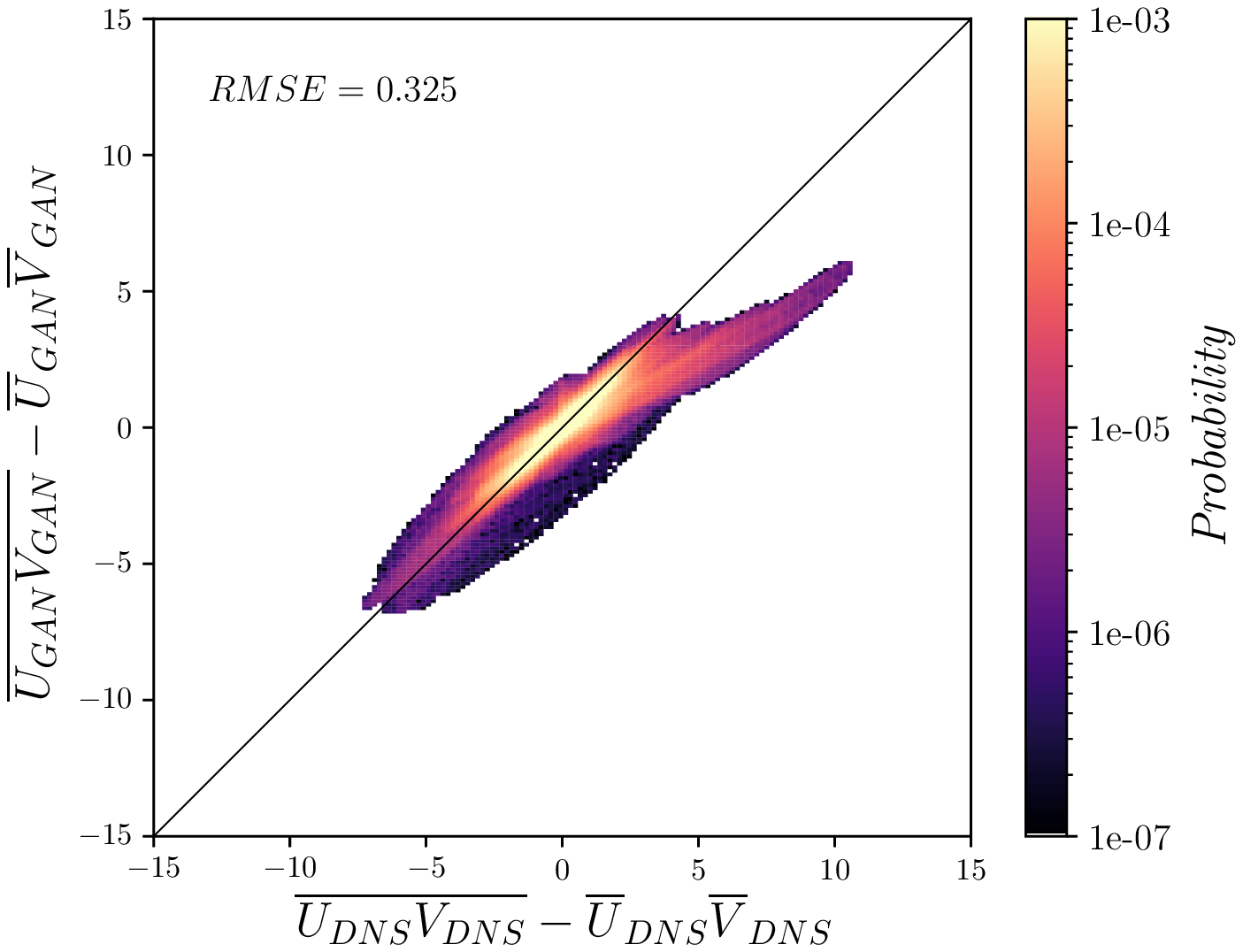}}\hfill
  \subfloat[$K2,\: \tau_{11}^r$]{\adjincludegraphics[width=0.45\linewidth, clip=true]{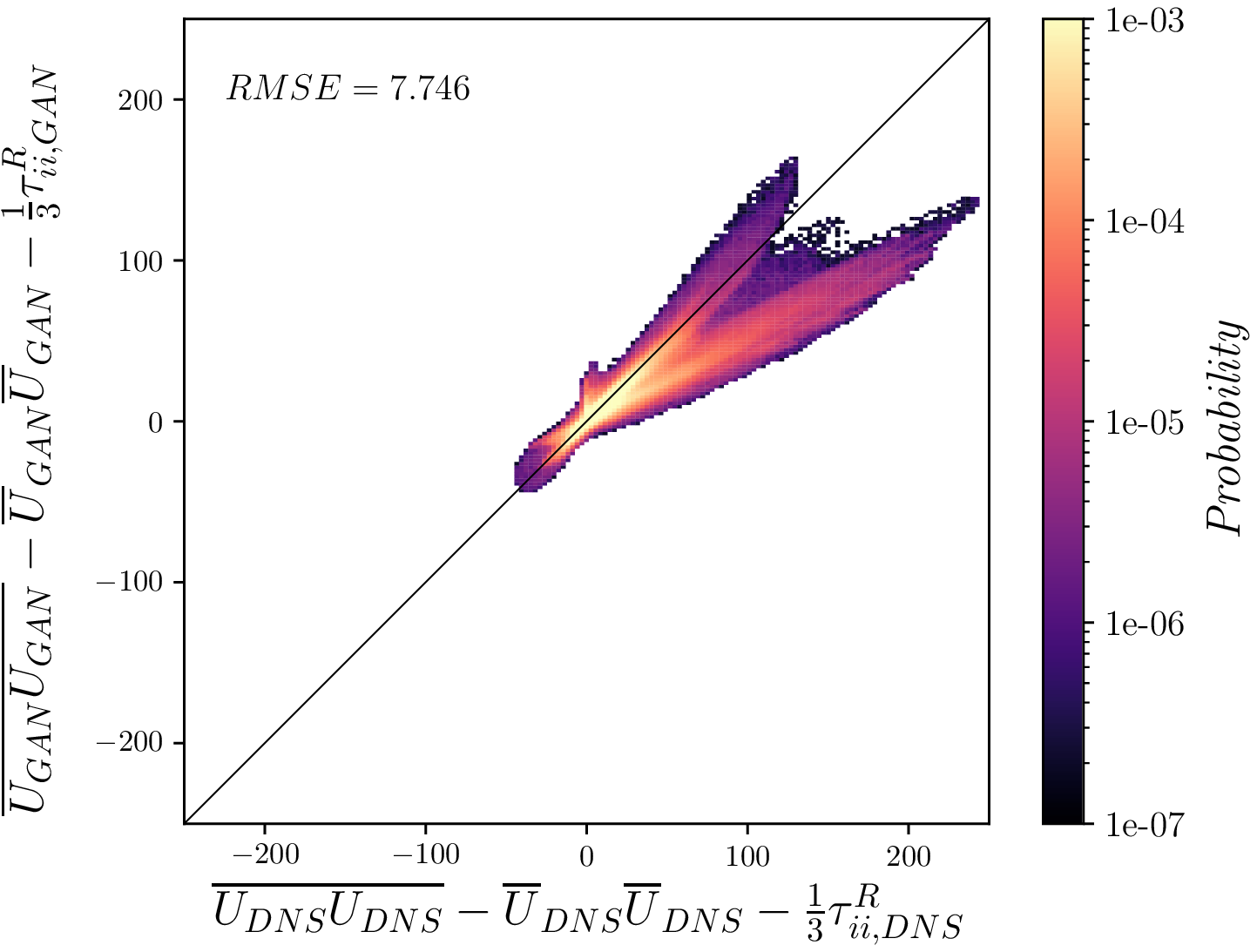}}\hspace{1mm}
  \subfloat[$K2,\: \tau_{12}^r$]{\adjincludegraphics[width=0.45\linewidth, clip=true]{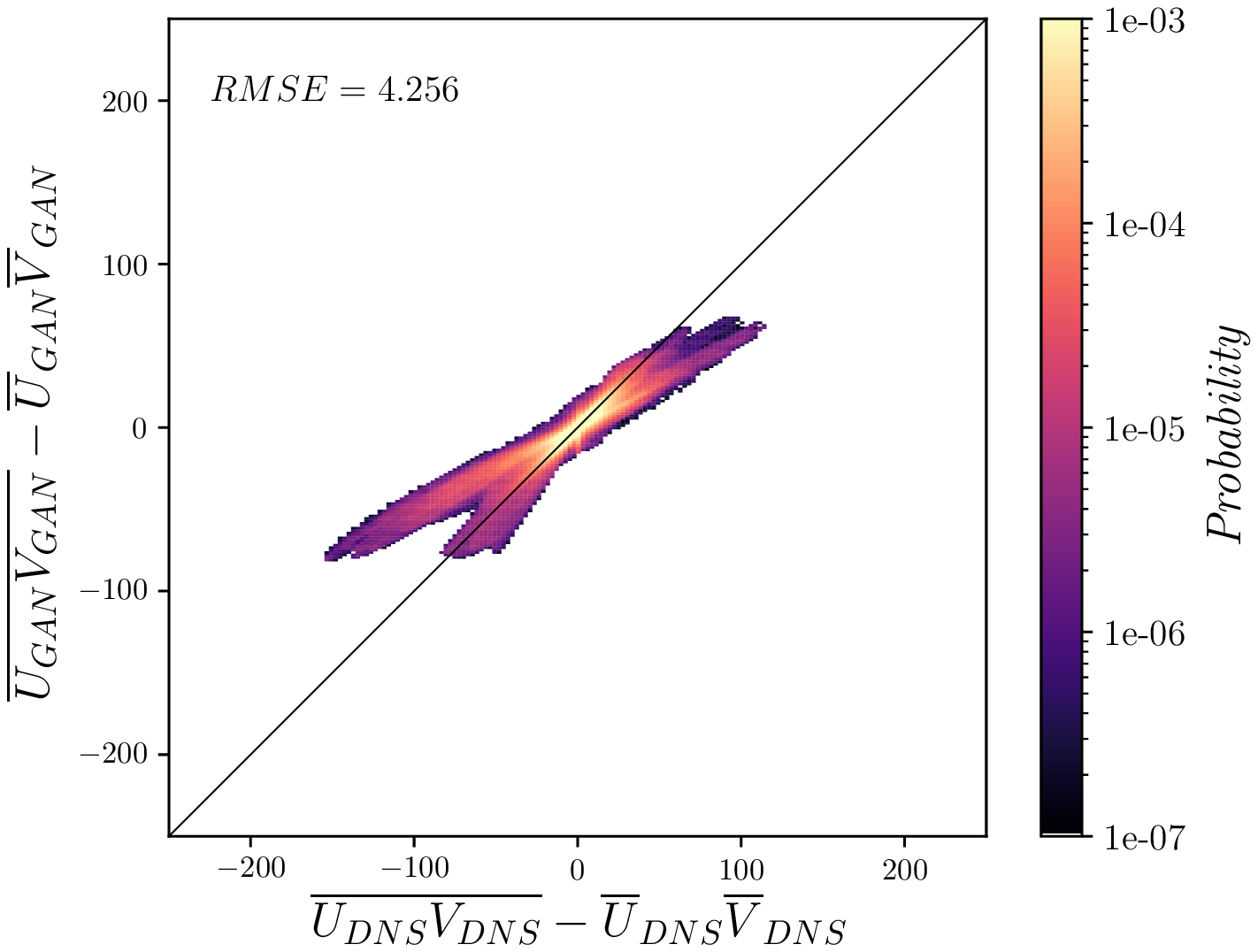}}\hfill
  \caption[jPDF subfilter-scales, mixed training]{Joint PDF plots of the subfilter-scales stress tensor components $\tau_{11}^r$ and $\tau_{12}^r$ for the network trained on a mixture of $K1$ and $K2$ data. 
  The subfilter-scale stresses are evaluated from the DNS data are reported on the horizontal axis, while on the vertical axis the same quantities are evaluated with the GAN data. RMSE indicates the Root Mean Squared Error.}
  \label{fig:jPDF_vel_Mixed}
\end{figure}

Figure \ref{fig:jPDF_vel_Mixed} depicts the jPDF plots of selected subfilter stress-tensor components. Qualitatively, the subfilter stresses are similar to the individual training. This is exemplified by the mean cross-correlation of the stress components, which decreases by $0.23\%$-points for the $K1$ case and by $1.45\%$-points for the $K2$ case relative to training and application on the same dataset. In both cases, the root-mean-square-error increases slightly but there are no artifacts or strong outliers in the jPDF. 

\begin{figure}[t]
  \centering
  \subfloat[$K1$]{\adjincludegraphics[width=0.485\linewidth, trim={{0.15\width} {.1\width} {0\width} {.2\width}}, clip=true]{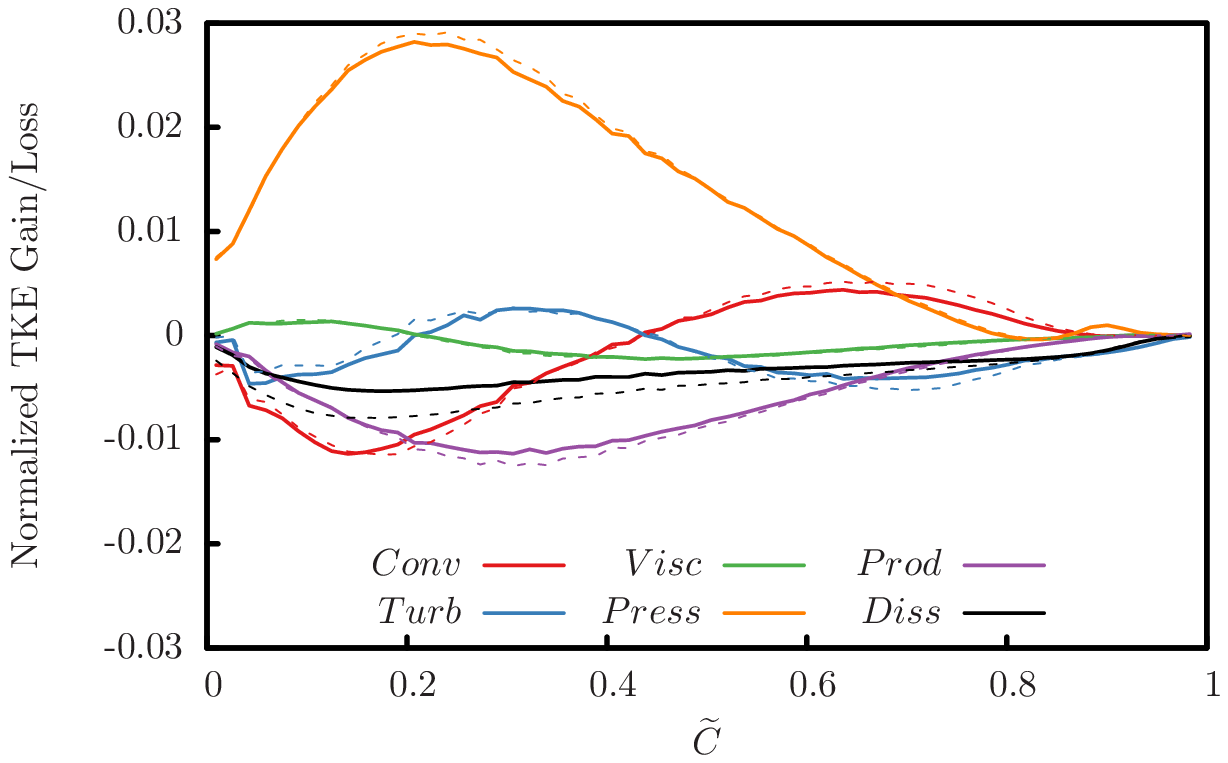}}
  \quad
  \subfloat[$K2$]{\adjincludegraphics[width=0.485\linewidth, trim={{0.15\width} {.1\width} {0\width} {.2\width}}, clip=true]{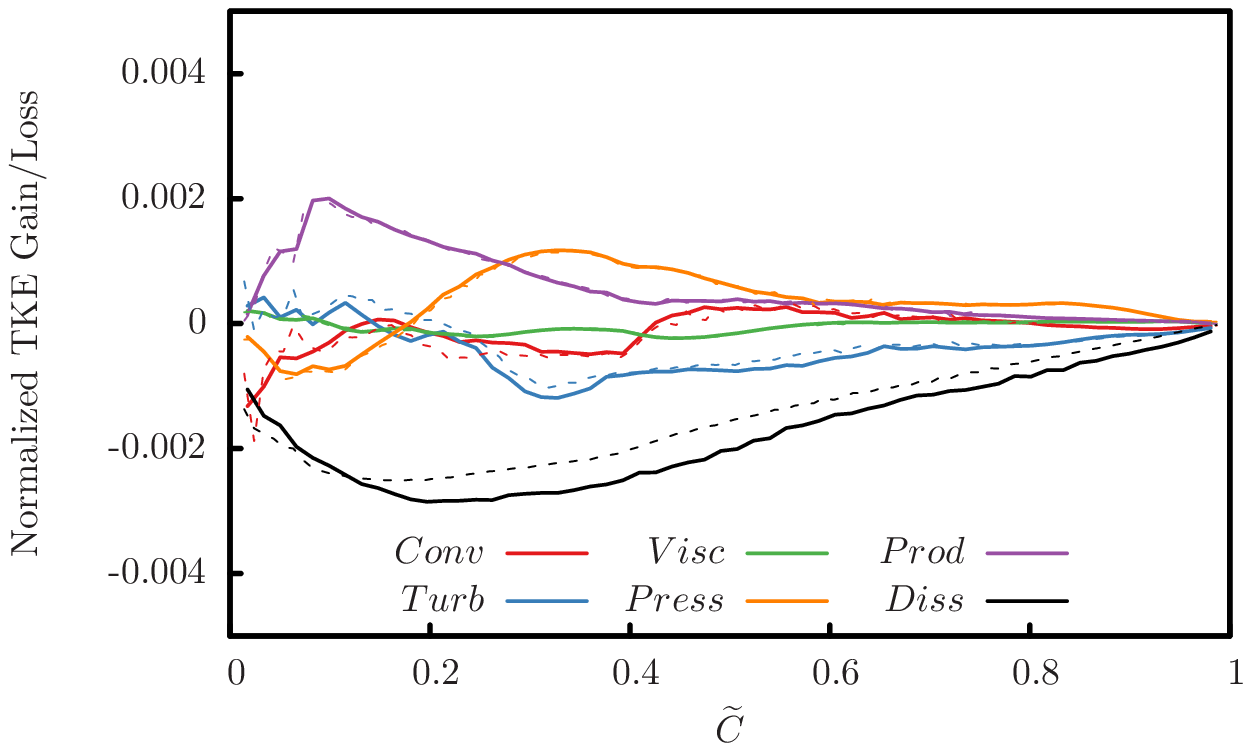}}
  \caption[Turbulent kinetic energy budgets mixed training]{Budgets of the turbulent kinetic energy for the model trained on a mixture of $K1$ and $K2$ data and applied to both datasets. $C$ is the mean convective transport, $T$ the turbulent transport, $V$ the viscous transport, $P$ the velocity-pressure gradient correlation, $\mathcal{P}$ the production by the mean shear and $\widetilde{\varepsilon}$ the viscous dissipation.} 
  \label{fig:TKE_Budgets_Mixed}
\end{figure}

In figure \ref{fig:TKE_Budgets_Mixed}, the normalized budgets of the turbulent kinetic energy obtained for both cases with the GAN model trained with a mixture of both datasets are shown. 
For the $K1$ case, all the terms are predicted nearly equally well to training exclusively with data below the critical $Ka$, making this approach a success. Small differences can be observed with respect to the DNS data, however, similar gaps can be found in the training with only $K1$ data (Fig.~\ref{fig:TKE_Eq_LowLow}). 
Also for the $K2$ case, all the terms are predicted nearly equally well to training exclusively with data above the critical $Ka$. Mean shear is matched with the DNS and the velocity-pressure gradient correlation is captured equally well. Small differences can be observed for the convective and turbulent transport terms as well as for the dissipation. Just like the previous case, the same gaps can be found in the training with only $K2$ data (Fig.~\ref{fig:TKE_Eq_HighHigh}).
Therefore, the mixed training, when the datasets are normalized individually, is a viable approach yielding results nearly equivalent to individual training and application on the datasets.

 \begin{figure}[t]
  \centering
  \subfloat[$K1$, DNS]{\adjincludegraphics[width=0.49\linewidth, trim={{0.2\width} {0\width} {0.1\width} {0\width}}, clip=true]{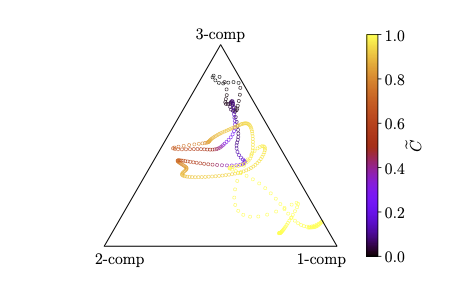}}\hspace{1mm}
  \subfloat[$K1$, Mixed training]{\adjincludegraphics[width=0.49\linewidth, trim={{0.2\width} {0\width} {0.1\width} {0\width}}, clip=true]{plot/Mixed_lowKa_x_Ho_3.png}}\hspace{1mm}
  \subfloat[$K2$, DNS]{\adjincludegraphics[width=0.49\linewidth, trim={{0.2\width} {0\width} {0.1\width} {0\width}}, clip=true]{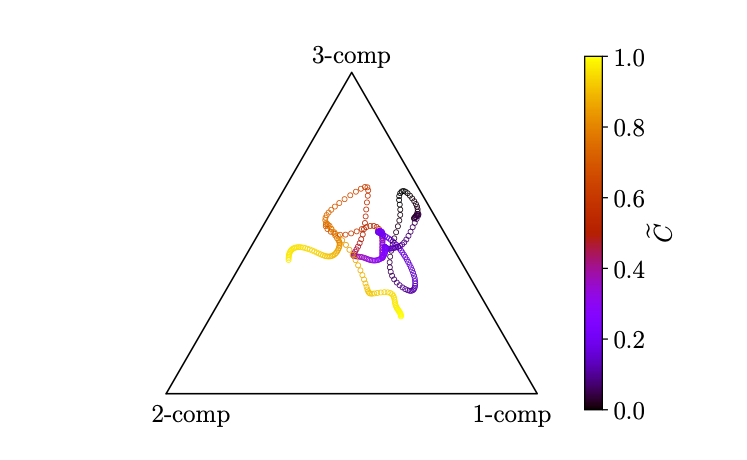}}\hfill
  \subfloat[$K2$, Mixed training]{\adjincludegraphics[width=0.49\linewidth, trim={{0.2\width} {0\width} {0.1\width} {0\width}}, clip=true]{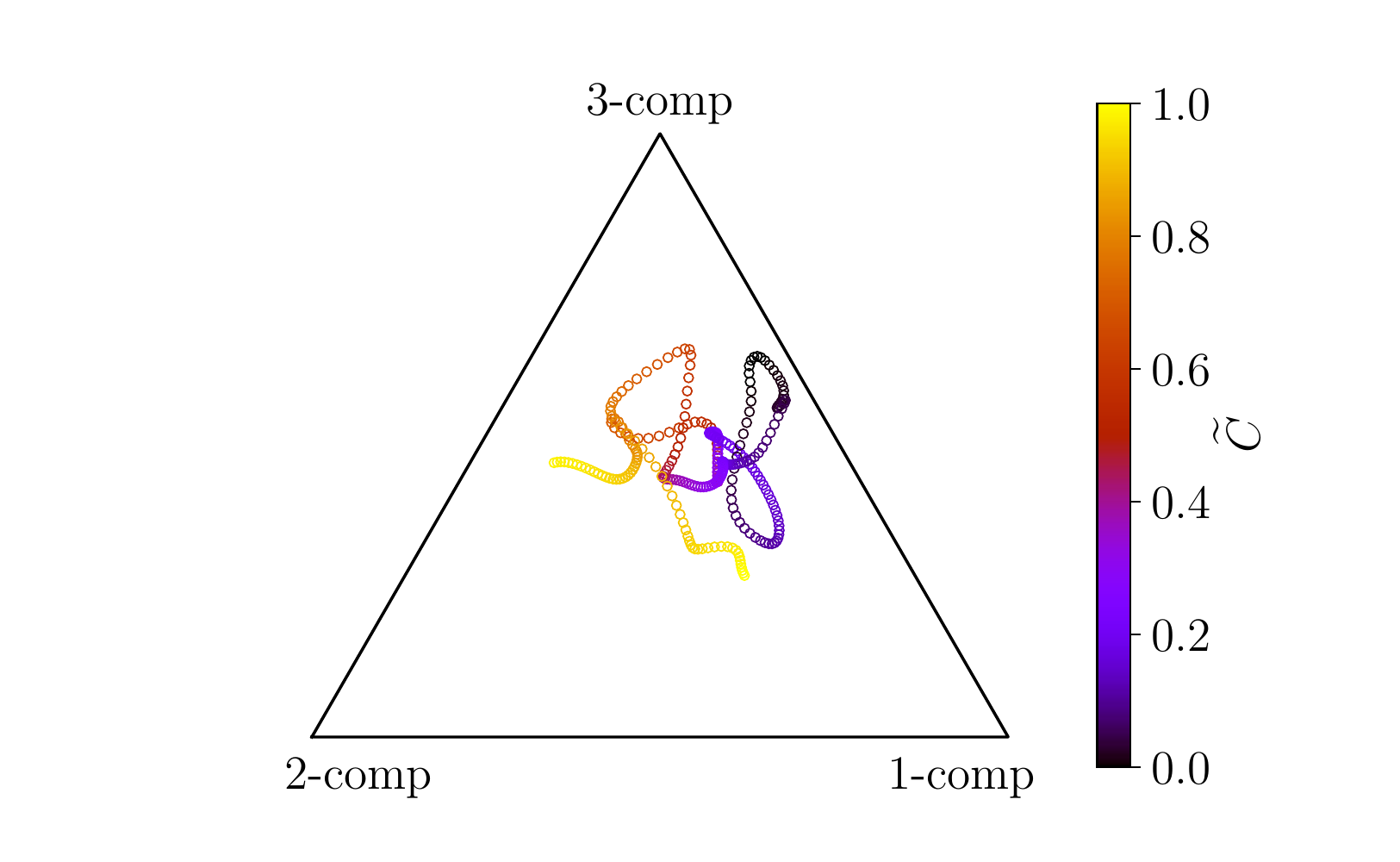}}\hfill
  \caption[map, mixed training]{Barycentric maps of Reynolds stress invariants for $K1$ case (left) and $K2$ case (right) for the network trained on a mixture of $K1$ and $K2$ data.}
  \label{fig:Map_vel_Mixed}
\end{figure}

In order to verify the validity of the GAN model, a comparison of the Reynolds stress invariant is shown in Fig.~\ref{fig:Map_vel_Mixed} on the Lumley triangles~\citep{lumley1977return}. For the case $K1$, the turbulence in the unburned reactants ($\tilde{C} =0$) is close to the three-components limit; it is, then, modified by the flame and the shear becoming two-components within the flame and ending in the one-component limit in the burned products.
Conversely, the case $K2$ never approaches one of the limit conditions: the unburned reactants are preferentially one-component, while the burned products are two-components. 
For both cases, there is a profound agreement between the DNS data and the ones for the GAN model, showing that the turbulence features are totally recovered.
  
In conclusion, training the network with mixed data from the $K1$ and $K2$ datasets can result in quality rivaling individual training and consistent application of the network. The mixed prediction is not only of the same accuracy for the instantaneous field, but also for subfilter stress and the recovery of turbulent kinetic energy budgets. The additional data from the respectively different regimes is therefore conducive to the prediction. 
 
\section*{Conclusions}

The recent progress on ML-based architectures has enabled deep neural networks to become a powerful tool for estimating complex non-linear relations, where classical approaches often have relevant limitations.
In this context, super-resolution GANs, originally developed for images reconstruction, should be able to seek the relation between the phenomena occurring at different scales. Thus, when applied to turbulent non-reacting or reacting flows, these networks may offer the possibility to reconstruct the subfilter-scales from the large-scales (e.g. LES) data.

It has been demonstrated that in premixed flame, the interactions between heat release and turbulence occur in two asymptotic regimes. Although these regimes have been largely investigated with DNS and experiments, there still does not exist a universal model able to accurately include both.  

Considering two premixed hydrogen planar flames datasets with $Ka$ below and above the critical value, the capabilities of the GAN architecture to include the elements of both asymptotic regimes have been investigated.

At the first stage, a super-resolution approach employing a GAN was trained and applied to the same flame condition to verify the general prediction capabilities and assess its accuracy.
Even when the training and application datasets are consistent, meaning the network was trained below (above) $Ka_{cr}$ and applied below (above) $Ka_{cr}$, the instantaneous velocity fields procured by the network do not match exactly those of the DNS. The larger structures of the subfilter-scales are well recovered, while the features at the smallest scale are still missing.
However, the subfilter-scale stress and scalar fluxes are strongly aligned with the DNS data, exceeding correlations of 90\% and exhibiting a low error. These are promising results highlighting the potential of a valid GAN-based closure model for LES.
 
Mixed training with data of both physical regimes was realized in order to verify the possibility to obtain a universal model-based GAN. It was found that such a model is able to achieve performance very similar to consistent, individual training and application on one of the datasets only. 
However, this is only the case if the datasets are normalized individually and not with a normalization consistent with a combined set of the datasets. 
The subfilter stresses predicted by the network are very close to individual training and application on the same dataset achieving similar correlations and errors. Moreover, the budgets of the turbulent kinetic energy and the barycentric maps of the Reynolds stress invariant indicate that the peculiarities of both regimes are learned by the network, as the DNS data was matched nearly perfectly.

\newpage
\section*{Acknowledgment}

The research leading to these results has received funding from the European Union’s Horizon 2020 research and innovation program under the Center of Excellence in Combustion (CoEC) project, grant agreement no.~\textit{952181}.
\\
Simulations were performed with computing resources granted by RWTH Aachen University under project \textit{rwth0658}. The authors gratefully acknowledge the computing resources from the DEEP-EST project, which received funding from the European Union’s Horizon 2020 research and innovation program under the grant agreement no.~\textit{754304}. We thank Mr. Sedona for the support in the porting of the application to DEEP-EST.
\\
This article has been accepted for publication in Combustion Science and Technology, published by Taylor \& Francis.

\bibliographystyle{apacite}
\bibliography{Libby_paper_ref}

\begin{thebibliography}{}

\bibitem [\protect \citeauthoryear {%
Abadi%
\ \protect \BOthers {.}}{%
Abadi%
\ \protect \BOthers {.}}{%
{\protect \APACyear {2016}}%
}]{%
abadi2016tensorflow}
\APACinsertmetastar {%
abadi2016tensorflow}%
\begin{APACrefauthors}%
Abadi, M.%
, Agarwal, A.%
, Barham, P.%
, Brevdo, E.%
, Chen, Z.%
, Citro, C.%
\BDBL {}others%
\end{APACrefauthors}%
\unskip\
\newblock
\APACrefYearMonthDay{2016}{}{}.
\newblock
{\BBOQ}\APACrefatitle {Tensorflow: Large-scale machine learning on
  heterogeneous distributed systems} {Tensorflow: Large-scale machine learning
  on heterogeneous distributed systems}.{\BBCQ}
\newblock
\APACjournalVolNumPages{arXiv preprint arXiv:1603.04467}{}{}{}.
\PrintBackRefs{\CurrentBib}

\bibitem [\protect \citeauthoryear {%
Aspden%
, Day%
\BCBL {}\ \BBA {} Bell%
}{%
Aspden%
\ \protect \BOthers {.}}{%
{\protect \APACyear {2011}}%
}]{%
aspden2011turbulence}
\APACinsertmetastar {%
aspden2011turbulence}%
\begin{APACrefauthors}%
Aspden, A.%
, Day, M.%
\BCBL {}\ \BBA {} Bell, J.%
\end{APACrefauthors}%
\unskip\
\newblock
\APACrefYearMonthDay{2011}{}{}.
\newblock
{\BBOQ}\APACrefatitle {Turbulence--flame interactions in lean premixed
  hydrogen: transition to the distributed burning regime} {Turbulence--flame
  interactions in lean premixed hydrogen: transition to the distributed burning
  regime}.{\BBCQ}
\newblock
\APACjournalVolNumPages{Journal of Fluid mechanics}{680}{}{287--320}.
\PrintBackRefs{\CurrentBib}

\bibitem [\protect \citeauthoryear {%
Batchelor%
}{%
Batchelor%
}{%
{\protect \APACyear {1952}}%
}]{%
batchelor1952effect}
\APACinsertmetastar {%
batchelor1952effect}%
\begin{APACrefauthors}%
Batchelor, G\BPBI K.%
\end{APACrefauthors}%
\unskip\
\newblock
\APACrefYearMonthDay{1952}{}{}.
\newblock
{\BBOQ}\APACrefatitle {The effect of homogeneous turbulence on material lines
  and surfaces} {The effect of homogeneous turbulence on material lines and
  surfaces}.{\BBCQ}
\newblock
\APACjournalVolNumPages{Proceedings of the Royal Society of London. Series A.
  Mathematical and Physical Sciences}{213}{1114}{349--366}.
\PrintBackRefs{\CurrentBib}

\bibitem [\protect \citeauthoryear {%
Bilger%
}{%
Bilger%
}{%
{\protect \APACyear {2004}}%
}]{%
bilger2004}
\APACinsertmetastar {%
bilger2004}%
\begin{APACrefauthors}%
Bilger, R.%
\end{APACrefauthors}%
\unskip\
\newblock
\APACrefYearMonthDay{2004}{}{}.
\newblock
{\BBOQ}\APACrefatitle {Some aspects of scalar dissipation} {Some aspects of
  scalar dissipation}.{\BBCQ}
\newblock
\APACjournalVolNumPages{Flow, turbulence and combustion}{72}{2}{93--114}.
\PrintBackRefs{\CurrentBib}

\bibitem [\protect \citeauthoryear {%
Bobbitt%
, Lapointe%
\BCBL {}\ \BBA {} Blanquart%
}{%
Bobbitt%
\ \protect \BOthers {.}}{%
{\protect \APACyear {2016}}%
}]{%
bobbitt2016vorticity}
\APACinsertmetastar {%
bobbitt2016vorticity}%
\begin{APACrefauthors}%
Bobbitt, B.%
, Lapointe, S.%
\BCBL {}\ \BBA {} Blanquart, G.%
\end{APACrefauthors}%
\unskip\
\newblock
\APACrefYearMonthDay{2016}{}{}.
\newblock
{\BBOQ}\APACrefatitle {Vorticity transformation in high Karlovitz number
  premixed flames} {Vorticity transformation in high karlovitz number premixed
  flames}.{\BBCQ}
\newblock
\APACjournalVolNumPages{Physics of Fluids}{28}{1}{015101}.
\PrintBackRefs{\CurrentBib}

\bibitem [\protect \citeauthoryear {%
Bode%
\ \protect \BOthers {.}}{%
Bode%
\ \protect \BOthers {.}}{%
{\protect \APACyear {2021}}%
}]{%
bode2021using}
\APACinsertmetastar {%
bode2021using}%
\begin{APACrefauthors}%
Bode, M.%
, Gauding, M.%
, Lian, Z.%
, Denker, D.%
, Davidovic, M.%
, Kleinheinz, K.%
\BDBL {}Pitsch, H.%
\end{APACrefauthors}%
\unskip\
\newblock
\APACrefYearMonthDay{2021}{}{}.
\newblock
{\BBOQ}\APACrefatitle {Using physics-informed enhanced super-resolution
  generative adversarial networks for subfilter modeling in turbulent reactive
  flows} {Using physics-informed enhanced super-resolution generative
  adversarial networks for subfilter modeling in turbulent reactive
  flows}.{\BBCQ}
\newblock
\APACjournalVolNumPages{Proceedings of the Combustion
  Institute}{38}{2}{2617--2625}.
\PrintBackRefs{\CurrentBib}

\bibitem [\protect \citeauthoryear {%
Bray%
, Libby%
, Masuya%
\BCBL {}\ \BBA {} Moss%
}{%
Bray%
\ \protect \BOthers {.}}{%
{\protect \APACyear {1981}}%
}]{%
bray1981turbulence}
\APACinsertmetastar {%
bray1981turbulence}%
\begin{APACrefauthors}%
Bray, K.%
, Libby, P\BPBI A.%
, Masuya, G.%
\BCBL {}\ \BBA {} Moss, J.%
\end{APACrefauthors}%
\unskip\
\newblock
\APACrefYearMonthDay{1981}{}{}.
\newblock
{\BBOQ}\APACrefatitle {Turbulence production in premixed turbulent flames}
  {Turbulence production in premixed turbulent flames}.{\BBCQ}
\newblock
\APACjournalVolNumPages{Combustion Science and Technology}{}{}{}.
\PrintBackRefs{\CurrentBib}

\bibitem [\protect \citeauthoryear {%
Bray%
, Libby%
\BCBL {}\ \BBA {} Moss%
}{%
Bray%
\ \protect \BOthers {.}}{%
{\protect \APACyear {1985}}%
}]{%
BRAY198587}
\APACinsertmetastar {%
BRAY198587}%
\begin{APACrefauthors}%
Bray, K.%
, Libby, P\BPBI A.%
\BCBL {}\ \BBA {} Moss, J.%
\end{APACrefauthors}%
\unskip\
\newblock
\APACrefYearMonthDay{1985}{}{}.
\newblock
{\BBOQ}\APACrefatitle {Unified modeling approach for premixed turbulent
  combustion—Part I: General formulation} {Unified modeling approach for
  premixed turbulent combustion—part i: General formulation}.{\BBCQ}
\newblock
\APACjournalVolNumPages{Combustion and flame}{61}{1}{87--102}.
\PrintBackRefs{\CurrentBib}

\bibitem [\protect \citeauthoryear {%
Brenner%
, Eldredge%
\BCBL {}\ \BBA {} Freund%
}{%
Brenner%
\ \protect \BOthers {.}}{%
{\protect \APACyear {2019}}%
}]{%
Brenner2019}
\APACinsertmetastar {%
Brenner2019}%
\begin{APACrefauthors}%
Brenner, M.%
, Eldredge, J.%
\BCBL {}\ \BBA {} Freund, J.%
\end{APACrefauthors}%
\unskip\
\newblock
\APACrefYearMonthDay{2019}{}{}.
\newblock
{\BBOQ}\APACrefatitle {Perspective on machine learning for advancing fluid
  mechanics} {Perspective on machine learning for advancing fluid
  mechanics}.{\BBCQ}
\newblock
\APACjournalVolNumPages{Physical Review Fluids}{4}{10}{100501}.
\PrintBackRefs{\CurrentBib}

\bibitem [\protect \citeauthoryear {%
Cao%
, Nishino%
, Mizuno%
\BCBL {}\ \BBA {} Torii%
}{%
Cao%
\ \protect \BOthers {.}}{%
{\protect \APACyear {2000}}%
}]{%
cao2000piv}
\APACinsertmetastar {%
cao2000piv}%
\begin{APACrefauthors}%
Cao, Z\BHBI M.%
, Nishino, K.%
, Mizuno, S.%
\BCBL {}\ \BBA {} Torii, K.%
\end{APACrefauthors}%
\unskip\
\newblock
\APACrefYearMonthDay{2000}{}{}.
\newblock
{\BBOQ}\APACrefatitle {PIV measurement of internal structure of diesel fuel
  spray} {Piv measurement of internal structure of diesel fuel spray}.{\BBCQ}
\newblock
\APACjournalVolNumPages{Experiments in fluids}{29}{1}{S211--S219}.
\PrintBackRefs{\CurrentBib}

\bibitem [\protect \citeauthoryear {%
Commission%
}{%
Commission%
}{%
{\protect \APACyear {2021}}%
}]{%
EUgreen}
\APACinsertmetastar {%
EUgreen}%
\begin{APACrefauthors}%
Commission, E.%
\end{APACrefauthors}%
\unskip\
\newblock
\APACrefYearMonthDay{2021}{}{}.
\newblock
\APACrefbtitle {A European Green Deal.} {A european green deal.}
\newblock
\begin{APACrefURL}
  [{2021-11-01}]\url{https://ec.europa.eu/info/strategy/priorities-2019-2024/european-green-deal}
  \end{APACrefURL}
\PrintBackRefs{\CurrentBib}

\bibitem [\protect \citeauthoryear {%
Davis%
, Joshi%
, Wang%
\BCBL {}\ \BBA {} Egolfopoulos%
}{%
Davis%
\ \protect \BOthers {.}}{%
{\protect \APACyear {2005}}%
}]{%
davis2005optimized}
\APACinsertmetastar {%
davis2005optimized}%
\begin{APACrefauthors}%
Davis, S\BPBI G.%
, Joshi, A\BPBI V.%
, Wang, H.%
\BCBL {}\ \BBA {} Egolfopoulos, F.%
\end{APACrefauthors}%
\unskip\
\newblock
\APACrefYearMonthDay{2005}{}{}.
\newblock
{\BBOQ}\APACrefatitle {An optimized kinetic model of H2/CO combustion} {An
  optimized kinetic model of h2/co combustion}.{\BBCQ}
\newblock
\APACjournalVolNumPages{Proceedings of the combustion
  Institute}{30}{1}{1283--1292}.
\PrintBackRefs{\CurrentBib}

\bibitem [\protect \citeauthoryear {%
Deng%
, He%
, Liu%
\BCBL {}\ \BBA {} Kim%
}{%
Deng%
\ \protect \BOthers {.}}{%
{\protect \APACyear {2019}}%
}]{%
deng2019super}
\APACinsertmetastar {%
deng2019super}%
\begin{APACrefauthors}%
Deng, Z.%
, He, C.%
, Liu, Y.%
\BCBL {}\ \BBA {} Kim, K\BPBI C.%
\end{APACrefauthors}%
\unskip\
\newblock
\APACrefYearMonthDay{2019}{}{}.
\newblock
{\BBOQ}\APACrefatitle {Super-resolution reconstruction of turbulent velocity
  fields using a generative adversarial network-based artificial intelligence
  framework} {Super-resolution reconstruction of turbulent velocity fields
  using a generative adversarial network-based artificial intelligence
  framework}.{\BBCQ}
\newblock
\APACjournalVolNumPages{Physics of Fluids}{31}{12}{125111}.
\PrintBackRefs{\CurrentBib}

\bibitem [\protect \citeauthoryear {%
Desjardins%
, Blanquart%
, Balarac%
\BCBL {}\ \BBA {} Pitsch%
}{%
Desjardins%
\ \protect \BOthers {.}}{%
{\protect \APACyear {2008}}%
}]{%
Desjardins2008}
\APACinsertmetastar {%
Desjardins2008}%
\begin{APACrefauthors}%
Desjardins, O.%
, Blanquart, G.%
, Balarac, G.%
\BCBL {}\ \BBA {} Pitsch, H.%
\end{APACrefauthors}%
\unskip\
\newblock
\APACrefYearMonthDay{2008}{}{}.
\newblock
{\BBOQ}\APACrefatitle {High order conservative finite difference scheme for
  variable density low Mach number turbulent flows} {High order conservative
  finite difference scheme for variable density low mach number turbulent
  flows}.{\BBCQ}
\newblock
\APACjournalVolNumPages{Journal of Computational Physics}{227}{15}{7125--7159}.
\PrintBackRefs{\CurrentBib}

\bibitem [\protect \citeauthoryear {%
Driscoll%
\ \BBA {} Gulati%
}{%
Driscoll%
\ \BBA {} Gulati%
}{%
{\protect \APACyear {1988}}%
}]{%
driscoll1988measurement}
\APACinsertmetastar {%
driscoll1988measurement}%
\begin{APACrefauthors}%
Driscoll, J\BPBI F.%
\BCBT {}\ \BBA {} Gulati, A.%
\end{APACrefauthors}%
\unskip\
\newblock
\APACrefYearMonthDay{1988}{}{}.
\newblock
{\BBOQ}\APACrefatitle {Measurement of various terms in the turbulent kinetic
  energy balance within a flame and comparison with theory} {Measurement of
  various terms in the turbulent kinetic energy balance within a flame and
  comparison with theory}.{\BBCQ}
\newblock
\APACjournalVolNumPages{Combustion and flame}{72}{2}{131--152}.
\PrintBackRefs{\CurrentBib}

\bibitem [\protect \citeauthoryear {%
Flemming%
, Sadiki%
\BCBL {}\ \BBA {} Janicka%
}{%
Flemming%
\ \protect \BOthers {.}}{%
{\protect \APACyear {2005}}%
}]{%
flemming2005using}
\APACinsertmetastar {%
flemming2005using}%
\begin{APACrefauthors}%
Flemming, F.%
, Sadiki, A.%
\BCBL {}\ \BBA {} Janicka, J.%
\end{APACrefauthors}%
\unskip\
\newblock
\APACrefYearMonthDay{2005}{}{}.
\newblock
{\BBOQ}\APACrefatitle {LES using artificial neural networks for chemistry
  representation} {Les using artificial neural networks for chemistry
  representation}.{\BBCQ}
\newblock
\APACjournalVolNumPages{Progress in Computational Fluid Dynamics, An
  International Journal}{5}{7}{375--385}.
\PrintBackRefs{\CurrentBib}

\bibitem [\protect \citeauthoryear {%
Fukami%
, Fukagata%
\BCBL {}\ \BBA {} Taira%
}{%
Fukami%
\ \protect \BOthers {.}}{%
{\protect \APACyear {2019}}%
}]{%
fukami2019super}
\APACinsertmetastar {%
fukami2019super}%
\begin{APACrefauthors}%
Fukami, K.%
, Fukagata, K.%
\BCBL {}\ \BBA {} Taira, K.%
\end{APACrefauthors}%
\unskip\
\newblock
\APACrefYearMonthDay{2019}{}{}.
\newblock
{\BBOQ}\APACrefatitle {Super-resolution reconstruction of turbulent flows with
  machine learning} {Super-resolution reconstruction of turbulent flows with
  machine learning}.{\BBCQ}
\newblock
\APACjournalVolNumPages{Journal of Fluid Mechanics}{870}{}{106--120}.
\PrintBackRefs{\CurrentBib}

\bibitem [\protect \citeauthoryear {%
Fureby%
}{%
Fureby%
}{%
{\protect \APACyear {2005}}%
}]{%
fureby2005fractal}
\APACinsertmetastar {%
fureby2005fractal}%
\begin{APACrefauthors}%
Fureby, C.%
\end{APACrefauthors}%
\unskip\
\newblock
\APACrefYearMonthDay{2005}{}{}.
\newblock
{\BBOQ}\APACrefatitle {A fractal flame-wrinkling large eddy simulation model
  for premixed turbulent combustion} {A fractal flame-wrinkling large eddy
  simulation model for premixed turbulent combustion}.{\BBCQ}
\newblock
\APACjournalVolNumPages{Proceedings of the Combustion
  Institute}{30}{1}{593--601}.
\PrintBackRefs{\CurrentBib}

\bibitem [\protect \citeauthoryear {%
Germano%
, Piomelli%
, Moin%
\BCBL {}\ \BBA {} Cabot%
}{%
Germano%
\ \protect \BOthers {.}}{%
{\protect \APACyear {1991}}%
}]{%
germano1991dynamic}
\APACinsertmetastar {%
germano1991dynamic}%
\begin{APACrefauthors}%
Germano, M.%
, Piomelli, U.%
, Moin, P.%
\BCBL {}\ \BBA {} Cabot, W\BPBI H.%
\end{APACrefauthors}%
\unskip\
\newblock
\APACrefYearMonthDay{1991}{}{}.
\newblock
{\BBOQ}\APACrefatitle {A dynamic subgrid-scale eddy viscosity model} {A dynamic
  subgrid-scale eddy viscosity model}.{\BBCQ}
\newblock
\APACjournalVolNumPages{Physics of Fluids A: Fluid Dynamics}{3}{7}{1760--1765}.
\PrintBackRefs{\CurrentBib}

\bibitem [\protect \citeauthoryear {%
G{\'e}ron%
}{%
G{\'e}ron%
}{%
{\protect \APACyear {2019}}%
}]{%
geron2019hands}
\APACinsertmetastar {%
geron2019hands}%
\begin{APACrefauthors}%
G{\'e}ron, A.%
\end{APACrefauthors}%
\unskip\
\newblock
\APACrefYear{2019}.
\newblock
\APACrefbtitle {Hands-on machine learning with Scikit-Learn, Keras, and
  TensorFlow: Concepts, tools, and techniques to build intelligent systems}
  {Hands-on machine learning with scikit-learn, keras, and tensorflow:
  Concepts, tools, and techniques to build intelligent systems}.
\newblock
\APACaddressPublisher{}{O'Reilly Media}.
\PrintBackRefs{\CurrentBib}

\bibitem [\protect \citeauthoryear {%
Grenga%
, MacArt%
\BCBL {}\ \BBA {} Mueller%
}{%
Grenga%
\ \protect \BOthers {.}}{%
{\protect \APACyear {2018}}%
}]{%
grenga2018dynamic}
\APACinsertmetastar {%
grenga2018dynamic}%
\begin{APACrefauthors}%
Grenga, T.%
, MacArt, J\BPBI F.%
\BCBL {}\ \BBA {} Mueller, M\BPBI E.%
\end{APACrefauthors}%
\unskip\
\newblock
\APACrefYearMonthDay{2018}{}{}.
\newblock
{\BBOQ}\APACrefatitle {Dynamic mode decomposition of a direct numerical
  simulation of a turbulent premixed planar jet flame: convergence of the
  modes} {Dynamic mode decomposition of a direct numerical simulation of a
  turbulent premixed planar jet flame: convergence of the modes}.{\BBCQ}
\newblock
\APACjournalVolNumPages{Combustion Theory and Modelling}{22}{4}{795--811}.
\PrintBackRefs{\CurrentBib}

\bibitem [\protect \citeauthoryear {%
Grenga%
\ \BBA {} Mueller%
}{%
Grenga%
\ \BBA {} Mueller%
}{%
{\protect \APACyear {2020}}%
}]{%
grenga_dmdbook}
\APACinsertmetastar {%
grenga_dmdbook}%
\begin{APACrefauthors}%
Grenga, T.%
\BCBT {}\ \BBA {} Mueller, M.%
\end{APACrefauthors}%
\unskip\
\newblock
\APACrefYearMonthDay{2020}{}{}.
\newblock
{\BBOQ}\APACrefatitle {Dynamic Mode Decomposition: A Tool to Extract Structures
  Hidden in Massive Datasets} {Dynamic mode decomposition: A tool to extract
  structures hidden in massive datasets}.{\BBCQ}
\newblock
\BIn{} \APACrefbtitle {Data Analysis for Direct Numerical Simulations of
  Turbulent Combustion} {Data analysis for direct numerical simulations of
  turbulent combustion}\ (\BPGS\ 157--176).
\newblock
\APACaddressPublisher{}{Springer}.
\PrintBackRefs{\CurrentBib}

\bibitem [\protect \citeauthoryear {%
Hamlington%
, Poludnenko%
\BCBL {}\ \BBA {} Oran%
}{%
Hamlington%
\ \protect \BOthers {.}}{%
{\protect \APACyear {2011}}%
}]{%
hamlington2011interactions}
\APACinsertmetastar {%
hamlington2011interactions}%
\begin{APACrefauthors}%
Hamlington, P\BPBI E.%
, Poludnenko, A\BPBI Y.%
\BCBL {}\ \BBA {} Oran, E\BPBI S.%
\end{APACrefauthors}%
\unskip\
\newblock
\APACrefYearMonthDay{2011}{}{}.
\newblock
{\BBOQ}\APACrefatitle {Interactions between turbulence and flames in premixed
  reacting flows} {Interactions between turbulence and flames in premixed
  reacting flows}.{\BBCQ}
\newblock
\APACjournalVolNumPages{Physics of Fluids}{23}{12}{125111}.
\PrintBackRefs{\CurrentBib}

\bibitem [\protect \citeauthoryear {%
He%
, Zhang%
, Ren%
\BCBL {}\ \BBA {} Sun%
}{%
He%
\ \protect \BOthers {.}}{%
{\protect \APACyear {2016}}%
}]{%
he2016deep}
\APACinsertmetastar {%
he2016deep}%
\begin{APACrefauthors}%
He, K.%
, Zhang, X.%
, Ren, S.%
\BCBL {}\ \BBA {} Sun, J.%
\end{APACrefauthors}%
\unskip\
\newblock
\APACrefYearMonthDay{2016}{}{}.
\newblock
{\BBOQ}\APACrefatitle {Deep residual learning for image recognition} {Deep
  residual learning for image recognition}.{\BBCQ}
\newblock
\BIn{} \APACrefbtitle {Proceedings of the IEEE conference on computer vision
  and pattern recognition} {Proceedings of the ieee conference on computer
  vision and pattern recognition}\ (\BPGS\ 770--778).
\PrintBackRefs{\CurrentBib}

\bibitem [\protect \citeauthoryear {%
Ihme%
}{%
Ihme%
}{%
{\protect \APACyear {2010}}%
}]{%
Ihme_NC}
\APACinsertmetastar {%
Ihme_NC}%
\begin{APACrefauthors}%
Ihme, M.%
\end{APACrefauthors}%
\unskip\
\newblock
\APACrefYearMonthDay{2010}{01}{}.
\newblock
{\BBOQ}\APACrefatitle {Topological optimization of artificial neural networks
  using a pattern search method} {Topological optimization of artificial neural
  networks using a pattern search method}.{\BBCQ}
\newblock
\BIn{} (\BPG~323-344).
\PrintBackRefs{\CurrentBib}

\bibitem [\protect \citeauthoryear {%
Ihme%
, Marsden%
\BCBL {}\ \BBA {} Pitsch%
}{%
Ihme%
\ \protect \BOthers {.}}{%
{\protect \APACyear {2008}}%
}]{%
ihme2008generation}
\APACinsertmetastar {%
ihme2008generation}%
\begin{APACrefauthors}%
Ihme, M.%
, Marsden, A\BPBI L.%
\BCBL {}\ \BBA {} Pitsch, H.%
\end{APACrefauthors}%
\unskip\
\newblock
\APACrefYearMonthDay{2008}{}{}.
\newblock
{\BBOQ}\APACrefatitle {Generation of optimal artificial neural networks using a
  pattern search algorithm: Application to approximation of chemical systems}
  {Generation of optimal artificial neural networks using a pattern search
  algorithm: Application to approximation of chemical systems}.{\BBCQ}
\newblock
\APACjournalVolNumPages{Neural Computation}{20}{2}{573--601}.
\PrintBackRefs{\CurrentBib}

\bibitem [\protect \citeauthoryear {%
Ihme%
, Pitsch%
\BCBL {}\ \BBA {} Bodony%
}{%
Ihme%
, Pitsch%
\BCBL {}\ \BBA {} Bodony%
}{%
{\protect \APACyear {2009}}%
}]{%
IHME20091545}
\APACinsertmetastar {%
IHME20091545}%
\begin{APACrefauthors}%
Ihme, M.%
, Pitsch, H.%
\BCBL {}\ \BBA {} Bodony, D.%
\end{APACrefauthors}%
\unskip\
\newblock
\APACrefYearMonthDay{2009}{}{}.
\newblock
{\BBOQ}\APACrefatitle {Radiation of noise in turbulent non-premixed flames}
  {Radiation of noise in turbulent non-premixed flames}.{\BBCQ}
\newblock
\APACjournalVolNumPages{Proceedings of the Combustion
  Institute}{32}{1}{1545--1553}.
\PrintBackRefs{\CurrentBib}

\bibitem [\protect \citeauthoryear {%
Ihme%
, Schmitt%
\BCBL {}\ \BBA {} Pitsch%
}{%
Ihme%
, Schmitt%
\BCBL {}\ \BBA {} Pitsch%
}{%
{\protect \APACyear {2009}}%
}]{%
ihme2009optimal}
\APACinsertmetastar {%
ihme2009optimal}%
\begin{APACrefauthors}%
Ihme, M.%
, Schmitt, C.%
\BCBL {}\ \BBA {} Pitsch, H.%
\end{APACrefauthors}%
\unskip\
\newblock
\APACrefYearMonthDay{2009}{}{}.
\newblock
{\BBOQ}\APACrefatitle {Optimal artificial neural networks and tabulation
  methods for chemistry representation in LES of a bluff-body swirl-stabilized
  flame} {Optimal artificial neural networks and tabulation methods for
  chemistry representation in les of a bluff-body swirl-stabilized
  flame}.{\BBCQ}
\newblock
\APACjournalVolNumPages{Proceedings of the Combustion
  Institute}{32}{1}{1527--1535}.
\PrintBackRefs{\CurrentBib}

\bibitem [\protect \citeauthoryear {%
Jolicoeur-Martineau%
}{%
Jolicoeur-Martineau%
}{%
{\protect \APACyear {2018}}%
}]{%
jolicoeur2018relativistic}
\APACinsertmetastar {%
jolicoeur2018relativistic}%
\begin{APACrefauthors}%
Jolicoeur-Martineau, A.%
\end{APACrefauthors}%
\unskip\
\newblock
\APACrefYearMonthDay{2018}{}{}.
\newblock
{\BBOQ}\APACrefatitle {The relativistic discriminator: a key element missing
  from standard GAN} {The relativistic discriminator: a key element missing
  from standard gan}.{\BBCQ}
\newblock
\APACjournalVolNumPages{arXiv preprint arXiv:1807.00734}{}{}{}.
\PrintBackRefs{\CurrentBib}

\bibitem [\protect \citeauthoryear {%
Kim%
, Kim%
, Won%
\BCBL {}\ \BBA {} Lee%
}{%
Kim%
\ \protect \BOthers {.}}{%
{\protect \APACyear {2021}}%
}]{%
Kim2021}
\APACinsertmetastar {%
Kim2021}%
\begin{APACrefauthors}%
Kim, H.%
, Kim, J.%
, Won, S.%
\BCBL {}\ \BBA {} Lee, C.%
\end{APACrefauthors}%
\unskip\
\newblock
\APACrefYearMonthDay{2021}{}{}.
\newblock
{\BBOQ}\APACrefatitle {Unsupervised deep learning for super-resolution
  reconstruction of turbulence} {Unsupervised deep learning for
  super-resolution reconstruction of turbulence}.{\BBCQ}
\newblock
\APACjournalVolNumPages{Journal of Fluid Mechanics}{910}{}{}.
\PrintBackRefs{\CurrentBib}

\bibitem [\protect \citeauthoryear {%
Kolla%
, Hawkes%
, Kerstein%
, Swaminathan%
\BCBL {}\ \BBA {} Chen%
}{%
Kolla%
\ \protect \BOthers {.}}{%
{\protect \APACyear {2014}}%
}]{%
kolla2014velocity}
\APACinsertmetastar {%
kolla2014velocity}%
\begin{APACrefauthors}%
Kolla, H.%
, Hawkes, E.%
, Kerstein, A.%
, Swaminathan, N.%
\BCBL {}\ \BBA {} Chen, J.%
\end{APACrefauthors}%
\unskip\
\newblock
\APACrefYearMonthDay{2014}{}{}.
\newblock
{\BBOQ}\APACrefatitle {On velocity and reactive scalar spectra in turbulent
  premixed flames} {On velocity and reactive scalar spectra in turbulent
  premixed flames}.{\BBCQ}
\newblock
\APACjournalVolNumPages{Journal of fluid mechanics}{754}{}{456--487}.
\PrintBackRefs{\CurrentBib}

\bibitem [\protect \citeauthoryear {%
Kong%
, Chang%
, Li%
\BCBL {}\ \BBA {} Chen%
}{%
Kong%
\ \protect \BOthers {.}}{%
{\protect \APACyear {2020}}%
}]{%
Kong2020}
\APACinsertmetastar {%
Kong2020}%
\begin{APACrefauthors}%
Kong, C.%
, Chang, J\BHBI T.%
, Li, Y\BHBI F.%
\BCBL {}\ \BBA {} Chen, R\BHBI Y.%
\end{APACrefauthors}%
\unskip\
\newblock
\APACrefYearMonthDay{2020}{}{}.
\newblock
{\BBOQ}\APACrefatitle {Deep learning methods for super-resolution
  reconstruction of temperature fields in a supersonic combustor} {Deep
  learning methods for super-resolution reconstruction of temperature fields in
  a supersonic combustor}.{\BBCQ}
\newblock
\APACjournalVolNumPages{AIP Advances}{10}{11}{115021}.
\PrintBackRefs{\CurrentBib}

\bibitem [\protect \citeauthoryear {%
Lapeyre%
, Misdariis%
, Cazard%
, Veynante%
\BCBL {}\ \BBA {} Poinsot%
}{%
Lapeyre%
\ \protect \BOthers {.}}{%
{\protect \APACyear {2019}}%
}]{%
lapeyre2019training}
\APACinsertmetastar {%
lapeyre2019training}%
\begin{APACrefauthors}%
Lapeyre, C\BPBI J.%
, Misdariis, A.%
, Cazard, N.%
, Veynante, D.%
\BCBL {}\ \BBA {} Poinsot, T.%
\end{APACrefauthors}%
\unskip\
\newblock
\APACrefYearMonthDay{2019}{}{}.
\newblock
{\BBOQ}\APACrefatitle {Training convolutional neural networks to estimate
  turbulent sub-grid scale reaction rates} {Training convolutional neural
  networks to estimate turbulent sub-grid scale reaction rates}.{\BBCQ}
\newblock
\APACjournalVolNumPages{Combustion and Flame}{203}{}{255--264}.
\PrintBackRefs{\CurrentBib}

\bibitem [\protect \citeauthoryear {%
Ledig%
\ \protect \BOthers {.}}{%
Ledig%
\ \protect \BOthers {.}}{%
{\protect \APACyear {2017}}%
}]{%
SRGAN}
\APACinsertmetastar {%
SRGAN}%
\begin{APACrefauthors}%
Ledig, C.%
, Theis, L.%
, Husz{\'a}r, F.%
, Caballero, J.%
, Cunningham, A.%
, Acosta, A.%
\BDBL {}others%
\end{APACrefauthors}%
\unskip\
\newblock
\APACrefYearMonthDay{2017}{}{}.
\newblock
{\BBOQ}\APACrefatitle {Photo-realistic single image super-resolution using a
  generative adversarial network} {Photo-realistic single image
  super-resolution using a generative adversarial network}.{\BBCQ}
\newblock
\BIn{} \APACrefbtitle {Proceedings of the IEEE conference on computer vision
  and pattern recognition} {Proceedings of the ieee conference on computer
  vision and pattern recognition}\ (\BPGS\ 4681--4690).
\PrintBackRefs{\CurrentBib}

\bibitem [\protect \citeauthoryear {%
Leith%
}{%
Leith%
}{%
{\protect \APACyear {1990}}%
}]{%
leith1990stochastic}
\APACinsertmetastar {%
leith1990stochastic}%
\begin{APACrefauthors}%
Leith, C.%
\end{APACrefauthors}%
\unskip\
\newblock
\APACrefYearMonthDay{1990}{}{}.
\newblock
{\BBOQ}\APACrefatitle {Stochastic backscatter in a subgrid-scale model: Plane
  shear mixing layer} {Stochastic backscatter in a subgrid-scale model: Plane
  shear mixing layer}.{\BBCQ}
\newblock
\APACjournalVolNumPages{Physics of Fluids A: Fluid Dynamics}{2}{3}{297--299}.
\PrintBackRefs{\CurrentBib}

\bibitem [\protect \citeauthoryear {%
Li%
\ \protect \BOthers {.}}{%
Li%
\ \protect \BOthers {.}}{%
{\protect \APACyear {2008}}%
}]{%
JHTDB_1}
\APACinsertmetastar {%
JHTDB_1}%
\begin{APACrefauthors}%
Li, Y.%
, Perlman, E.%
, Wan, M.%
, Yang, Y.%
, Meneveau, C.%
, Burns, R.%
\BDBL {}Eyink, G.%
\end{APACrefauthors}%
\unskip\
\newblock
\APACrefYearMonthDay{2008}{}{}.
\newblock
{\BBOQ}\APACrefatitle {A public turbulence database cluster and applications to
  study Lagrangian evolution of velocity increments in turbulence} {A public
  turbulence database cluster and applications to study lagrangian evolution of
  velocity increments in turbulence}.{\BBCQ}
\newblock
\APACjournalVolNumPages{Journal of Turbulence}{}{9}{N31}.
\PrintBackRefs{\CurrentBib}

\bibitem [\protect \citeauthoryear {%
Libby%
\ \BBA {} Bray%
}{%
Libby%
\ \BBA {} Bray%
}{%
{\protect \APACyear {1981}}%
}]{%
libby1981countergradient}
\APACinsertmetastar {%
libby1981countergradient}%
\begin{APACrefauthors}%
Libby, P\BPBI A.%
\BCBT {}\ \BBA {} Bray, K.%
\end{APACrefauthors}%
\unskip\
\newblock
\APACrefYearMonthDay{1981}{}{}.
\newblock
{\BBOQ}\APACrefatitle {Countergradient diffusion in premixed turbulent flames}
  {Countergradient diffusion in premixed turbulent flames}.{\BBCQ}
\newblock
\APACjournalVolNumPages{AIAA journal}{19}{2}{205--213}.
\PrintBackRefs{\CurrentBib}

\bibitem [\protect \citeauthoryear {%
Lilly%
}{%
Lilly%
}{%
{\protect \APACyear {1992}}%
}]{%
lilly1992proposed}
\APACinsertmetastar {%
lilly1992proposed}%
\begin{APACrefauthors}%
Lilly, D\BPBI K.%
\end{APACrefauthors}%
\unskip\
\newblock
\APACrefYearMonthDay{1992}{}{}.
\newblock
{\BBOQ}\APACrefatitle {A proposed modification of the Germano subgrid-scale
  closure method} {A proposed modification of the germano subgrid-scale closure
  method}.{\BBCQ}
\newblock
\APACjournalVolNumPages{Physics of Fluids A: Fluid Dynamics}{4}{3}{633--635}.
\PrintBackRefs{\CurrentBib}

\bibitem [\protect \citeauthoryear {%
Lipatnikov%
\ \BBA {} Chomiak%
}{%
Lipatnikov%
\ \BBA {} Chomiak%
}{%
{\protect \APACyear {2010}}%
}]{%
lipatnikov2010effects}
\APACinsertmetastar {%
lipatnikov2010effects}%
\begin{APACrefauthors}%
Lipatnikov, A.%
\BCBT {}\ \BBA {} Chomiak, J.%
\end{APACrefauthors}%
\unskip\
\newblock
\APACrefYearMonthDay{2010}{}{}.
\newblock
{\BBOQ}\APACrefatitle {Effects of premixed flames on turbulence and turbulent
  scalar transport} {Effects of premixed flames on turbulence and turbulent
  scalar transport}.{\BBCQ}
\newblock
\APACjournalVolNumPages{Progress in Energy and Combustion
  Science}{36}{1}{1--102}.
\PrintBackRefs{\CurrentBib}

\bibitem [\protect \citeauthoryear {%
Liu%
, Tang%
, Huang%
\BCBL {}\ \BBA {} Lu%
}{%
Liu%
\ \protect \BOthers {.}}{%
{\protect \APACyear {2020}}%
}]{%
Liu2020}
\APACinsertmetastar {%
Liu2020}%
\begin{APACrefauthors}%
Liu, B.%
, Tang, J.%
, Huang, H.%
\BCBL {}\ \BBA {} Lu, X\BHBI Y.%
\end{APACrefauthors}%
\unskip\
\newblock
\APACrefYearMonthDay{2020}{}{}.
\newblock
{\BBOQ}\APACrefatitle {Deep learning methods for super-resolution
  reconstruction of turbulent flows} {Deep learning methods for
  super-resolution reconstruction of turbulent flows}.{\BBCQ}
\newblock
\APACjournalVolNumPages{Physics of Fluids}{32}{2}{025105}.
\PrintBackRefs{\CurrentBib}

\bibitem [\protect \citeauthoryear {%
Lumley%
\ \BBA {} Newman%
}{%
Lumley%
\ \BBA {} Newman%
}{%
{\protect \APACyear {1977}}%
}]{%
lumley1977return}
\APACinsertmetastar {%
lumley1977return}%
\begin{APACrefauthors}%
Lumley, J\BPBI L.%
\BCBT {}\ \BBA {} Newman, G\BPBI R.%
\end{APACrefauthors}%
\unskip\
\newblock
\APACrefYearMonthDay{1977}{}{}.
\newblock
{\BBOQ}\APACrefatitle {The return to isotropy of homogeneous turbulence} {The
  return to isotropy of homogeneous turbulence}.{\BBCQ}
\newblock
\APACjournalVolNumPages{Journal of Fluid Mechanics}{82}{1}{161--178}.
\PrintBackRefs{\CurrentBib}

\bibitem [\protect \citeauthoryear {%
Maas%
, Hannun%
\BCBL {}\ \BBA {} Ng%
}{%
Maas%
\ \protect \BOthers {.}}{%
{\protect \APACyear {2013}}%
}]{%
Maas2013}
\APACinsertmetastar {%
Maas2013}%
\begin{APACrefauthors}%
Maas, A\BPBI L.%
, Hannun, A\BPBI Y.%
\BCBL {}\ \BBA {} Ng, A\BPBI Y.%
\end{APACrefauthors}%
\unskip\
\newblock
\APACrefYearMonthDay{2013}{}{}.
\newblock
\APACrefbtitle {Rectifier Nonlinearities Improve Neural Network Acoustic
  Models} {Rectifier nonlinearities improve neural network acoustic models}\
  \APACbVolEdTR{}{\BTR{}}.
\newblock
\APACaddressInstitution{}{Computer Science Department, Stanford University, CA
  94305 USA}.
\PrintBackRefs{\CurrentBib}

\bibitem [\protect \citeauthoryear {%
MacArt%
, Grenga%
\BCBL {}\ \BBA {} Mueller%
}{%
MacArt%
\ \protect \BOthers {.}}{%
{\protect \APACyear {2018}}%
}]{%
macart17a}
\APACinsertmetastar {%
macart17a}%
\begin{APACrefauthors}%
MacArt, J\BPBI F.%
, Grenga, T.%
\BCBL {}\ \BBA {} Mueller, M\BPBI E.%
\end{APACrefauthors}%
\unskip\
\newblock
\APACrefYearMonthDay{2018}{}{}.
\newblock
{\BBOQ}\APACrefatitle {Effects of combustion heat release on velocity and
  scalar statistics in turbulent premixed jet flames at low and high Karlovitz
  numbers} {Effects of combustion heat release on velocity and scalar
  statistics in turbulent premixed jet flames at low and high karlovitz
  numbers}.{\BBCQ}
\newblock
\APACjournalVolNumPages{Combustion and Flame}{191}{}{468--485}.
\PrintBackRefs{\CurrentBib}

\bibitem [\protect \citeauthoryear {%
MacArt%
, Grenga%
\BCBL {}\ \BBA {} Mueller%
}{%
MacArt%
\ \protect \BOthers {.}}{%
{\protect \APACyear {2019}}%
}]{%
macart17b}
\APACinsertmetastar {%
macart17b}%
\begin{APACrefauthors}%
MacArt, J\BPBI F.%
, Grenga, T.%
\BCBL {}\ \BBA {} Mueller, M\BPBI E.%
\end{APACrefauthors}%
\unskip\
\newblock
\APACrefYearMonthDay{2019}{}{}.
\newblock
{\BBOQ}\APACrefatitle {Evolution of flame-conditioned velocity statistics in
  turbulent premixed jet flames at low and high Karlovitz numbers} {Evolution
  of flame-conditioned velocity statistics in turbulent premixed jet flames at
  low and high karlovitz numbers}.{\BBCQ}
\newblock
\APACjournalVolNumPages{Proceedings of the Combustion
  Institute}{37}{2}{2503--2510}.
\PrintBackRefs{\CurrentBib}

\bibitem [\protect \citeauthoryear {%
MacArt%
\ \BBA {} Mueller%
}{%
MacArt%
\ \BBA {} Mueller%
}{%
{\protect \APACyear {2016}}%
}]{%
MacArt2016}
\APACinsertmetastar {%
MacArt2016}%
\begin{APACrefauthors}%
MacArt, J\BPBI F.%
\BCBT {}\ \BBA {} Mueller, M\BPBI E.%
\end{APACrefauthors}%
\unskip\
\newblock
\APACrefYearMonthDay{2016}{}{}.
\newblock
{\BBOQ}\APACrefatitle {Semi-implicit iterative methods for low Mach number
  turbulent reacting flows: Operator splitting versus approximate
  factorization} {Semi-implicit iterative methods for low mach number turbulent
  reacting flows: Operator splitting versus approximate factorization}.{\BBCQ}
\newblock
\APACjournalVolNumPages{Journal of Computational Physics}{326}{}{569--595}.
\PrintBackRefs{\CurrentBib}

\bibitem [\protect \citeauthoryear {%
Mason%
\ \BBA {} Thomson%
}{%
Mason%
\ \BBA {} Thomson%
}{%
{\protect \APACyear {1992}}%
}]{%
mason1992stochastic}
\APACinsertmetastar {%
mason1992stochastic}%
\begin{APACrefauthors}%
Mason, P\BPBI J.%
\BCBT {}\ \BBA {} Thomson, D\BPBI J.%
\end{APACrefauthors}%
\unskip\
\newblock
\APACrefYearMonthDay{1992}{}{}.
\newblock
{\BBOQ}\APACrefatitle {Stochastic backscatter in large-eddy simulations of
  boundary layers} {Stochastic backscatter in large-eddy simulations of
  boundary layers}.{\BBCQ}
\newblock
\APACjournalVolNumPages{Journal of Fluid Mechanics}{242}{}{51--78}.
\PrintBackRefs{\CurrentBib}

\bibitem [\protect \citeauthoryear {%
Moin%
, Squires%
, Cabot%
\BCBL {}\ \BBA {} Lee%
}{%
Moin%
\ \protect \BOthers {.}}{%
{\protect \APACyear {1991}}%
}]{%
moin1991dynamic}
\APACinsertmetastar {%
moin1991dynamic}%
\begin{APACrefauthors}%
Moin, P.%
, Squires, K.%
, Cabot, W.%
\BCBL {}\ \BBA {} Lee, S.%
\end{APACrefauthors}%
\unskip\
\newblock
\APACrefYearMonthDay{1991}{}{}.
\newblock
{\BBOQ}\APACrefatitle {A dynamic subgrid-scale model for compressible
  turbulence and scalar transport} {A dynamic subgrid-scale model for
  compressible turbulence and scalar transport}.{\BBCQ}
\newblock
\APACjournalVolNumPages{Physics of Fluids A: Fluid
  Dynamics}{3}{11}{2746--2757}.
\PrintBackRefs{\CurrentBib}

\bibitem [\protect \citeauthoryear {%
Nations%
}{%
Nations%
}{%
{\protect \APACyear {2021}}%
}]{%
UNgreen}
\APACinsertmetastar {%
UNgreen}%
\begin{APACrefauthors}%
Nations, U.%
\end{APACrefauthors}%
\unskip\
\newblock
\APACrefYearMonthDay{2021}{}{}.
\newblock
\APACrefbtitle {United Nations sustainable development goals.} {United nations
  sustainable development goals.}
\newblock
\begin{APACrefURL}
  [{2021-11-01}]\url{https://www.un.org/sustainabledevelopment/sustainable-development-goals}
  \end{APACrefURL}
\PrintBackRefs{\CurrentBib}

\bibitem [\protect \citeauthoryear {%
Z.~Nikolaou%
, Chrysostomou%
, Vervisch%
\BCBL {}\ \BBA {} Cant%
}{%
Z.~Nikolaou%
\ \protect \BOthers {.}}{%
{\protect \APACyear {2019}}%
}]{%
nikolaou2019progress}
\APACinsertmetastar {%
nikolaou2019progress}%
\begin{APACrefauthors}%
Nikolaou, Z.%
, Chrysostomou, C.%
, Vervisch, L.%
\BCBL {}\ \BBA {} Cant, S.%
\end{APACrefauthors}%
\unskip\
\newblock
\APACrefYearMonthDay{2019}{}{}.
\newblock
{\BBOQ}\APACrefatitle {Progress variable variance and filtered rate modelling
  using convolutional neural networks and flamelet methods} {Progress variable
  variance and filtered rate modelling using convolutional neural networks and
  flamelet methods}.{\BBCQ}
\newblock
\APACjournalVolNumPages{Flow, Turbulence and Combustion}{103}{2}{485--501}.
\PrintBackRefs{\CurrentBib}

\bibitem [\protect \citeauthoryear {%
Z\BPBI M.~Nikolaou%
, Chrysostomou%
, Vervisch%
\BCBL {}\ \BBA {} Cant%
}{%
Z\BPBI M.~Nikolaou%
\ \protect \BOthers {.}}{%
{\protect \APACyear {2018}}%
}]{%
nikolaou2018modelling}
\APACinsertmetastar {%
nikolaou2018modelling}%
\begin{APACrefauthors}%
Nikolaou, Z\BPBI M.%
, Chrysostomou, C.%
, Vervisch, L.%
\BCBL {}\ \BBA {} Cant, S.%
\end{APACrefauthors}%
\unskip\
\newblock
\APACrefYearMonthDay{2018}{}{}.
\newblock
{\BBOQ}\APACrefatitle {Modelling turbulent premixed flames using convolutional
  neural networks: application to sub-grid scale variance and filtered reaction
  rate} {Modelling turbulent premixed flames using convolutional neural
  networks: application to sub-grid scale variance and filtered reaction
  rate}.{\BBCQ}
\newblock
\APACjournalVolNumPages{arXiv preprint arXiv:1810.07944}{}{}{}.
\PrintBackRefs{\CurrentBib}

\bibitem [\protect \citeauthoryear {%
Nista%
\ \protect \BOthers {.}}{%
Nista%
\ \protect \BOthers {.}}{%
{\protect \APACyear {2021}}%
}]{%
nista2021turbulent}
\APACinsertmetastar {%
nista2021turbulent}%
\begin{APACrefauthors}%
Nista, L.%
, Schumann, C.%
, Grenga, T.%
, Karimi, A\BPBI N.%
, Scialabba, G.%
, Bode, M.%
\BDBL {}Pitsch, H.%
\end{APACrefauthors}%
\unskip\
\newblock
\APACrefYearMonthDay{2021}{}{}.
\newblock
{\BBOQ}\APACrefatitle {Turbulent mixing predictive model with physics-based
  Generative Adversarial Network} {Turbulent mixing predictive model with
  physics-based generative adversarial network}.{\BBCQ}
\newblock
\BIn{} \APACrefbtitle {10th European combustion meeting} {10th european
  combustion meeting}\ (\BPGS\ 460--465).
\PrintBackRefs{\CurrentBib}

\bibitem [\protect \citeauthoryear {%
O'Brien%
\ \protect \BOthers {.}}{%
O'Brien%
\ \protect \BOthers {.}}{%
{\protect \APACyear {2017}}%
}]{%
o2017cross}
\APACinsertmetastar {%
o2017cross}%
\begin{APACrefauthors}%
O'Brien, J.%
, Towery, C\BPBI A.%
, Hamlington, P\BPBI E.%
, Ihme, M.%
, Poludnenko, A\BPBI Y.%
\BCBL {}\ \BBA {} Urzay, J.%
\end{APACrefauthors}%
\unskip\
\newblock
\APACrefYearMonthDay{2017}{}{}.
\newblock
{\BBOQ}\APACrefatitle {The cross-scale physical-space transfer of kinetic
  energy in turbulent premixed flames} {The cross-scale physical-space transfer
  of kinetic energy in turbulent premixed flames}.{\BBCQ}
\newblock
\APACjournalVolNumPages{Proceedings of the Combustion
  Institute}{36}{2}{1967--1975}.
\PrintBackRefs{\CurrentBib}

\bibitem [\protect \citeauthoryear {%
O’Brien%
, Urzay%
, Ihme%
, Moin%
\BCBL {}\ \BBA {} Saghafian%
}{%
O’Brien%
\ \protect \BOthers {.}}{%
{\protect \APACyear {2014}}%
}]{%
OBrien2014}
\APACinsertmetastar {%
OBrien2014}%
\begin{APACrefauthors}%
O’Brien, J.%
, Urzay, J.%
, Ihme, M.%
, Moin, P.%
\BCBL {}\ \BBA {} Saghafian, A.%
\end{APACrefauthors}%
\unskip\
\newblock
\APACrefYearMonthDay{2014}{}{}.
\newblock
{\BBOQ}\APACrefatitle {Subgrid-scale backscatter in reacting and inert
  supersonic hydrogen--air turbulent mixing layers} {Subgrid-scale backscatter
  in reacting and inert supersonic hydrogen--air turbulent mixing
  layers}.{\BBCQ}
\newblock
\APACjournalVolNumPages{Journal of Fluid Mechanics}{743}{}{554--584}.
\PrintBackRefs{\CurrentBib}

\bibitem [\protect \citeauthoryear {%
Pant%
\ \BBA {} Farimani%
}{%
Pant%
\ \BBA {} Farimani%
}{%
{\protect \APACyear {2020}}%
}]{%
pant2020deep}
\APACinsertmetastar {%
pant2020deep}%
\begin{APACrefauthors}%
Pant, P.%
\BCBT {}\ \BBA {} Farimani, A\BPBI B.%
\end{APACrefauthors}%
\unskip\
\newblock
\APACrefYearMonthDay{2020}{}{}.
\newblock
{\BBOQ}\APACrefatitle {Deep Learning for Efficient Reconstruction of
  High-Resolution Turbulent DNS Data} {Deep learning for efficient
  reconstruction of high-resolution turbulent dns data}.{\BBCQ}
\newblock
\APACjournalVolNumPages{arXiv preprint arXiv:2010.11348}{}{}{}.
\PrintBackRefs{\CurrentBib}

\bibitem [\protect \citeauthoryear {%
Perlman%
, Burns%
, Li%
\BCBL {}\ \BBA {} Meneveau%
}{%
Perlman%
\ \protect \BOthers {.}}{%
{\protect \APACyear {2007}}%
}]{%
JHTDB_2}
\APACinsertmetastar {%
JHTDB_2}%
\begin{APACrefauthors}%
Perlman, E.%
, Burns, R.%
, Li, Y.%
\BCBL {}\ \BBA {} Meneveau, C.%
\end{APACrefauthors}%
\unskip\
\newblock
\APACrefYearMonthDay{2007}{}{}.
\newblock
{\BBOQ}\APACrefatitle {Data exploration of turbulence simulations using a
  database cluster} {Data exploration of turbulence simulations using a
  database cluster}.{\BBCQ}
\newblock
\BIn{} \APACrefbtitle {Proceedings of the 2007 ACM/IEEE Conference on
  Supercomputing} {Proceedings of the 2007 acm/ieee conference on
  supercomputing}\ (\BPGS\ 1--11).
\PrintBackRefs{\CurrentBib}

\bibitem [\protect \citeauthoryear {%
Pope%
}{%
Pope%
}{%
{\protect \APACyear {2000}}%
}]{%
pope2000turbulent}
\APACinsertmetastar {%
pope2000turbulent}%
\begin{APACrefauthors}%
Pope, S\BPBI B.%
\end{APACrefauthors}%
\unskip\
\newblock
\APACrefYear{2000}.
\newblock
\APACrefbtitle {Turbulent flows} {Turbulent flows}.
\newblock
\APACaddressPublisher{}{Cambridge university press}.
\PrintBackRefs{\CurrentBib}

\bibitem [\protect \citeauthoryear {%
Richardson%
}{%
Richardson%
}{%
{\protect \APACyear {2007}}%
}]{%
richardson2007weather}
\APACinsertmetastar {%
richardson2007weather}%
\begin{APACrefauthors}%
Richardson, L\BPBI F.%
\end{APACrefauthors}%
\unskip\
\newblock
\APACrefYear{2007}.
\newblock
\APACrefbtitle {Weather prediction by numerical process} {Weather prediction by
  numerical process}.
\newblock
\APACaddressPublisher{}{Cambridge university press}.
\PrintBackRefs{\CurrentBib}

\bibitem [\protect \citeauthoryear {%
Rogerson%
, Swaminathan%
, Tanahashi%
\BCBL {}\ \BBA {} Shiwaku%
}{%
Rogerson%
\ \protect \BOthers {.}}{%
{\protect \APACyear {2007}}%
}]{%
rogerson2007analysis}
\APACinsertmetastar {%
rogerson2007analysis}%
\begin{APACrefauthors}%
Rogerson, J.%
, Swaminathan, N.%
, Tanahashi, M.%
\BCBL {}\ \BBA {} Shiwaku, N.%
\end{APACrefauthors}%
\unskip\
\newblock
\APACrefYearMonthDay{2007}{}{}.
\newblock
{\BBOQ}\APACrefatitle {Analysis of progress variable variance equations using
  DNS data} {Analysis of progress variable variance equations using dns
  data}.{\BBCQ}
\newblock
\BIn{} \APACrefbtitle {Proceedings of the European Combustion Meeting.}
  {Proceedings of the european combustion meeting.}
\PrintBackRefs{\CurrentBib}

\bibitem [\protect \citeauthoryear {%
Rotunno%
\ \protect \BOthers {.}}{%
Rotunno%
\ \protect \BOthers {.}}{%
{\protect \APACyear {2009}}%
}]{%
rotunno2009large}
\APACinsertmetastar {%
rotunno2009large}%
\begin{APACrefauthors}%
Rotunno, R.%
, Chen, Y.%
, Wang, W.%
, Davis, C.%
, Dudhia, J.%
\BCBL {}\ \BBA {} Holland, G.%
\end{APACrefauthors}%
\unskip\
\newblock
\APACrefYearMonthDay{2009}{}{}.
\newblock
{\BBOQ}\APACrefatitle {Large-eddy simulation of an idealized tropical cyclone}
  {Large-eddy simulation of an idealized tropical cyclone}.{\BBCQ}
\newblock
\APACjournalVolNumPages{Bulletin of the American Meteorological
  Society}{90}{12}{1783--1788}.
\PrintBackRefs{\CurrentBib}

\bibitem [\protect \citeauthoryear {%
Schumann%
}{%
Schumann%
}{%
{\protect \APACyear {1995}}%
}]{%
schumann1995stochastic}
\APACinsertmetastar {%
schumann1995stochastic}%
\begin{APACrefauthors}%
Schumann, U.%
\end{APACrefauthors}%
\unskip\
\newblock
\APACrefYearMonthDay{1995}{}{}.
\newblock
{\BBOQ}\APACrefatitle {Stochastic backscatter of turbulence energy and scalar
  variance by random subgrid-scale fluxes} {Stochastic backscatter of
  turbulence energy and scalar variance by random subgrid-scale fluxes}.{\BBCQ}
\newblock
\APACjournalVolNumPages{Proceedings of the Royal Society of London. Series A:
  Mathematical and Physical Sciences}{451}{1941}{293--318}.
\PrintBackRefs{\CurrentBib}

\bibitem [\protect \citeauthoryear {%
Seltz%
, Domingo%
, Vervisch%
\BCBL {}\ \BBA {} Nikolaou%
}{%
Seltz%
\ \protect \BOthers {.}}{%
{\protect \APACyear {2019}}%
}]{%
seltz2019direct}
\APACinsertmetastar {%
seltz2019direct}%
\begin{APACrefauthors}%
Seltz, A.%
, Domingo, P.%
, Vervisch, L.%
\BCBL {}\ \BBA {} Nikolaou, Z\BPBI M.%
\end{APACrefauthors}%
\unskip\
\newblock
\APACrefYearMonthDay{2019}{}{}.
\newblock
{\BBOQ}\APACrefatitle {Direct mapping from LES resolved scales to
  filtered-flame generated manifolds using convolutional neural networks}
  {Direct mapping from les resolved scales to filtered-flame generated
  manifolds using convolutional neural networks}.{\BBCQ}
\newblock
\APACjournalVolNumPages{Combustion and Flame}{210}{}{71--82}.
\PrintBackRefs{\CurrentBib}

\bibitem [\protect \citeauthoryear {%
Sergeev%
\ \BBA {} Del~Balso%
}{%
Sergeev%
\ \BBA {} Del~Balso%
}{%
{\protect \APACyear {2018}}%
}]{%
sergeev2018horovod}
\APACinsertmetastar {%
sergeev2018horovod}%
\begin{APACrefauthors}%
Sergeev, A.%
\BCBT {}\ \BBA {} Del~Balso, M.%
\end{APACrefauthors}%
\unskip\
\newblock
\APACrefYearMonthDay{2018}{}{}.
\newblock
{\BBOQ}\APACrefatitle {Horovod: fast and easy distributed deep learning in
  TensorFlow} {Horovod: fast and easy distributed deep learning in
  tensorflow}.{\BBCQ}
\newblock
\APACjournalVolNumPages{arXiv preprint arXiv:1802.05799}{}{}{}.
\PrintBackRefs{\CurrentBib}

\bibitem [\protect \citeauthoryear {%
Steinberg%
, Driscoll%
\BCBL {}\ \BBA {} Swaminathan%
}{%
Steinberg%
\ \protect \BOthers {.}}{%
{\protect \APACyear {2012}}%
}]{%
steinberg2012statistics}
\APACinsertmetastar {%
steinberg2012statistics}%
\begin{APACrefauthors}%
Steinberg, A\BPBI M.%
, Driscoll, J\BPBI F.%
\BCBL {}\ \BBA {} Swaminathan, N.%
\end{APACrefauthors}%
\unskip\
\newblock
\APACrefYearMonthDay{2012}{}{}.
\newblock
{\BBOQ}\APACrefatitle {Statistics and dynamics of turbulence--flame alignment
  in premixed combustion} {Statistics and dynamics of turbulence--flame
  alignment in premixed combustion}.{\BBCQ}
\newblock
\APACjournalVolNumPages{Combustion and flame}{159}{8}{2576--2588}.
\PrintBackRefs{\CurrentBib}

\bibitem [\protect \citeauthoryear {%
Subramaniam%
, Wong%
, Borker%
, Nimmagadda%
\BCBL {}\ \BBA {} Lele%
}{%
Subramaniam%
\ \protect \BOthers {.}}{%
{\protect \APACyear {2020}}%
}]{%
subramaniam2020turbulence}
\APACinsertmetastar {%
subramaniam2020turbulence}%
\begin{APACrefauthors}%
Subramaniam, A.%
, Wong, M\BPBI L.%
, Borker, R\BPBI D.%
, Nimmagadda, S.%
\BCBL {}\ \BBA {} Lele, S\BPBI K.%
\end{APACrefauthors}%
\unskip\
\newblock
\APACrefYearMonthDay{2020}{}{}.
\newblock
{\BBOQ}\APACrefatitle {Turbulence enrichment using physics-informed generative
  adversarial networks} {Turbulence enrichment using physics-informed
  generative adversarial networks}.{\BBCQ}
\newblock
\APACjournalVolNumPages{arXiv preprint arXiv:2003.01907}{}{}{}.
\PrintBackRefs{\CurrentBib}

\bibitem [\protect \citeauthoryear {%
Veynante%
, Trouv{\'e}%
, Bray%
\BCBL {}\ \BBA {} Mantel%
}{%
Veynante%
\ \protect \BOthers {.}}{%
{\protect \APACyear {1997}}%
}]{%
veynante1997gradient}
\APACinsertmetastar {%
veynante1997gradient}%
\begin{APACrefauthors}%
Veynante, D.%
, Trouv{\'e}, A.%
, Bray, K.%
\BCBL {}\ \BBA {} Mantel, T.%
\end{APACrefauthors}%
\unskip\
\newblock
\APACrefYearMonthDay{1997}{}{}.
\newblock
{\BBOQ}\APACrefatitle {Gradient and counter-gradient scalar transport in
  turbulent premixed flames} {Gradient and counter-gradient scalar transport in
  turbulent premixed flames}.{\BBCQ}
\newblock
\APACjournalVolNumPages{Journal of Fluid Mechanics}{332}{}{263--293}.
\PrintBackRefs{\CurrentBib}

\bibitem [\protect \citeauthoryear {%
H.~Wang%
, Hawkes%
\BCBL {}\ \BBA {} Chen%
}{%
H.~Wang%
\ \protect \BOthers {.}}{%
{\protect \APACyear {2016}}%
}]{%
wang2016turbulence}
\APACinsertmetastar {%
wang2016turbulence}%
\begin{APACrefauthors}%
Wang, H.%
, Hawkes, E\BPBI R.%
\BCBL {}\ \BBA {} Chen, J\BPBI H.%
\end{APACrefauthors}%
\unskip\
\newblock
\APACrefYearMonthDay{2016}{}{}.
\newblock
{\BBOQ}\APACrefatitle {Turbulence-flame interactions in DNS of a laboratory
  high Karlovitz premixed turbulent jet flame} {Turbulence-flame interactions
  in dns of a laboratory high karlovitz premixed turbulent jet flame}.{\BBCQ}
\newblock
\APACjournalVolNumPages{Physics of Fluids}{28}{9}{095107}.
\PrintBackRefs{\CurrentBib}

\bibitem [\protect \citeauthoryear {%
X.~Wang%
\ \protect \BOthers {.}}{%
X.~Wang%
\ \protect \BOthers {.}}{%
{\protect \APACyear {2018}}%
}]{%
ESRGAN}
\APACinsertmetastar {%
ESRGAN}%
\begin{APACrefauthors}%
Wang, X.%
, Yu, K.%
, Wu, S.%
, Gu, J.%
, Liu, Y.%
, Dong, C.%
\BDBL {}Change~Loy, C.%
\end{APACrefauthors}%
\unskip\
\newblock
\APACrefYearMonthDay{2018}{}{}.
\newblock
{\BBOQ}\APACrefatitle {Esrgan: Enhanced super-resolution generative adversarial
  networks} {Esrgan: Enhanced super-resolution generative adversarial
  networks}.{\BBCQ}
\newblock
\BIn{} \APACrefbtitle {Proceedings of the European conference on computer
  vision (ECCV) workshops} {Proceedings of the european conference on computer
  vision (eccv) workshops}\ (\BPGS\ 0--0).
\PrintBackRefs{\CurrentBib}

\bibitem [\protect \citeauthoryear {%
Zhang%
\ \BBA {} Rutland%
}{%
Zhang%
\ \BBA {} Rutland%
}{%
{\protect \APACyear {1995}}%
}]{%
zhang1995premixed}
\APACinsertmetastar {%
zhang1995premixed}%
\begin{APACrefauthors}%
Zhang, S.%
\BCBT {}\ \BBA {} Rutland, C\BPBI J.%
\end{APACrefauthors}%
\unskip\
\newblock
\APACrefYearMonthDay{1995}{}{}.
\newblock
{\BBOQ}\APACrefatitle {Premixed flame effects on turbulence and
  pressure-related terms} {Premixed flame effects on turbulence and
  pressure-related terms}.{\BBCQ}
\newblock
\APACjournalVolNumPages{Combustion and Flame}{102}{4}{447--461}.
\PrintBackRefs{\CurrentBib}

\end{thebibliography}


\end{document}